\documentclass{aa} 
\usepackage{graphicx}
\usepackage{txfonts}
\usepackage{natbib,twoopt}
\usepackage[breaklinks=true]{hyperref} 
\bibpunct{(}{)}{;}{a}{}{,}             

\newcommand{\lya}{Ly$\alpha$}

\def\lesssim{\mathrel{\hbox{\rlap{\hbox{%
 \lower4pt\hbox{$\sim$}}}\hbox{$<$}}}}
\def\gtrsim{\mathrel{\hbox{\rlap{\hbox{%
 \lower4pt\hbox{$\sim$}}}\hbox{$>$}}}}

\def\arcsec{\hbox{$^{\prime\prime}$}}

\begin{document} 

   \title{Simulating the escaping atmospheres\\
          of hot gas planets in the solar neighborhood\thanks{
                Simulated atmospheres are available in tabulated
                form via CDS.}}
   \titlerunning{Simulating the escaping atmospheres of hot gas planets
                 in the solar neighborhood}

   \author{M. Salz\inst{1},
           S. Czesla\inst{1},
           P. C. Schneider\inst{2}$^,$\inst{1},
           J. H. M. M. Schmitt\inst{1}
          }
   \authorrunning {M. Salz et al.}

   \institute{Hamburger Sternwarte, Universit\"at Hamburg,
               Gojenbergsweg 112, 21029 Hamburg, Germany\\
              \email{msalz@hs.uni-hamburg.de}
         \and
              European Space Research and Technology Centre (ESA/ESTEC),
              Keplerlaan 1, 2201 AZ  Noordwijk, The Netherlands
             }

   \date{Submitted 16 March 2015 / Accepted 5 November 2015}

   \abstract{
Absorption of high-energy radiation in planetary thermospheres is generally believed to lead to the formation of planetary winds. The resulting mass-loss rates can affect the evolution, particularly of small gas planets.
We present 1D, spherically symmetric hydrodynamic simulations of the escaping atmospheres of 18 hot gas planets in the solar neighborhood. Our sample only includes strongly irradiated planets, whose expanded atmospheres may be detectable via transit spectroscopy using current instrumentation. The simulations were performed with the PLUTO-CLOUDY interface, which couples a detailed photoionization and plasma simulation code with a general MHD code. We study the thermospheric escape and derive improved estimates for the planetary mass-loss rates.
Our simulations reproduce the temperature-pressure profile measured via sodium D absorption in HD\,189733\,b, but show still unexplained differences in the case of HD\,209458\,b. In contrast to general assumptions, we find that the gravitationally more tightly bound thermospheres of massive and compact planets, such as HAT-P-2\,b are hydrodynamically stable. Compact planets dispose of the radiative energy input through hydrogen \lya{} and free-free emission. Radiative cooling is also important in HD\,189733\,b, but it decreases toward smaller planets  like GJ\,436\,b.
Computing the planetary \lya{} absorption and emission signals from the simulations, we find that the strong and cool winds of smaller planets mainly cause strong \lya{} absorption but little emission. Compact and massive planets with hot, stable thermospheres cause small absorption signals but are strong \lya{} emitters, possibly detectable with the current instrumentation. The absorption and emission signals provide a possible distinction between these two classes of thermospheres in hot gas planets. According to our results, WASP-80 and GJ 3470 are currently the most promising targets for observational follow-up aimed at detecting atmospheric \lya{} absorption signals. 
   }


   \maketitle

\section{Introduction}\label{SectIntro}

Gas giants are ubiquitous among the known extrasolar planets, because they produce rather strong transit signals that can be easily detected. Today, hot gas planets with semimajor axes smaller than 0.1~au constitute more than 15\,\% of all verified planets \citep[e.g., exoplanets.org,][]{Wright2011}. Such hot planets have been found as close as two stellar radii above the photosphere of their host stars \citep{Hebb2009} and some of them also orbit highly active host stars in close proximity \citep{Alonso2008, Schroeter2011}. This places their high-energy irradiation level as much as $10^5$ times higher than the irradiation of Earth. The detection of gas planets in such extreme environments immediately raises questions about the stability of their atmospheres, about possible evaporation, and about the resulting lifetime of these planets \citep[e.g.,][]{Schneider1998, Lammer2003, Lecavelier2007}.

In the case of close-in planets, the absorption of high-energy radiation causes temperatures of several thousand degrees in the planetary thermosphere\footnote{Hot atmospheric layers of a planet where (extreme) ultraviolet radiation is absorbed.}. The onset of the temperature rise at the base of the  thermosphere was confirmed in several observational studies of HD\,209458\,b and HD\,189733\,b \citep[][]{Ballester2007, Sing2008, Vidal2011, Huitson2012}. Even at 1~au the thermosphere of Earth is heated to up to 2000~K \citep[e.g.,][]{Banks1973}. Under these conditions the upper atmospheric layers of hot gas planets are probably unstable and expand continuously, unless there is a confining outer pressure \citep{Watson1981, Chassefiere1996}. 
The formation of such a hydrodynamic planetary wind is similar to the early solar wind models by \citet{Parker1958}, with the major difference that the energy source for the wind is the external irradiation. \citet{Watson1981} also showed that under certain circumstances irradiated atmospheres undergo an energy-limited escape, which simply means that the radiative energy input is completely used to drive the planetary wind. Naturally one has to account for the fact that the heating efficiency of the absorption is less than unity. Values ranging from 0.1 to 1.0 have been used for the heating efficiency in the literature. Recently, \citet{Shematovich2014} performed a detailed analysis of the atmosphere of HD\,209458\,b and found a height-dependent value between 0.1 and 0.25.

In the case of energy-limited escape the mass-loss rate is proportional to the XUV flux\footnote{XUV stands for X-ray ($\lambda<100$~\AA{}) plus extreme ultraviolet radiation (EUV, $100<\lambda<912$~\AA{}).} impinging on the atmosphere and inversely proportional to the planetary density \citep{Sanz2011}. Since the mass-loss rate only depends on the planetary density but not on its mass, small planets with low densities are most heavily affected by the fractional mass-loss rate. While for most detected hot gas planets the predicted mass-loss rates do not result in the loss of large portions of their total mass over their lifetime \citep{Ehrenreich2011}, a complete evaporation of volatile elements is possible for smaller gas planets. Indeed this might be the reason for the lack of detected hot mini-Neptunes \citep{Lecavelier2007, Carter2012}.

The existence of expanded thermospheres in hot gas planets was first confirmed in the exoplanet HD\,209458\,b, where \citet{Vidal2003} found absorption of up to 15\% in the line wings of the stellar hydrogen \lya{} line during the transit of the hot Jupiter. Since the optical transit depth in this system is only 1.5\% \citep{Henry2000, Charbonneau2000}, the excess absorption is most likely caused by a neutral hydrogen cloud around the planet. As \citeauthor{Vidal2003} also noted, the variability of solar-like stars is not sufficient to mimic this absorption signal. Although the \lya{} emission of the Sun is strongly inhomogeneously distributed over the solar disk (e.g., see images from The Multi-Spectral Solar Telescope Array, \href{http://solarscience.msfc.nasa.gov/MSSTA.shtml}{MSSTA}), this is only true for the line core, which is produced at temperatures of around 20\,000~K in the solar chromosphere \citep{Vernazza1973}. 
The line wings originate deeper within the chromosphere at around 6000 to 7000~K. Emission at these temperatures is more homogeneously distributed over the solar disk, as seen for example by images in H$\alpha$ or \ion{Ca}{ii} K filters -- lines that are produced at similar temperatures in the chromosphere. In analogy to the Sun, it is reasonable to assume that for solar-like stars the variability in the calcium line cores is also indicative of the variability in the wings of the \lya{} line, where the planetary absorption is detected.

The variability of stars in the \ion{Ca}{ii} H\&K emission lines has been studied in detail for many stars \citep[e.g.][]{Vaughan1981}. In particular, the inactive star HD\,209458 \citep{Henry2000} shows a fractional standard deviation of only 1.0\% over several years compared to the solar variation of 14.4\% \citep{Hall2007}, and the short term variability is on the same order \citep[using data from the California Planet Search program, ][]{Isaacson2010}. For this star the variability in the \lya{} line wings has been studied directly by \citet{Ben2007}, who also found a variability of about 1\% (with an uncertainty of 3\%) in their three out-of-transit time bins. Compared to the \lya{} absorption signal this is small, therefore, variability or spatial inhomogeneity of the stellar emission is unlikely to be the dominating source of the absorption signal seen during the planetary transit.

In the case of HD\,209458\,b the confidence in the detection of an expanded atmosphere was further increased by several transit observations, which revealed excess absorption in spectral lines of \ion{H}{i}, \ion{O}{I}, \ion{C}{II}, \ion{Si}{III}, and \ion{Mg}{I} \citep{Vidal2004, Ballester2007, Ehrenreich2008, Linsky2010, Jensen2012, Vidal2013}, although the detection of silicon was recently questioned \citep{Ballester2015}. There is also an indication of possible absorption in a \ion{Si}{IV} line during the transit of this planet \citep{Schlawin2010}.
Today, three additional systems exist with a comparable coverage of observations. In the more active host star HD\,189733 excess absorption was detected in \ion{H}{i}, \ion{O}{I}, and possibly in \ion{C}{II} \citep{Lecavelier2010,Lecavelier2012,Jensen2012,Ben2013} during the transit of the hot Jupiter. The superposition of several X-ray observations also indicates an excess transit depth in this system \citep{Poppenhaeger2013}. In WASP-12 three observations confirmed the existence of an expanded atmosphere \citep{Fossati2010, Haswell2012, Nichols2015}. In GJ\,436\,b a large cometary tail like hydrogen cloud was detected, absorbing as much as 50\% of the \lya{} emission of the host star in three individual transit observations \citep{Kulow2014, Ehrenreich2015}. In addition, a tentative detection of excess absorption was obtained in the grazing transit of the hot Jupiter 55\,Cancri\,b, while in the same system the transit of the super-Earth 55\,Cancri\,e yielded a non-detection of atmospheric excess absorption in the \lya{} line \citep{Ehrenreich2012}.

\section{Approach}\label{SectApproach}

Although observations show that an extended atmosphere exists in at least the four planets HD\,209458\,b, HD\,189733\,b, WASP-12\,b, and GJ\,436\,b there is no generally accepted theory for the observed absorption signals. The planetary atmospheres are affected by photoevaporation, by interactions with the stellar wind, by radiation pressure, and by planetary magnetic fields. All four processes also affect the absorption signals, for example, the observed signal of HD\,209458\,b has been explained by several models based on different physical assumptions \citep{Koskinen2013b, Bourrier2013, Tremblin2013, Trammell2014, Kislyakova2014}. In our view, one of the most promising ways to disentangle the different influences is to observe the planetary atmospheres in different systems with a wide parameter range \citep{Fossati2015}. Since all the processes depend on different parameters (e.g., \lya{} emission line strength, stellar wind density and velocity, magnetic field strength) the comparison of absorption signals from different systems can reveal the 
dominating processes in the atmospheres.

In this context, we identified the most promising targets in the search for yet undetected expanded atmospheres. Here, we present coupled radiative-hydrodynamical simulations of the planetary winds in 18 of these systems. Our simulations were performed with the PLUTO-CLOUDY interface \citep[][]{Salz2015b}. In contrast to simple energy-limited estimates, our simulations identify the best targets for future transit observations with high certainty.
The spherically symmetric simulations are comparable to several wind simulations in previous studies \citep[e.g.,][]{Yelle2004, Tian2005, Garcia2007, Penz2008-2, Murray2009, Koskinen2013a, Shaikhislamov2014}, but for the first time we applied them to possible future targets. Furthermore, our use of a detailed photoionization and plasma simulation code is the most important improvement compared to previous simulations. Additionally, we used a detailed reconstruction of the planetary irradiation levels in four spectral ranges for our simulations.

We first introduce our code and verify our assumptions (Sect.~\ref{SectSims}) and then compare the results for the different systems, explaining in detail the atmospheric structures (Sect.~\ref{SectResults}). We explain the transition from strong planetary winds in small planets to hydrodynamically stable thermospheres of compact, massive planets. Finally, we compute the \lya{} absorption and emission signals from the simulated atmospheres and investigate their detectability (Sect.~\ref{Sect:AbsSignals}).

\begin{table*}
\setlength{\tabcolsep}{4pt}
\small
\caption{System parameters of the complete sample (see Sect.~\ref{SectSample}). Sorted according to the strength of the simulated mass loss.}             
\label{tabSysPara}      
\centering
\begin{tabular}{l@{\hspace{-1pt}}l ccccccr@{}lc  l cc@{\hspace{3pt}}c@{\hspace{3pt}}ccccc }
\hline\hline\vspace{-5pt}\\
    &
    &
  \multicolumn{9}{c}{Host star} & 
    &
  \multicolumn{8}{c}{Planet} \\
  \vspace{-7pt}\\ \cline{3-11}\cline{13-20} \vspace{-5pt}\\
  System & 
    &
  Sp. type & 
  $T_{\mathrm{eff}}$ & 
  $V$ & 
  $B\!-\!V$ & 
  $J\!-\!K$ &
  $d$ & 
  \multicolumn{2}{c}{$P_{\mathrm{rot}}$} & 
  Age  &
    &
  $R_{\mathrm{pl}}$ & 
  $M_{\mathrm{pl}}$ &  
  $\rho_{\mathrm{pl}}$  &  
  $T_{\mathrm{eq}}$  & 
  $P_{\mathrm{orb}}$  & 
  $a$  & 
  $e$  &  
  TD \\ 
  \vspace{-9pt}\\
    &
    & 
    & 
  (K)  & 
  (mag) & 
  (mag) & 
  (mag) & 
  (pc) & 
  \multicolumn{2}{c}{(d)}  & 
  (Ga)  &
    &
  $(R_{\mathrm{jup}})$ & 
  $(M_{\mathrm{jup}})$ &  
   (g\,cm$^{-3}$) & 
   (K) &  
   (d) & 
  (au) &
    & 
  (\%)  \\
  \vspace{-7pt}\\ \hline\vspace{-5pt}\\ 
  \object{WASP-12}      &&  G0V   & 6300 &             11.6  &  0.57 &  0.29 &            380 &  37.5\,&            &             13.2  &&  1.8\hphantom{5}  &             1.4\hphantom{11} &  \hphantom{1}0.32             &            2900 & \hphantom{1}1.1 & 0.023 & 0.05  &   1.4\hphantom{15}  \\
  \vspace{-8pt}\\
  \object{GJ 3470}      &&  M1.5  & 3600 &             12.3  &  1.17 &  0.81 & \hphantom{1}29 &  20.7\,&$\pm\, 0.2$ &  \hphantom{1}1.2  &&  0.37             &             0.044            &  \hphantom{1}1.1\hphantom{3}  & \hphantom{1}650 & \hphantom{1}3.3 & 0.036 &    0  &  0.57\hphantom{1} \\
  \vspace{-8pt}\\
  \object{WASP-80}      && K7-M0V & 4150 &             11.9  &  0.94 &  0.87 & \hphantom{1}60 &  8.1\,&$\pm\, 0.8$  &  \hphantom{1}0.2  &&  0.95             & \hphantom{1}0.55\hphantom{5} &  \hphantom{1}0.73             & \hphantom{1}800 & \hphantom{1}3.1 & 0.034 &    0  &   2.9\hphantom{15} \\
  \vspace{-8pt}\\
  \object{HD 149026}    &&  G0IV  & 6150 &  \hphantom{1}8.1  &  0.61 &  ---  & \hphantom{1}79 &  11.5\,&            &  \hphantom{1}1.2  &&  0.65             &             0.36\hphantom{5} &  \hphantom{1}1.6\hphantom{3}  &            1440 & \hphantom{1}2.9 & 0.043 &    0  &  0.29\hphantom{1}  \\
  \vspace{-8pt}\\
  \object{HAT-P-11}     &&  K4V   & 4750 &  \hphantom{1}9.5  &  1.19 &  0.60 & \hphantom{1}37 &  29.2\,&            &  \hphantom{1}2.9  &&  0.42             &             0.083            &  \hphantom{1}1.3\hphantom{3}  & \hphantom{1}850 & \hphantom{1}4.9 & 0.053 &  0.2  &  0.33\hphantom{1} \\
  \vspace{-8pt}\\
  \object{HD 209458}    &&  G0V   & 6065 &  \hphantom{1}7.6  &  0.58 &  0.28 & \hphantom{1}50 &  11.4\,&            &  \hphantom{1}1.5  &&  1.4\hphantom{5}  & \hphantom{1}0.69\hphantom{5} &  \hphantom{1}0.34             &            1320 & \hphantom{1}3.5 & 0.047 &    0  &   1.5\hphantom{15}  \\
  \vspace{-8pt}\\
  \object{55 Cnc} (e)   && K0IV-V & 5200 &  \hphantom{1}6.0  &  0.87 &  0.58 & \hphantom{1}12 &  42.7\,&$\pm\, 2.5$ &  \hphantom{1}6.7  &&  0.19             &             0.026            &  \hphantom{1}4.2\hphantom{3}  &            1950 & \hphantom{1}0.7 & 0.015 &    0  & 0.045 \\
  \vspace{-8pt}\\
  \object{GJ\,1214}     &&  M4.5  & 3050 &             14.7  &  1.73 &  0.97 & \hphantom{1}13 &  44.3\,&$\pm\, 1.2$ &  \hphantom{1}3.4  &&  0.24             &             0.020            &  \hphantom{1}1.9\hphantom{3}  & \hphantom{1}550 & \hphantom{1}1.6 & 0.014 &    0  &   1.3\hphantom{15}  \\
  \vspace{-8pt}\\                                              
  \object{GJ 436}       && M2.5V  & 3350 &             10.6  &  1.47 &  0.83 & \hphantom{1}10 &  56.5\,&            &  \hphantom{1}6.5  &&  0.38             &             0.073            &  \hphantom{1}1.7\hphantom{3}  & \hphantom{1}650 & \hphantom{1}2.6 & 0.029 &  0.2  &  0.70\hphantom{1}   \\
  \vspace{-8pt}\\
  \object{HD\,189733}   && K0-2V  & 5040 &  \hphantom{1}7.6  &  0.93 &  0.53 & \hphantom{1}19 &  12.0\,&$\pm\, 0.1$ &  \hphantom{1}0.7  &&  1.1\hphantom{5}  &             1.1\hphantom{15} &  \hphantom{1}0.96             &            1200 & \hphantom{1}2.2 & 0.031 &    0  &   2.4\hphantom{15} \\
  \vspace{-8pt}\\
  \object{HD 97658}     &&  K1V   & 5100 &  \hphantom{1}7.7  &  0.86 &  0.47 & \hphantom{1}21 &  38.5\,&$\pm\, 1.0$ &  \hphantom{1}6.6  &&  0.21             &             0.025            &  \hphantom{1}3.4\hphantom{3}  & \hphantom{1}750 & \hphantom{1}9.5 & 0.080 & 0.06  &  0.085 \\
  \vspace{-8pt}\\
  \object{WASP-77}      &&  G8V   & 5500 &             10.3  &  0.75 &  0.37 & \hphantom{1}93 & 15.4\,&$\pm\, 0.4$  &  \hphantom{1}1.7  &&  1.2\hphantom{5}  & \hphantom{1}1.8\hphantom{15} &  \hphantom{1}1.3\hphantom{3}  &            1650 & \hphantom{1}1.4 & 0.024 &    0  &   1.7\hphantom{15} \\
  \vspace{-8pt}\\
  \object{WASP-43}      &&  K7V   & 4400 &             12.4  &  1.00 &  0.73 & \hphantom{1}80 & 15.6\,&$\pm\, 0.4$  &  \hphantom{1}0.8  &&  0.93             & \hphantom{1}1.8\hphantom{15} &  \hphantom{1}2.9\hphantom{3}  &            1350 & \hphantom{1}0.8 & 0.014 &    0  &   2.6\hphantom{15} \\
  \vspace{-8pt}\\
  \object{CoRoT-2}      &&  G7V   & 5650 &             12.6  &  0.85 &  0.47 &            270 &   4.5\,&$\pm\, 0.1$ &  \hphantom{1}0.1  &&  1.5\hphantom{5}  &             3.3\hphantom{15} &  \hphantom{1}1.5\hphantom{3}  &            1550 & \hphantom{1}1.7 & 0.028 & 0.01  &  2.8\hphantom{15}  \\
  \vspace{-8pt}\\
  \object{WASP-8}       &&  G8V   & 5600 &  \hphantom{1}9.9  &  0.82 &  0.41 & \hphantom{1}87 & 16.4\,&$\pm\, 1.0$  &  \hphantom{1}1.6  &&  1.0\hphantom{5}  & \hphantom{1}2.2\hphantom{15} &  \hphantom{1}2.6\hphantom{3}  & \hphantom{1}950 & \hphantom{1}8.2 & 0.080 &  0.3  &   1.3\hphantom{15}   \\
  \vspace{-8pt}\\
  \object{WASP-10}      &&  K5V   & 4700 &             12.7  &  1.15 &  0.62 & \hphantom{1}90 & 11.9\,&$\pm\, 0.9$  &  \hphantom{1}0.6  &&  1.1\hphantom{5}  & \hphantom{1}3.2\hphantom{15} &  \hphantom{1}3.1\hphantom{3}  & \hphantom{1}950 & \hphantom{1}3.1 & 0.038 & 0.05  &   2.5\hphantom{15} \\
  \vspace{-8pt}\\
  \object{HAT-P-2}      &&  F8V   & 6300 &  \hphantom{1}8.7  &  0.46 &  0.19 &            114 &  3.7\,&$\pm\, 0.4$  &  \hphantom{1}0.4  &&  1.2\hphantom{5}  & \hphantom{1}8.9\hphantom{15} &  \hphantom{1}7.3\hphantom{3}  &            1700 & \hphantom{1}5.6 & 0.068 &  0.5  &  0.52\hphantom{1} \\ 
  \vspace{-8pt}\\
  \object{HAT-P-20}     &&  K3V   & 4600 &             11.3  &  0.99 &  0.67 & \hphantom{1}70 & 14.6\,&$\pm\, 0.9$  &  \hphantom{1}0.8  &&  0.87             & \hphantom{1}7.3\hphantom{15} &             13.8\hphantom{3}  & \hphantom{1}950 & \hphantom{1}2.9 & 0.036 & 0.01  &   1.6\hphantom{15}\\
  \vspace{-8pt}\\
  \object{WASP-38}      &&  F8V   & 6200 &  \hphantom{1}9.4  &  0.48 &  0.29 &            110 &  7.5\,&$\pm\, 1.0$  &  \hphantom{1}1.0  &&  1.1\hphantom{5}  & \hphantom{1}2.7\hphantom{15} &  \hphantom{1}2.1\hphantom{3}  &            1250 & \hphantom{1}6.9 & 0.076 & 0.03  &  0.69\hphantom{1} \\
  \vspace{-8pt}\\
  \object{WASP-18}      && F6IV-V & 6400 &  \hphantom{1}9.3  &  0.44 &  0.28 & \hphantom{1}99 &  5.0\,&$\pm\, 1.0$  &  \hphantom{1}0.7  &&  1.3\hphantom{5}  &            10.2\hphantom{15} &             10.3\hphantom{3}  &            2400 & \hphantom{1}0.9 & 0.020 & 0.01  &  0.92\hphantom{1} \\
  \vspace{-8pt}\\
  \object{55 Cnc} (b)   && K0IV-V & 5200 &  \hphantom{1}6.0  &  0.87 &  0.58 & \hphantom{1}12 &  42.7\,&$\pm\, 2.5$ &  \hphantom{1}6.7  &&   ---             &             0.80\hphantom{5} &  ---                          & \hphantom{1}700 &            14.6 & 0.113 &    0  &  ---  \\
  \vspace{-7pt}\\ \hline                  
\end{tabular}                                                                                                                                            
\tablefoot{Explanation of the columns: 
           name of the system,
           spectral type, 
           effective temperature,
           visual magnitude, 
           colors (SIMBAD),
           distance, 
           stellar rotation period,
           gyrochronological age according to \citet{Brown2014} or mean of the three color based age estimates from the same author,
           planetary radius, mass, density, 
           equilibrium temperature or where cited 
           average dayside brightness temperature, 
           orbital period, semimajor axis,
           orbit eccentricity, and transit depth.   
           }
\tablebib{
           The data were compiled using exoplanets.org \citep{Wright2011}                                                                      
           and the following publications: 
           HAT-P-2: \citet{Bakos2007, Hipparcos2007, Pal2010}, $P_{\mathrm{rot}}$ from v\,$\sin i$, $T_{\mathrm{eq}}$ varies due to eccentricity (1250 to 2150~K);
           WASP-38: \citet{Barros2011, Brown2012}; 
           WASP-77: \citet{Maxted2013};                                                                                                    
           WASP-10: \citet{Christian2009, Johnson2009, Smith2009}, $P_{\mathrm{rot}}$ from \citep{Salz2015a};
           HAT-P-20: \citet{Bakos2011}, $P_{\mathrm{rot}}$ from \citep{Salz2015a};                                                                    
           WASP-8: \citet{Queloz2010, Cubillos2012}, $P_{\mathrm{rot}}$ from \citep{Salz2015a};
           WASP-80: \citet{Triaud2013}, $P_{\mathrm{rot}}$ from v\,$\sin i$;                                                            
           WASP-43: \citet{Hellier2011};              
           WASP-18: \citet{Hellier2009, Pillitteri2014}, $P_{\mathrm{rot}}$ from v\,$\sin i$;
           HD\,209458:\citet{Charbonneau2000, Henry2000, Torres2008, SilvaValio2008}, dayside brightness temperature from Spitzer observation \citep{Crossfield2012}; 
           HD\,189733: \citet{Bouchy2005, Henry2008, Southworth2010} , dayside brightness temperature from Spitzer observation \citep{Knutson2007}; 
           GJ\,1214: \citet{Charbonneau2009, Berta2011, Narita2013};
           GJ\,3470: \citet{Bonfils2012, Biddle2014};
           GJ\,436: \citet{Butler2004, Knutson2011};  
           55\,Cnc: \citet{Butler1997, McArthur2004, Gray2003, Fischer2008};
           HAT-P-11: \citet{Bakos2010};  
           HD\,149026: \citet{Sato2005}, $P_{\mathrm{rot}}$ from v\,$\sin i$, dayside brightness temperature from Spitzer observation \citep{Knutson2009};
           HD\,97658: \citet{Howard2011, Henry2011};  
           WASP-12: \citet{Hebb2009}, $P_{\mathrm{rot}}$ upper limit from v\,$\sin i$, dayside brightness temperature from HST WFC3 observation \citep{2013Swain};
           CoRoT-2: \citet{Alonso2008, Lanza2009, Schroeter2011}.
           }                                                                                                                                     
\end{table*}

\subsection{Target selection}\label{SectSample}

Our target selection has been described in \citet{Salz2015a}. Basically we select targets where expanded atmospheres can be detected by \lya{} transit spectroscopy. We predict the \lya{} luminosity of all host stars based on their X-ray luminosity, and if not available, based on spectral type and stellar rotation period using the relations presented by \citet{Linsky2013}. We further only select transiting systems with an orbital separation $<$\,0.1~au to ensure high levels of irradiation. Note that this is not necessarily a boundary for hydrodynamic escape. Earth-sized planets can harbor escaping atmospheres at even smaller irradiation levels, because their atmospheres are relatively weakly bound as a result of the low planetary mass; this was already studied by \citet{Watson1981} for Earth and Venus.
While we selected planets with a transit depth $>$\,0.5\% in \citet{Salz2015a}, we drop this criterion here, because recent observations indicate that also small planets can produce large absorption signals \citep{Kulow2014}. This extends our sample further toward planets with smaller masses and increases the parameter space.

Finally we limit the sample by selecting planets with a \lya{} flux stronger than 1/5 of the reconstructed flux of HD\,209458\,b. Considering the remaining uncertainties of the scaling relations and the interstellar absorption, these targets are potentially bright enough for transit spectroscopy.

In total we find 18 suitable targets (see Table~\ref{tabSysPara}~\&~\ref{tabSim}), all located within 120~pc, since the stellar distance has a strong impact on the detectability. This probably places the targets within the Local Bubble of hot ionized interstellar material \citep{Redfield2008}, limiting the uncertainty introduced by interstellar absorption. As expected, systems with detected absorption signals (HD\,189733, HD\,209458, GJ\,436) also lead the ranking according to our estimates. 55\,Cnc\,e is also among the top targets, but for this planet an expanded atmosphere was not detected \citep{Ehrenreich2012}. We add three further targets to the list, which do not strictly fulfill our selection criteria: WASP-12 is too distant for \lya{} transit spectroscopy, but excess absorption was measured in metal lines \citep{Fossati2010}. The distance of CoRoT-2 is too large, but the host star is extremely active \citep{Schroeter2011}, which makes the system interesting. 55\,Cnc\,b does not transit its host star, but a grazing transit of an expanded atmosphere has been proposed \citep{Ehrenreich2012}.

This finally leaves us with 21 planets in 20 systems. We present simulations for 18 of these targets. We did not simulate the atmospheres of WASP-38\,b and WASP-18\,b, because they host stable thermospheres (see Sect.~\ref{SectJeans}). 55\,Cnc\,b was not simulated, because the radius is unknown in this non-transiting planet. Planets similar to 55\,Cnc\,b show densities ranging from 0.1 to 2.4~g\,cm$^{-3}$ \citep[WASP-59\,b, Kepler-435\,b;][]{Hebrard2, Almenara2015}. Reasonable values for the radius of 55\,Cnc\,b range from 0.8 to 2.0 Jupiter-radii, which has a major impact on the planets atmosphere. Therefore, this system should be studied with a simulation grid in a dedicated study.

\subsection{Reconstruction of the XUV flux and the SED}\label{SectSED}

One of the main input parameters for our simulations is the irradiation strength and the spectral energy distribution (SED) of the host stars. Therefore, we first explain our reconstruction of the stellar SED, before going into details about the simulations.

\begin{figure}[t!]
  \centering
  \includegraphics[width=\hsize]{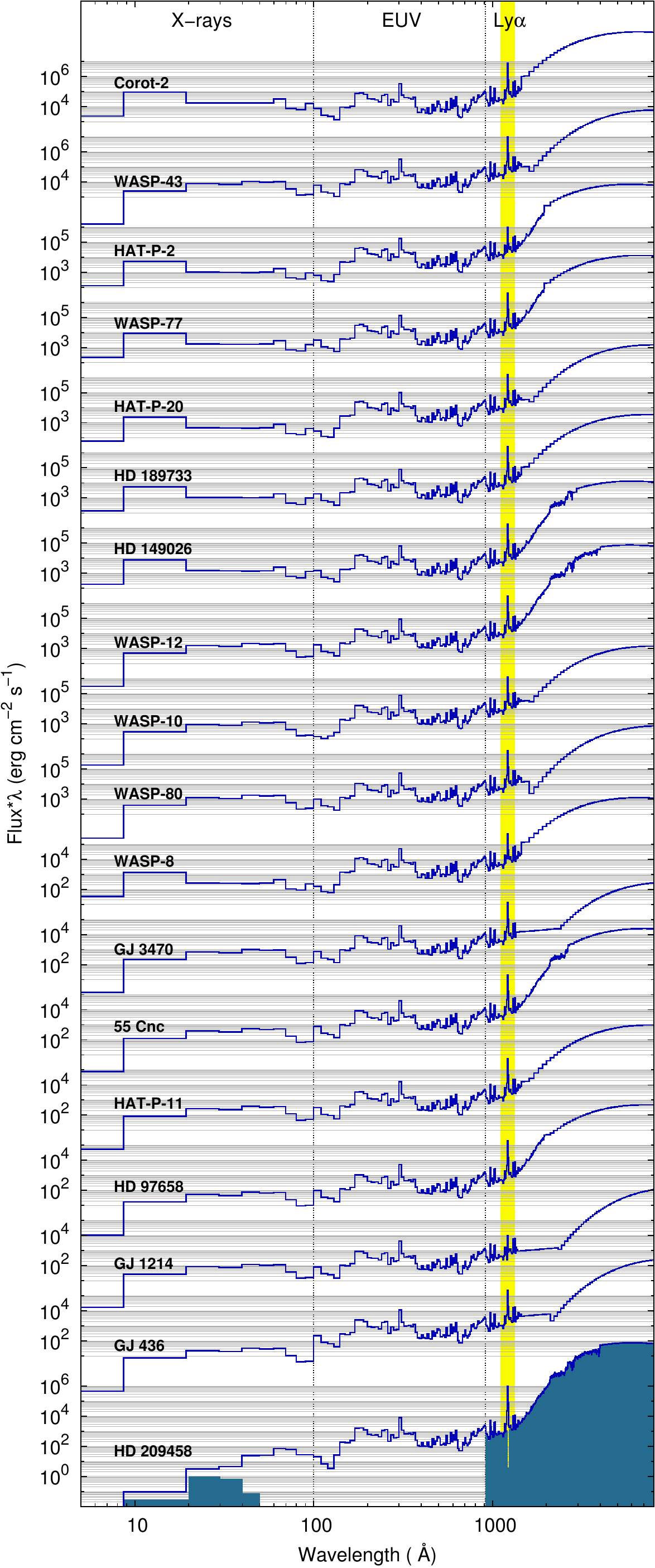}
  \caption{Reconstructed stellar flux sorted from top to
           bottom according to the
           irradiation level at the planetary distance.
           HD\,209458\,b is plotted separately at the bottom, because we also
           show the radiation that is transmitted through the simulated
           thermosphere (shaded region).
           The shape of the SEDs is similar, but the XUV irradiation strength drops by more than two orders of magnitude
           from CoRoT-2 to GJ\,436.
           }
  \label{fig:spec}
\end{figure}

Clearly, the strength of a planetary wind crucially depends on the available energy, which is mostly supplied by the EUV emission of the host stars, but X-rays also contribute a large percentage of the high-energy emission of active host stars like CoRoT-2 \citep{Schroeter2011}. Interstellar absorption extinguishes EUV emission even from close-by targets so that observations cannot be used to determine the irradiation level in this spectral range. 
The upper atmosphere, which is simulated here, is mostly translucent to FUV, NUV, and optical emission, except for line absorption as shown by the observations (see Sect.~\ref{SectIntro}). The simulations presented here only include hydrogen and helium (H+He), and any absorption beyond the \lya{} line is negligible in terms of radiative heating. At the bottom of Fig.~\ref{fig:spec}, we anticipate the spectrum of radiation that has passed through the simulated atmosphere of HD\,209458\,b to justify this statement. This radiation would be absorbed in lower atmospheric layers or close to the planetary photosphere.
The atmospheric absorption from 10 to 912~\AA{} is clear as well as the transmission of the optical emission, and furthermore the complete extinction of the \lya{} line.

The reconstruction of the SED must be sufficiently accurate in the spectral range up to and including the \lya{} line. For this purpose, we developed a piecewise reconstruction, which is mainly based on the prediction of the EUV luminosity of late type dwarfs by \citet{Linsky2014}. Note that several methods exist to predict the EUV luminosity of stars \citep{Salz2015a}. The results of these methods differ by up to one order of magnitude for active stars, which directly translates into the uncertainty of the mass-loss rates in the simulations (see Sect.~\ref{SectUncertainties}).

The reconstruction of the SEDs is split into four spectral regions and for each range we need a spectral shape and a luminosity to assemble the full SED:
\begin{enumerate}
\item
For the X-ray spectrum (0 to 100~\AA{}) we use a 2 or $4\times10^6$~K plasma emission model from CHIANTI \citep{Dere1997, Dere2009} for inactive or active host stars respectively (active: $L_{\mathrm{X}}>10^{28}$erg\,s$^{-1}$). The SED is normalized to the observed X-ray luminosities of the host stars. Only if observations are unavailable, we use an estimate based on the stellar rotation period and the stellar mass \citep{Pizzolato2003}. The models are computed with a low resolution (bin width =~10~\AA{}) adapted to the resolution in the EUV range. Since the absorption of X-rays causes ionizations in the planetary atmospheres, resolving emission lines is irrelevant in the X-ray regime and the low resolution is sufficient.

\item
For the hydrogen \lya{} emission line of the host stars a Gaussian with a full width at half maximum (FWHM) of 9.4~\AA{} is used. The irradiation strength is normalized according to the host star's total \lya{} luminosity. Only for four host stars has the luminosity been reconstructed based on HST observations (see references in Table~\ref{tabSim}). For most of the targets the \lya{} luminosity is predicted based on the X-ray luminosity following \citet{Linsky2013}.
The line width is adapted to the resolution of our photoionization solver, 
which is 6.1~\AA{} at the \lya{} line. Our approach places about 90\% of the line flux in the spectral bin at the \lya{} line center.
The line width is about a factor of 10 broader than usual \lya{} stellar emission lines \citep[compare][]{Wood2005}.
Based on test simulations, we confirm that this only affects the strength of the radiation pressure significantly (see Sect.~\ref{SectRadPres}).

\item
For the EUV range (100~\AA{} to $\sim$\,912~\AA{}) the luminosity of the host stars is predicted based on the \lya{} luminosity given by step~2 \citep{Linsky2014}. For most targets this procedure reverts back to an X-ray based estimate, because the \lya{} luminosity is based on the X-ray luminosity. If a measurement of the \lya{} luminosity is available, we obtain two predictions for the EUV luminosities (X-ray and \lya{} based) and use the average. The shape of the SED is taken from an active or inactive solar spectrum \citep{Woods2002}, depending on the activity of the host star defined in step 1.
This shape is continued to a connecting point with the photospheric blackbody, where we use a visual best fit for the connection point between 1500 to 4000~\AA{}. Certainly the SED of the host stars will differ from the solar type emission in the EUV range, however, only the relative strength of broader ranges is relevant, because the radiation causes ionizations and the presence of spectral lines is unimportant (except for the \ion{He}{ii} \lya{} see below).

\item
For the remaining part of the spectrum up to $5\times10^4$~\AA{} we choose a blackbody according to the host star's effective temperature. This range does not affect the simulations, but can be used to check absorption in the hydrogen Balmer lines for example.
\end{enumerate}

Our approach does not include a prediction of the \ion{He}{ii} \lya{} line (304~\AA{}), which accounts for 20\% of the EUV emission in the solar spectrum. The use of the solar SED shape introduces another error at this point (private communication Linsky). While a more detailed reconstruction of the SEDs is conceivable, it certainly has a smaller effect on the results than the uncertainty on the overall EUV luminosity of the host stars.

Figure\,\ref{fig:spec} shows the reconstructed SEDs for the host stars and the derived luminosities for each spectral range are given in Table.~\ref{tabSim} alongside the results from our simulations. The total XUV irradiation is a factor of 200 stronger for CoRoT-2\,b than in GJ\,436\,b. In these two systems X-rays contribute 60\% and 30\% of the total energy of hydrogen ionizing radiation. Also the relative \lya{} emission line strength varies strongly for the individual targets. HD\,209458 shows powerful \lya{} emission given the inactive state of the host star. In contrast, the upper limit for GJ\,1214 derived by \citet{France2013} is extremely low.

\section{Simulations}\label{SectSims}

We simulate the hydrodynamically expanding atmospheres of 18 hot gas planets in spherically symmetric, 1D simulations. Absorption and emission, and the corresponding heating and cooling is solved self-consistently using a photoionization and plasma simulation code, which incorporates most elementary processes from first principles. The computational effort of the presented simulations corresponds approximately to 300\,000~hours on a standard 1~GHz CPU.

The structure of this section is as follows. We introduce the code (\ref{Sect:code}) used for the simulations and explain the numerical setup (\ref{SectSetup}). We verify the convergence of the simulations (\ref{SectConverge}). Resolution and boundary effects are investigated in Sects.~\ref{SectResolution} to \ref{SectSimDepth}. The impact of metals and molecules on the results is checked in Sect.~\ref{SectMetalsMolec}. We conclude this section with an overview of the uncertainties that generally affect 1D simulations such as ours both from numerical and observational points of view (\ref{SectUncertainties}).

\subsection{Code}\label{Sect:code}

We use our photoionization-hydrodynamics simulation code the PLUTO-CLOUDY interface (TPCI) for the simulations of the escaping planetary atmospheres \citep[][]{Salz2015b}. The interface couples the photoionization and plasma simulation code CLOUDY to the magnetohydrodynamics code PLUTO. This allows simulations of steady-state photoevaporative flows. PLUTO is a versatile MHD code that includes different physical phenomena, among others, gravitational acceleration and thermal conduction \citep{Mignone2007, Mignone2012}. 
Our interface introduces radiative heating and cooling, as well as acceleration caused by radiation pressure from CLOUDY in PLUTO \citep{Salz2015b}. CLOUDY solves the microphysical state of a gas under a given irradiation in a static density structure \citep{Ferland1998, Ferland2013}. The user can choose to include all elements from hydrogen to zinc, and CLOUDY then solves the equilibrium state regarding the degree of ionization, the chemical state, and the level populations in model atoms. Radiative transfer is approximated by the escape probability mechanism \citep[][]{Castor1970, Elitzur1982}. The impact of the advection of species on the equilibrium state is included via a CLOUDY internal steady state solver as described in \citet[][]{Salz2015b}.

Recently \citet{Shematovich2014} have found that photoelectrons and suprathermal electron populations contribute significantly to excitation and ionization rates in the atmospheres of hot Jupiters. CLOUDY includes these processes assuming local energy deposition \citep{Ferland1998, Ferland2013}, a transport of photoelectrons is not included. In partially ionized gases, this assumption is valid as long as the column densities of the individual cells exceed 10$^{17}$~cm$^{-2}$ \citep{Dalgarno1999}, which is approximately fulfilled in neutral parts of the presented atmospheres. For highly ionized plasmas distant collision of the ionized species reduce the mean free path of photoelectrons by more than one order of magnitude \citep{Spitzer1962}.
The mean free path of photoelectrons becomes larger than one cell width only in the upper thermosphere of CoRoT-2\,b and the planets which have stable thermospheres at a height where the atmospheres also become collisionless for neutral hydrogen \citep[see Sect.~\ref{SectJeans}, using equations from][]{Goedbloed2004}.

The use of CLOUDY is one of the main improvements compared with previous models of escaping planetary atmospheres because it solves not only the absorption of ionizing radiation in detail but also the emission of the gas. Most previous studies have either chosen a fixed heating efficiency for the absorbed radiation to account for a general radiative cooling effect or they specifically included individual radiative cooling agents like \lya{} cooling \citep[e.g.,][]{Murray2009, Koskinen2013a, Shaikhislamov2014}. 
With such an approach it is impossible to find radiative equilibrium, i.e., a state where the complete radiative energy input is re-emitted, because not all relevant microphysical processes are included. Therefore, these simulations can only be used for strongly escaping atmospheres, where radiative cooling has a minor impact, but not for almost stable thermospheres, which are close to radiative equilibrium. TPCI solves this issue and can be used to simulate the transition from rapidly escaping atmospheres, where the radiative energy input is virtually completely used to drive the hydrodynamic wind, all the way to a situation with a hydrostatic atmospheres, where the absorbed energy is re-emitted.

\subsection{Simulation setup}\label{SectSetup}

We simulate the expanding planetary atmospheres on a spherical, 1D grid. A similar setup is chosen for all simulations. The planetary atmosphere is irradiated from the top. We simulate the substellar point, which results in a maximal outflow rate because of the high irradiation level at this point. Forces due to the effective gravitation potential in the rotating two-body system are included. The resulting force is a superposition of the planetary and stellar gravitational forces with the centrifugal force caused by the planetary orbital motion.
Furthermore, radiation pressure is included and thermal conduction is considered in a simplified manner by adding the thermal conductivity coefficients of an electron plasma and of a neutral hydrogen gas. A detailed explanation of the numerical implementation of these terms is given in \citet[][]{Salz2015b}.

Our simulations have several input parameters, which are fixed by observations, i.e., the planetary mass and radius, the semimajor axis, and the stellar mass. Additionally, we need to know the irradiating SED (see Sect.~\ref{SectSED}). The last two required parameters are not well defined: First, the temperature in the lower atmosphere, which is close to the equilibrium temperature but with a considerable uncertainty (see Sect.~\ref{Sect:BC_temp}); and second, the boundary density, which reflects how deep our simulations proceed into the lower atmosphere (see Sect.~\ref{SectSimDepth}). Fortunately, these two parameters have no strong impact on the outcome, therefore, the resulting atmosphere is unique for every planet and we do not need to sample a parameter space.

The simulated atmospheres consist of hydrogen and helium only; metals or molecules are neglected in our current setup (see Sect.~\ref{SectMetalsMolec}). CLOUDY solves a plane parallel atmosphere, allowing radiation to escape in both directions toward the host star and toward the lower atmosphere. We double the optical depth at the bottom of the simulated atmosphere to ensure that radiation does not escape freely into a vacuum in this direction. We further use a constant microturbulence of 1~km\,s$^{-1}$, which is adapted to the conditions close to our lower boundary; the microturbulence has little impact on the overall atmospheric structure.

A stretched grid with 500 grid cells spans the range from 1 to 12/15 planetary radii. The resolution is increased from  $2\times10^{-4}~R_{\mathrm{pl}}$ to  0.22~$R_{\mathrm{pl}}$ at the top boundary. We emphasize the high resolution at the lower boundary, which is not always visible in figures showing the complete atmospheres. The photoionization solver runs on an independent grid, which is autonomously chosen and varies throughout the simulation progress. It uses 600 to 1200 grid cells. The resolution in the dense atmosphere at the lower boundary can be up to a factor ten higher than in the hydrodynamic solver.

The number of boundary conditions that are fixed are two at the bottom (density, pressure) and none at the top boundary. This is the correct choice for a subsonic inflow and a supersonic outflow \citep[e.g. ][]{Aluru1995}. The bottom boundary condition is anchored in the lower atmosphere at a number density of $10^{14}$~cm$^{-3}$. This is sufficient to resolve the absorption of the EUV emission, which is mostly absorbed in atmospheric layers with a density of around $10^{11}$~cm$^{-3}$. The density at the planetary photosphere is usually argued to be around $10^{17}$ to $10^{18}$~cm$^{-3}$, which we only state for a comparison
\citep{Lecavelier2008a, Lecavelier2008b}. Furthermore, we fix the pressure according to the equilibrium temperature of the planets (see Table~\ref{tabSysPara}). 
This results in pressures around 14~dyn\,cm$^{-2}$ equivalent to 14~$\mu$bar. The velocity in the boundary cell is adapted at each time step according to the velocity in the first grid cell. However, we only allow inflow, setting the boundary velocity to zero otherwise. This choice dampens oscillations that appear in simulations with steep density gradients at the lower boundary. The impact of the lower boundary conditions on the outcome of the simulations is discussed in Sect.~\ref{Sect:BC_temp} and \ref{SectSimDepth}.

The chosen numerical setup of the PLUTO code is a third order Runge-Kutta integrator \citep{Gottlieb1996}, with the weighted essentially non-oscillatory finite difference scheme \citep[{\tt WENO3} ,][]{Jiang1996} for interpolation and the Harten, Lax and van Leer approximate Riemann solver with contact discontinuity \citep[{\tt HLLC},][]{Toro1994}. The radiative source term is not included in the Runge-Kutta integrator, but a simple forward Euler time stepping is used.

\subsection{Convergence and initial conditions}\label{SectConverge}

\begin{figure}[t]
  \centering
  \includegraphics[width=\hsize]{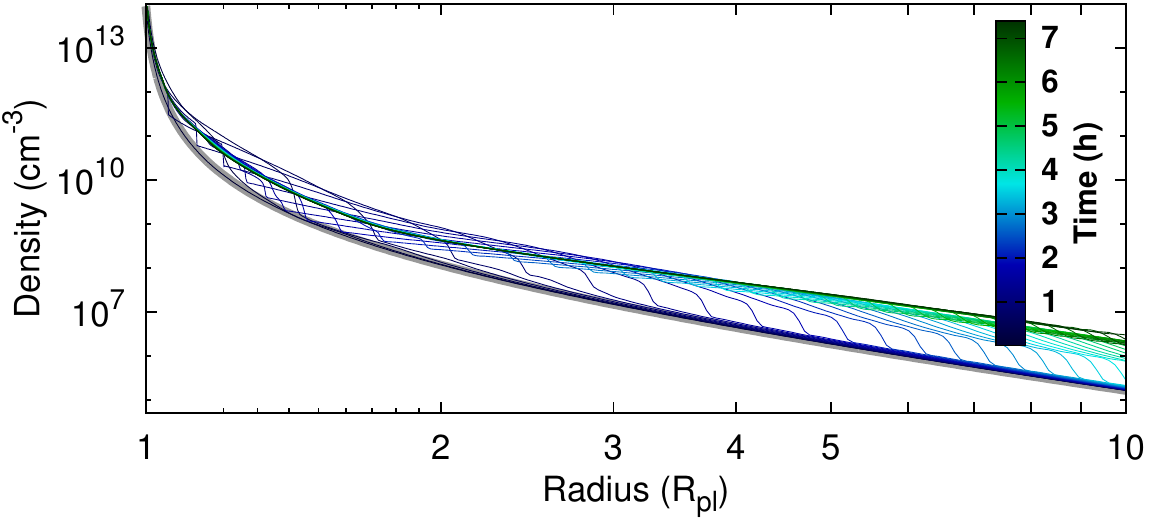}
  \caption{Convergence of the atmosphere of WASP-80.
           The density structure is plotted versus the radius.
           The initial structure is indicated by a thick gray line, the
           time evolution is given by the color scheme.}
  \label{fig:converge}
\end{figure}

All simulations were started from the same initial conditions, which are adjusted to an atmosphere with a medium mass-loss rate. Every atmosphere undergoes a series of shock waves in the first few hours of real time. This is depicted in Fig.~\ref{fig:converge} for the density structure in the atmosphere of WASP-80\,b. Two shock waves propagate from the lower atmosphere (lefthand side) into the upper atmosphere (righthand side), increasing the average density in the thermosphere by a factor of ten. The depicted 7~h evolution in the atmosphere corresponds to less than $10^{-3}$ of the final converging time. The final steady-state does not depend on the chosen initial conditions.

The convergence of the simulations was analyzed by checking the mass-loss rate and the temperature structure. We define a system specific time scale as
\begin{equation}
  t_{\mathrm{sys}} = \frac{R_{\mathrm{pl}}}{v_{\mathrm{pl}}} \,
\end{equation}
and choose $v_{\mathrm{pl}} = 1~\mathrm{km\,s}^{-1}$ as a typical velocity scale in the middle of the planetary thermospheres. 1000 of such time steps correspond to between half a year and four years in the systems. This time is sufficient for atmospheric material to propagate at least once through the computational domain. We ensured that the temperature structure and the total mass-loss rate have converged to a level of 0.1 on this time scale ($\Delta\dot{M}/\dot{M} < 0.1$). Individual simulations were advanced a factor of 10 longer and did not show any significant further changes.

\subsection{Resolution effects}\label{SectResolution}

We have checked that the chosen resolution does not affect the results of our simulations. The resolution of the stretched grid at the lower boundary has been increased in three steps by factors of two in the simulations of HD\,209458\,b and HD\,189733\,b. HD\,189733\,b has a steeper density gradient in the lower atmosphere and is more challenging for the numerical solver. Decreasing the resolution leads to larger oscillations in the atmospheres, which remain after the convergence of the simulations as defined above. 
The remaining oscillations are larger in the atmosphere of HD\,189733\,b than in HD\,209458\,b, hence, atmospheres with lower mass-loss rates are affected more strongly by such oscillations. They do not affect the time averaged density, temperature, or velocity structure and, therefore, also not the mass-loss rate (fractional change of the mass-loss rate $\Delta\dot{M}/\dot{M} < 10^{-2}$). 

The oscillations are numerical artifacts with two sources. First, the small acceleration of the planetary wind in the lower atmospheres results from a slight imbalance of the gravitational force and the force due to the hydrostatic pressure gradient. Subtraction of these large forces results in a small value, which is prone to numerical inaccuracy. Increasing the resolution results in better estimates of the pressure gradient, which in turn reduces the errors and thus the oscillations. The second source of oscillations in the lower atmosphere of planets with a small mass-loss rate is a consequence of these atmospheric layers being almost in radiative equilibrium. When the microphysics solver is called within TPCI the best level of equilibrium between radiative heating and cooling is about $10^{-2}$. This results in small transient heating and cooling events in regions of radiative equilibrium, which then induce oscillations.

In some cases the oscillations are large enough to be seen in our figures. If this is the case, we present time averaged structures. The convergence to the steady-state as defined above is unaffected by these oscillations, but the accuracy of the final atmospheric structure is worse than in simulations without oscillations. This is one reason for our conservative convergence level of 10\%. Furthermore, the computational effort of the individual simulation is increased by the oscillations.

The independence of the mass-loss rates on the resolution distinguishes our simulations from those of \citet{Tian2005}. The mass-loss rates in their simulations depended on the chosen number of grid points. However, the authors used a Lax-Friedrich solver with a higher numerical diffusion compared with the more advanced Godunov-type scheme with the approximate {\tt HLLC} Riemann solver. \citeauthor{Tian2005} identified the numerical diffusion to be responsible for the dependency of their mass-loss rates on the resolution, 
which is consistent with our findings.

\subsection{Boundary temperature}\label{Sect:BC_temp}

The temperature at the lower boundary is one of the free parameters in our simulations. It corresponds to a temperature in the lower atmosphere of the planet at a certain height. Basically, the equilibrium temperature $T_{\mathrm{eq}}$ of the planet provides a simple approximation for this temperature  \citep[with zero albedo, e.g.,][]{Charbonneau2005}:
\begin{equation}\label{}
  T_{\mathrm{eq}} = T_{\mathrm{eff}}\sqrt{R_{\mathrm{st}}/2a} \, .
\end{equation}
Here, $T_{\mathrm{eff}}$ is the effective temperature of the host star, $R_{\mathrm{st}}$ is the stellar radius, and $a$ is the semimajor axis of planetary orbit.
In some cases measurements of the average dayside brightness temperatures are available from infrared transit observations (see references of Table~\ref{tabSysPara}). However, neither the equilibrium temperature nor the brightness temperature necessarily represents the temperature at the atmospheric height, which corresponds to the lower boundary of our simulations. Detailed radiative transfer models are another method to obtain an accurate temperature structure for the lower atmosphere, but such models only exist for well studied systems like HD\,209458\,b. 
For HD\,209458\,b and HD\,189733\,b pressure-temperature profiles were reconstructed for the atmosphere above the terminator based on sodium absorption measurements \citep{Vidal2011, Huitson2012, Wyttenbach2015}, but these probably do not apply to the substellar point. Even for the probably best studied case, HD\,209458\,b, \citet{Koskinen2013a} conclude that despite several different observations and models the temperature structure in the lower atmosphere is uncertain. 

\begin{figure}[t]
  \centering
  \includegraphics[width=\hsize]{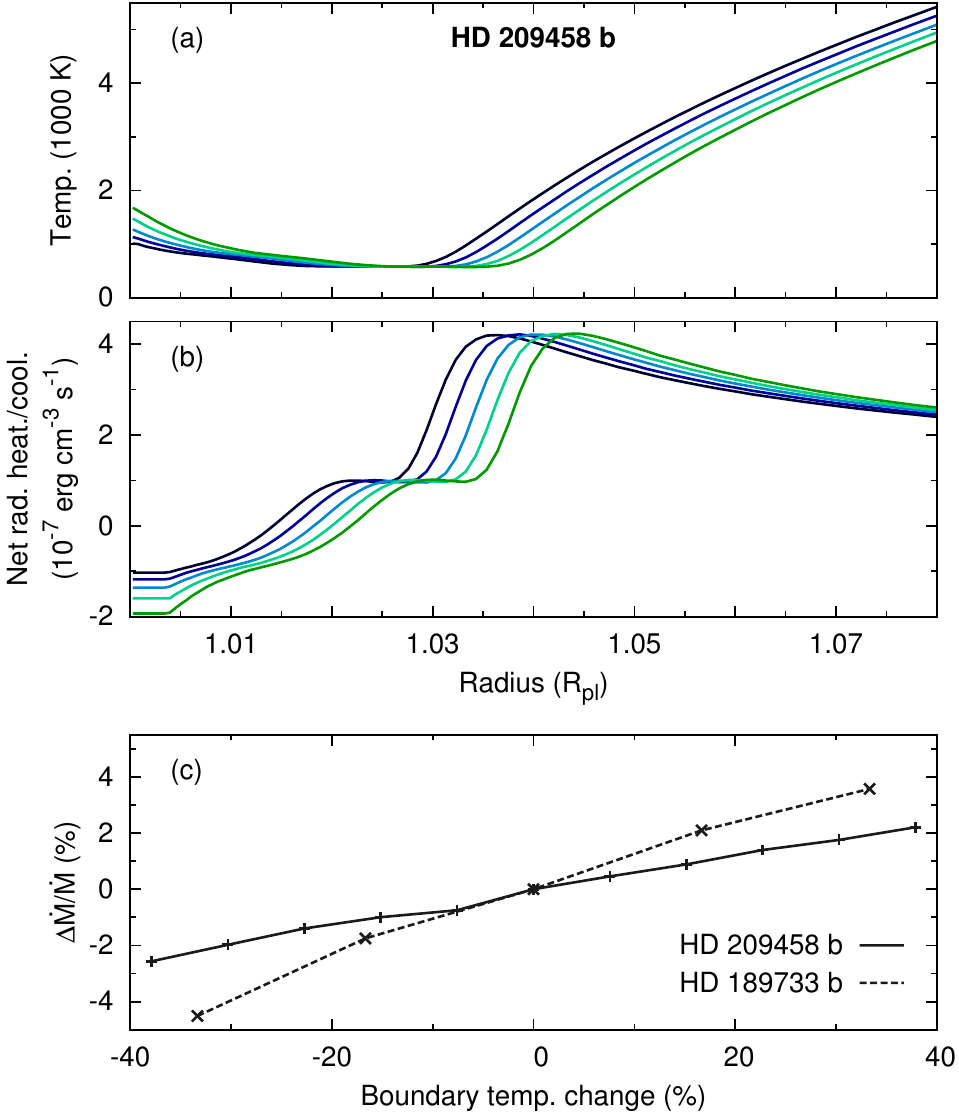}
  \caption{Influence of the temperature at the lower boundary
           on the mass-loss rates.
           Temperature structure (a) and the heating rate (b) 
           in the lower atmosphere of HD\,209458\,b are shown along
           with the change in the mass-loss rate (c) in the 
           planets HD\,209458\,b and HD\,189733\,b.
           The color scheme in panels (a) and (b) indicates the
           boundary temperature from $-400$~K (dark blue) to $+400$~K
           (green).
           A higher boundary temperature results in a slightly higher
           mass-loss rate. }
  \label{fig:BCtemp}
\end{figure}

Fortunately, we do not need to know the temperature at the lower boundary with high precision.
To verify this, we tested the impact of this temperature on our results. In the simulations of the atmospheres of HD\,209458\,b and HD\,189733\,b, which have boundary temperatures of 1320~K and 1200~K respectively, we increased and decreased the temperature in steps of 100~K/200~K. Figure~\ref{fig:BCtemp} shows that increasing the boundary temperature slightly expands the planetary atmosphere, but the large scale structure remains unaffected as does the temperature structure in the upper planetary thermosphere.
The change in the mass-loss rate scales linearly with the change of the boundary temperature over the tested range. The overall impact is small: A change of 40\% in the boundary temperature causes the mass-loss rate to change by less than 5\%. Our results are comparable to those of \citet{Murray2009}, who tested a wider range for the lower boundary temperature. This result is not surprising, because the gain in advected energy due to a 1000~K temperature change is about a factor ten smaller than the total radiative energy input. 
The change in the mass-loss rate is about twice as strong in the simulation of HD\,189733\,b. A temperature minimum in the atmosphere of HD\,209458\,b decouples the simulation better from the boundary conditions, because radiative processes have more time to dispose of additional energy. 

Even considerable changes in the lower boundary temperature affect the total mass-loss rates in our simulations by less than 10\%. With respect to the uncertainty on other parameters, especially the irradiation level, we can neglect the effect at this point.

\subsection{Boundary density}\label{SectSimDepth}
 
\begin{figure}[t]
  \centering
  \includegraphics[width=\hsize]{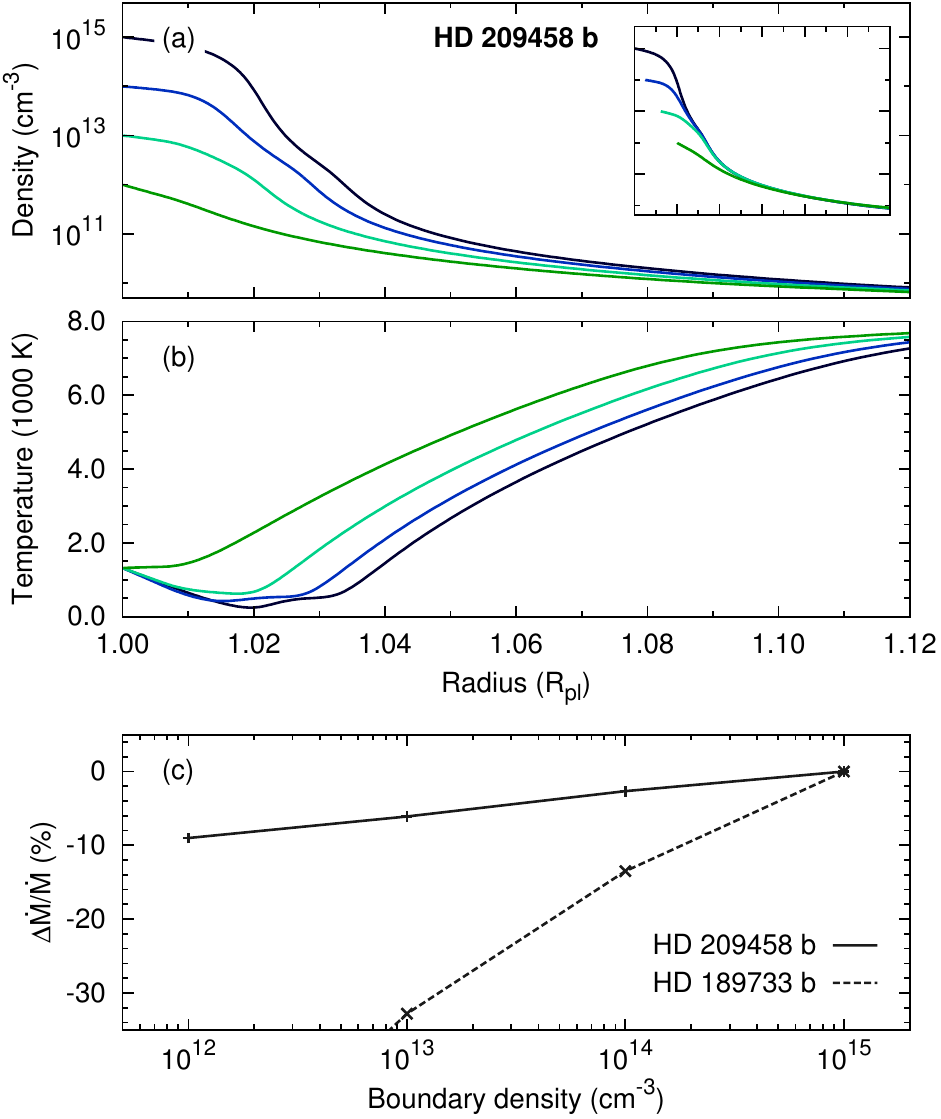}
  \caption{Influence of the density at the lower boundary on
           the mass-loss rates.
           The density (a) and the temperature (b)
           in the lower atmosphere of HD\,209458\,b are depicted.
           The insert shows the density structures shifted to
           merge the profiles.
           Panel (c) shows the change in the mass-loss rate in the 
           two test simulations (HD\,209458\,b, HD\,189733\,b).
           The color scheme indicates the boundary density ranging
           from $10^{15}$~cm$^{-3}$ (dark blue) to
           $10^{12}$~cm$^{-3}$ (green).
           Reducing the boundary density by two orders of
           magnitude decreases
           the mass-loss rate by less than 40\%.}
  \label{fig:BCdens}
\end{figure}

The density at the lower boundary is another free parameter in our simulations. Changing the boundary density has a larger effect on the mass-loss rate than changing the boundary temperature. This is a result of a slight inconsistency in the setup of our simulations. The density at the lower boundary is a density that is reached at a certain height above the planetary photosphere\footnote{Since a gas planet has no surface, we refer to the outer most layer that is opaque to visual light as photosphere.}. 
However, our simulation setup always starts at 1~$R_{\mathrm{pl}}$. If we start with a density two orders of magnitude smaller, this layer is located higher up in the planetary atmosphere; the change in height of the maximal volume heating rate is about 0.025~$R_{\mathrm{pl}}$ in the case of HD\,209458\,b. In the spherical simulation the irradiated area of the grid cells increases with $R_{\mathrm{pl}}^2$, thus  in this example the similar volume heating rate results in a 5\% increase of the total heating rate, which in turn affects the mass-loss rate.
This effect is depicted in Fig.~\ref{fig:BCdens}, which shows that the density structures of the individual simulations almost merge in the upper thermosphere of HD\,209458\,b, but larger differences occur in the lower atmosphere. These would be mitigated if we had started the simulations with a lower boundary density at the correct heights as depicted in the insert.

As a secondary effect, the temperature minimum in the lower atmosphere becomes deeper as the mass loss increases. In this region of the atmosphere, radiative heating is small and the increased adiabatic cooling caused by a higher mass-loss rate reduces the local temperature \citep{Watson1981, Garcia2007}. At even lower boundary densities ($10^{12}$~cm$^{-3}$) a significant fraction of the XUV radiation passes through our atmosphere without being absorbed in the computational domain, which further decreases the mass-loss rate.

For HD\,189733\,b the impact of the boundary density on the mass-loss rate is larger, because the steeper atmosphere shifts stronger in response to a change of the boundary density. Decreasing the boundary density to $10^{12}$~cm$^{-3}$ causes a discontinuity in the density structure from the boundary cell to the first grid point, and the total mass-loss rate is reduced by 60\%. We confirmed that none of the presented simulations is affected by such a density discontinuity.

Our standard simulation has a density of $10^{14}$~cm$^{-3}$ at the lower boundary.
The uncertainty in the correct height of our boundary layer above the planetary photosphere introduces an error on the mass-loss rates of less than 50\%, consistent with the results of \citet{Murray2009}. We further note that the observationally determined planetary radii are usually not accurate to the percent level, so that this error is inevitable.

\subsection{Metals and Molecules}\label{SectMetalsMolec}

Our current simulations include the chemistry of hydrogen and helium. Adding metals substantially increases the computational load and will be studied for individual systems in the future. To justify that we can neglect metals in this study, we anticipate results from a test simulation of the atmosphere of HD\,209458\,b including all 30 lightest metals with solar abundances. We find that the temperature structure of the lower atmosphere ($<1.05~R_{\mathrm{pl}}$) is dominated by line heating and cooling of metals, but the temperature in this region only has a small effect on the mass-loss rate (see Sect.~\ref{Sect:BC_temp}). 
Cooling of \ion{Ca}{ii} and \ion{Fe}{ii} dominates from 1.1 to $1.6~R_{\mathrm{pl}}$ and reduces the maximum temperature in the atmosphere by about 1000~K. Additionally, \ion{Mg}{ii} and \ion{Si}{ii} are important cooling agents. While such heavy metals can be advected by the strong planetary wind \citep{Yelle2008, Garcia2007}, their occurrence in the upper thermosphere depends on the formation of condensates in the lower atmosphere \citep[e.g.,][]{Sudarsky2000}.
Including metals reduces the mass-loss rate in our test simulation by a factor of two. While this value may be used for guidance, the result certainly depends on the metal abundance, because increasing the abundances to supersolar values will cool the planetary atmosphere more efficiently. Hence, metals have an effect on the results, but as long as they remain minor constituents their impact likely does not exceed the uncertainty in the EUV irradiation level \citep[][]{Salz2015a}.

Metals also efficiently absorb X-rays, which increases the available energy. 
Strongly expanded thermospheres like that of HD\,209458\,b are opaque to X-rays longward of 25~\AA{}, but more compact atmosphere only are opaque longward of 60~\AA{} in our simulations (defined as 50\% absorption). Including metals shifts these cutoff values to approximately 10 and 20~\AA{}. In the case of our most active host star CoRoT-2 metals would increase the absorbed energy in the planetary atmosphere by 13\%.

We also disable the formation of molecules in CLOUDY. This choice is unphysical, because photodissociation of H$_2$ would occur close to the lower boundary \citep{Yelle2004, Garcia2007, Koskinen2013a}. Molecules strongly affect the temperature structure of the lower atmospheres, but the impact of this temperature structure on the mass-loss rate is quite small (see Sect.~\ref{Sect:BC_temp}). Furthermore, planets with a moderate irradiation level can host a molecular outflow. A detailed study of this phenomenon is beyond the scope of this paper, but we performed a test simulation of one of the planets with the smallest irradiation level, GJ\,1214\,b, including molecules in the planetary atmosphere. The simulation shows an H$_2$ fraction of 10 to 25\% throughout the thermosphere of the planet. 
H$^-$ is a significant cooling agent, as for example also seen in the solar chromosphere at similar conditions \citep[][]{Vernazza1981}. The mass-loss rate is reduced by 15\%, thus, our simulations without molecules remain valid as long as hydrogen is the major constituent of the atmospheres.

\subsection{Overview of the uncertainty budget}\label{SectUncertainties}

\begin{table}[t]
  \caption{Uncertainties on the result of the simulations.}
  \label{tabUncertainties}
  \centering
  \begin{tabular}{l r l c }
    \hline\hline
      Source  &  Trend\tablefootmark{A} &  Impact & Ref.\\
    \hline
      Stellar wind                                 & ($-$)    & uncert.    & 4  \\
      Three dimensional structure\tablefootmark{B} & ($-$)    & $\times$ 4 & 1\\
      Irradiation strength                         & ($\pm$)  & $\times$ 3 & 2\\
      Magnetic fields                              & ($-$)    & $\times$ 3 & 3 \\
      Metals                                       & ($-$)    & $\times$ 2 &  \\
      Molecules\tablefootmark{C} (H$_2$)           & ($-$)    & $\times$ 2 &  \\
      Boundary density                             & ($\pm$)  & 50\% & \\
      Temperature in lower atmosphere              & ($\pm$)  & 10\% & \\
   \hline 
   \end{tabular}
\tablefoot{
\tablefoottext{A}{A negative trend indicates that the effect will
reduce the simulated mass loss rates.}
\tablefoottext{B}{Our mass-loss rates are reduced by this factor and 
the remaining uncertainty is smaller.}
\tablefoottext{C}{Only moderately irradiated smaller planets
host molecular winds.}
}
\tablebib{
          Phenomena without reference are discussed in the text.
          (1) \citet{Stone2009}; 
          (2) \citet[][]{Salz2015a};
          (3) \citet{Trammell2014};
          (4) \citet{Murray2009}.
}
\end{table}

One-dimensional simulations of the planetary winds are subject to several uncertainties. These are introduced by observational limitations and by the simplifications of the 1D model. We provide a short overview and estimate the impact of probably the most important effects that are neglected (see Table~\ref{tabUncertainties}).

A large uncertainty is introduced by neglecting the 3D structure of the system geometry, however, \citet{Stone2009} have shown that 1D models produce valid estimates for the planetary mass-loss rates, if the irradiation is averaged correctly over the planetary surface. In \citet[][]{Salz2015a} we indicated that the irradiation level of hot gas planets is uncertain, because the EUV spectral range is completely absorbed by interstellar hydrogen and different reconstruction methods differ by up to a factor of 10. Most recently, \citet{Chadney2015} revised one of the reconstruction methods, reducing the differences for active stars. Nevertheless, we adopt a conservative error estimate of a factor of 3 for the EUV reconstruction. The impact of metals and molecules on the wind has been explained in the previous section. Related to our simulations, the boundary conditions introduce a small uncertainty on the resulting mass-loss rates.

1D-HD simulations of the planetary winds ignore any interactions with possible stellar winds. This approximation is valid if the interaction with the stellar wind occurs after the planetary wind becomes supersonic. However, \citet{Murray2009} showed that the stellar wind can suppress the formation of a planetary wind at least on the dayside of the planet. There are likely two regimes of confinement: First, the planetary wind becomes too weak to sustain a supersonic expansion and a subsonic planetary wind persists with a lower mass-loss rate. Second, if the stellar wind is strong enough, the planetary wind is completely suppressed.

Furthermore, planetary winds can interact with planetary magnetic fields. This has been studied for example by \citet{Trammell2014}, who stated that, if the planet's magnetic field is strong enough, the outflow can be suppressed over larger fractions of the planetary surface. The authors also found that there are always open field lines along which a planetary wind can develop freely.

The formation of planetary winds is a complex physical system, which currently can only be studied by neglecting certain aspects. We focus on accurately solving the energy balance throughout the thermospheres regarding the absorption and emission of radiation by the atmospheric gas in detail. In the following sections we also show atmospheric structures above the planetary Roche lobe height, which is only for guidance. Our spherical approximation is invalid above this height and interactions with stellar winds can deform and accelerate the planetary wind material.

\section{Results and discussion of the simulations}\label{SectResults}

We now present the simulated photoevaporative winds of 18 exoplanets. The atmospheres of HD\,209458\,b and HD\,189733\,b are compared in Sect.~\ref{SectGenStruc}, where we explain in detail the individual processes affecting the planetary atmospheres. The complete sample of planets is presented and compared in Sect.~\ref{SectCompare}, where the impact of individual system parameters on the planetary atmospheres is investigated.  Here, we also explain the approach to radiative equilibrium in the thermospheres of massive and compact planets.
We then compute the mass-loss rates (\ref{Sect:mloss}), explain the case of WASP-12\,b (\ref{Sect:wasp12}), identify stable thermospheres (\ref{SectJeans}), consider the ram pressure of the planetary winds (\ref{Sect:Ram}), and estimate the strength of the radiation pressure (\ref{SectRadPres}).

\subsection{Heating fraction and heating efficiency}\label{SectHeatFrac}

For the analysis of our simulations we define the height-dependent heating fraction as
\begin{equation}\label{EqHeatFrac}
  \mathrm{heating\,fraction} = \frac{\mathrm{rad.\,heating} - \mathrm{rad.\,cooling}}{\mathrm{rad.\,heating}} \, .
\end{equation}
Note that the heating fraction differs from the often used heating efficiency, which is defined as:
\begin{equation}\label{EqHeatEff}
  \mathrm{heating\,eff.} = \eta = \frac{\mathrm{rad.\,heating} - \mathrm{rad.\,cooling}}{\mathrm{absorbed~rad.\,energy}} \, .
\end{equation}
Here, the denominator holds the total energy contained in the absorbed radiative flux. The difference of the two definitions is mainly the energy that is used to ionize hydrogen and helium; this energy fraction is not available for heating the planetary atmosphere. Equation~\ref{EqHeatFrac} excludes this fraction of the radiative energy input, hence, the heating fraction approaches one as radiative cooling becomes insignificant, and it is zero in radiative equilibrium. The heating efficiency is also zero in radiative equilibrium, but it never reaches a value of one. The heating efficiency is not directly available in our simulations. 

\subsection{Structure of the escaping atmospheres} \label{SectGenStruc}

\begin{figure}[tb]
  \centering
  \includegraphics[width=\hsize]{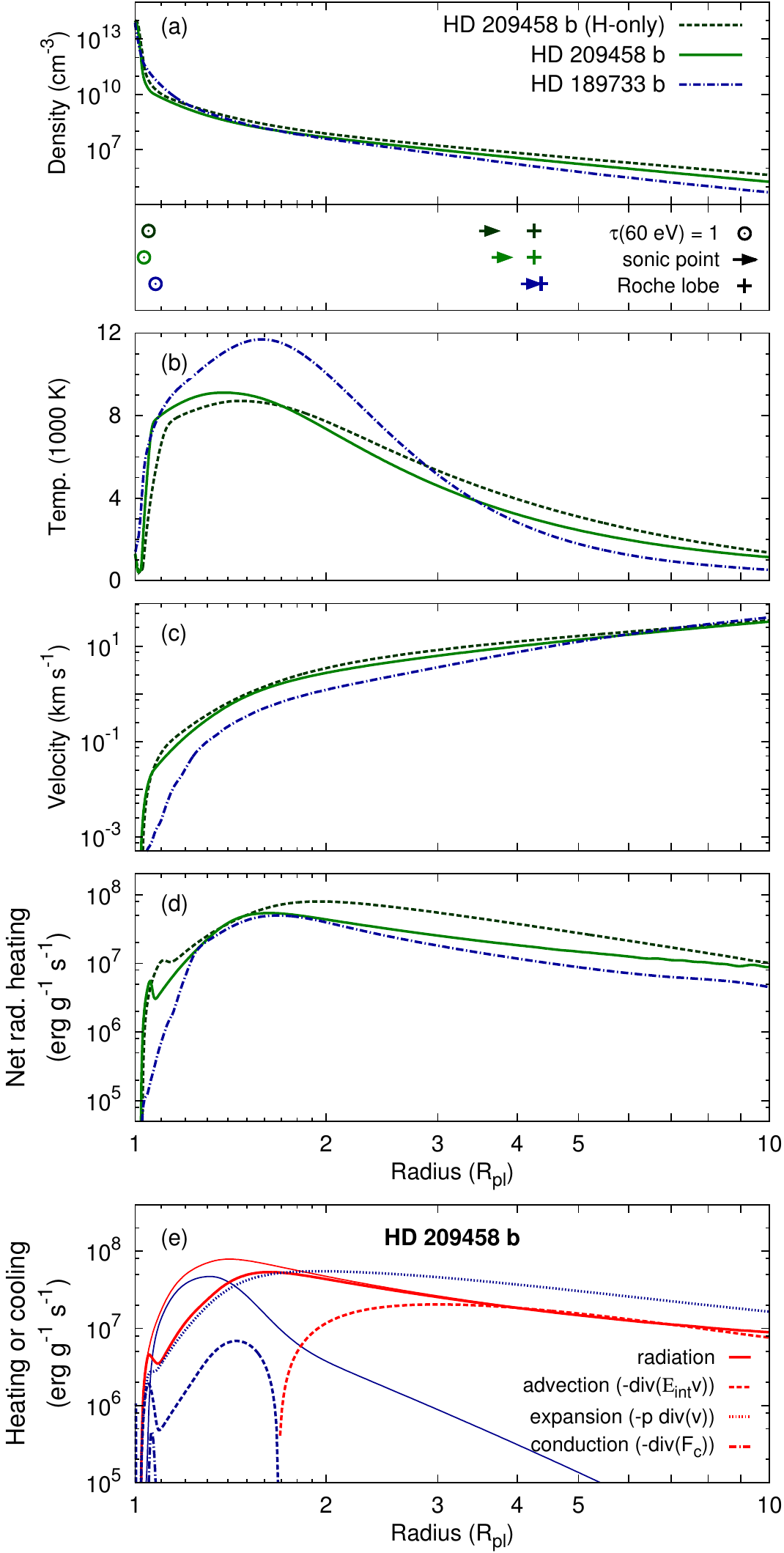}
  \caption{Atmospheres of HD\,209458\,b and HD\,189733\,b.
           For HD\,209458\,b both a hydrogen-only and a
           hydrogen and helium simulation are shown.
           We plot the density, temperature, velocity, and the specific
           heating rate. 
           The symbols in the lower panel (a) indicate the 
           $\tau (60~\mathrm{eV}) = 1$ levels (circles),
           the sonic points (arrows), and the Roche lobe heights (plus signs),
           with the order of the planets according to the legend.
           Panel (e) shows the individual heating (red) and cooling (multiplied by minus one, blue) agents in 
           the H+He atmosphere of HD\,209458\,b. 
           Radiative heating (red) and cooling (blue) are displayed individually by the thin solid lines
           and the thick solid line gives the net radiative-heating rate.
           }
  \label{fig:hd2vshd1}
\end{figure}

\begin{figure*}[tb]
  \centering
  {\bf\scriptsize \hspace{10mm} HD\,209458\,b (H-only) \hspace{35mm} HD\,209458\,b (H+He) \hspace{35mm} HD\,189733\,b (H+He) \\}
  \vspace{3pt}
  \includegraphics[width=0.361\hsize, trim=0cm 0cm 0.03cm 0cm, clip=true]{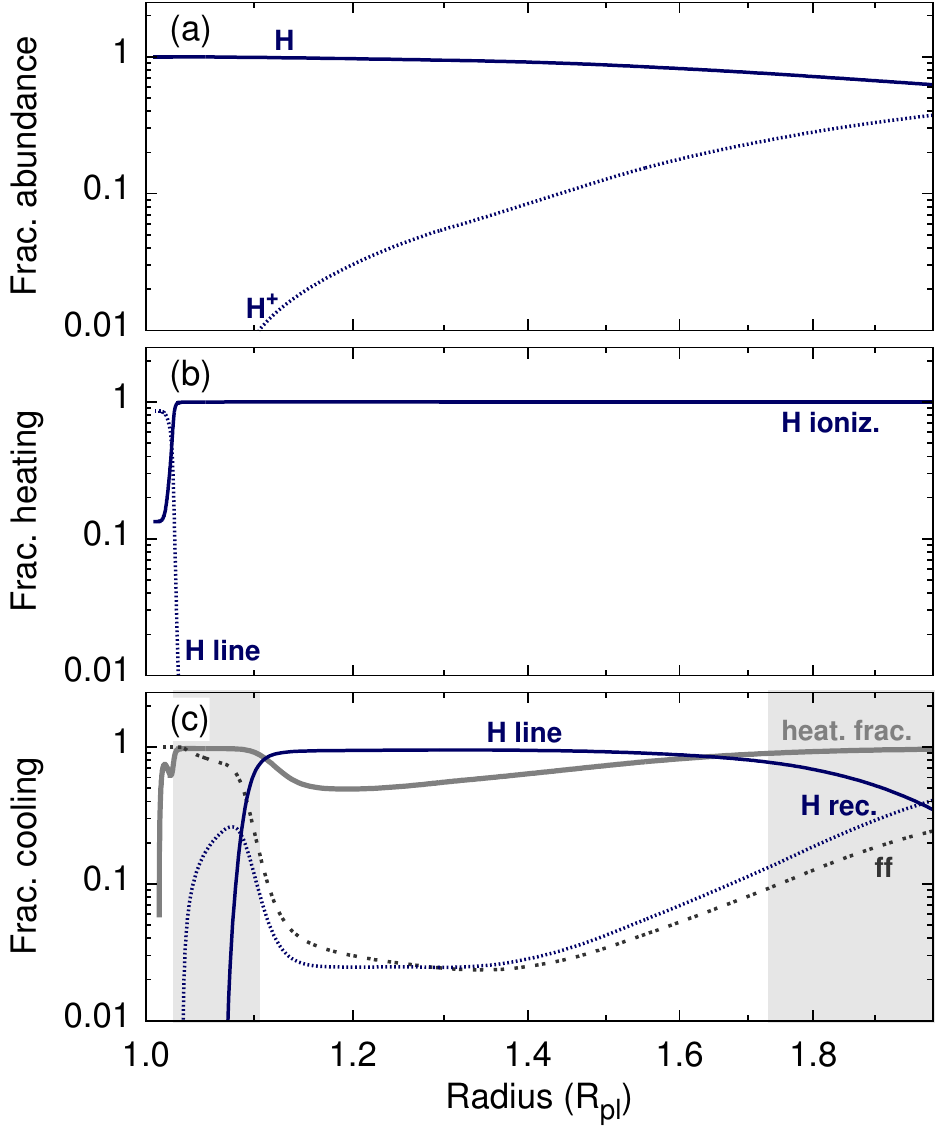}
  \hfill
  \includegraphics[width=0.313\hsize, trim=0cm 0cm 0.03cm 0cm, clip=true]{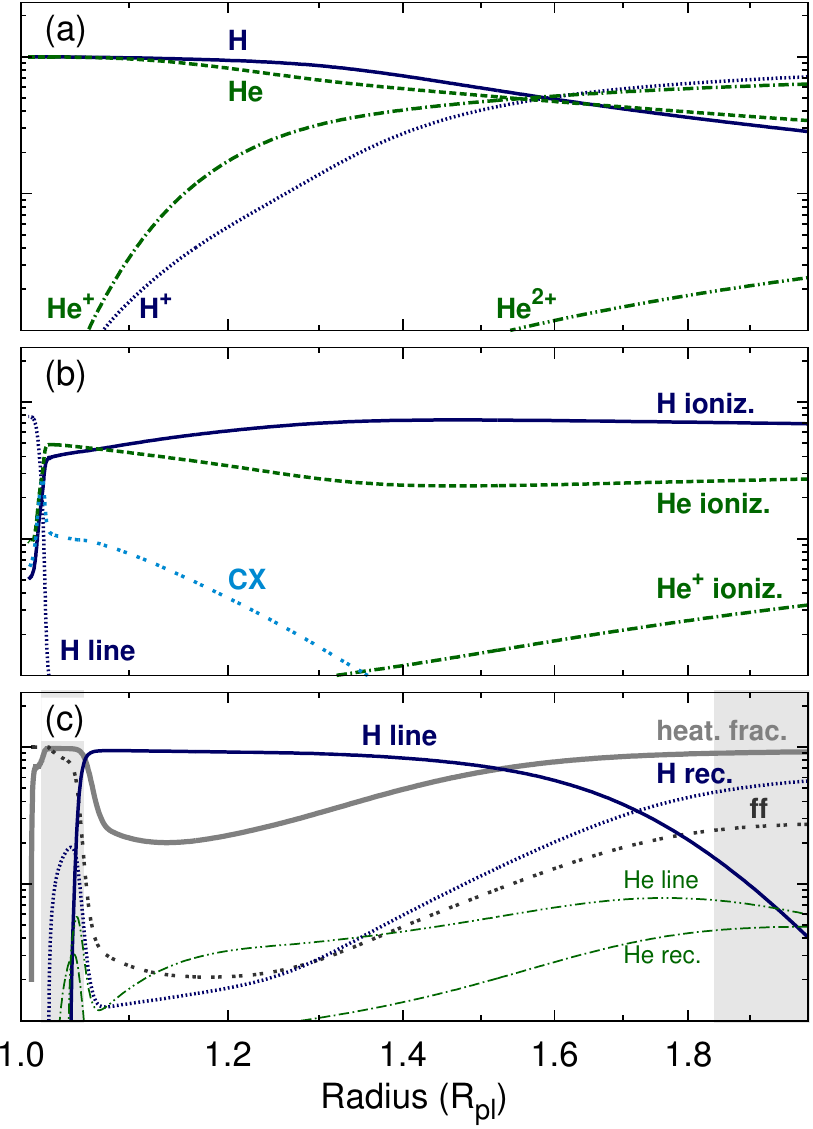}
  \hfill
  \includegraphics[width=0.313\hsize]{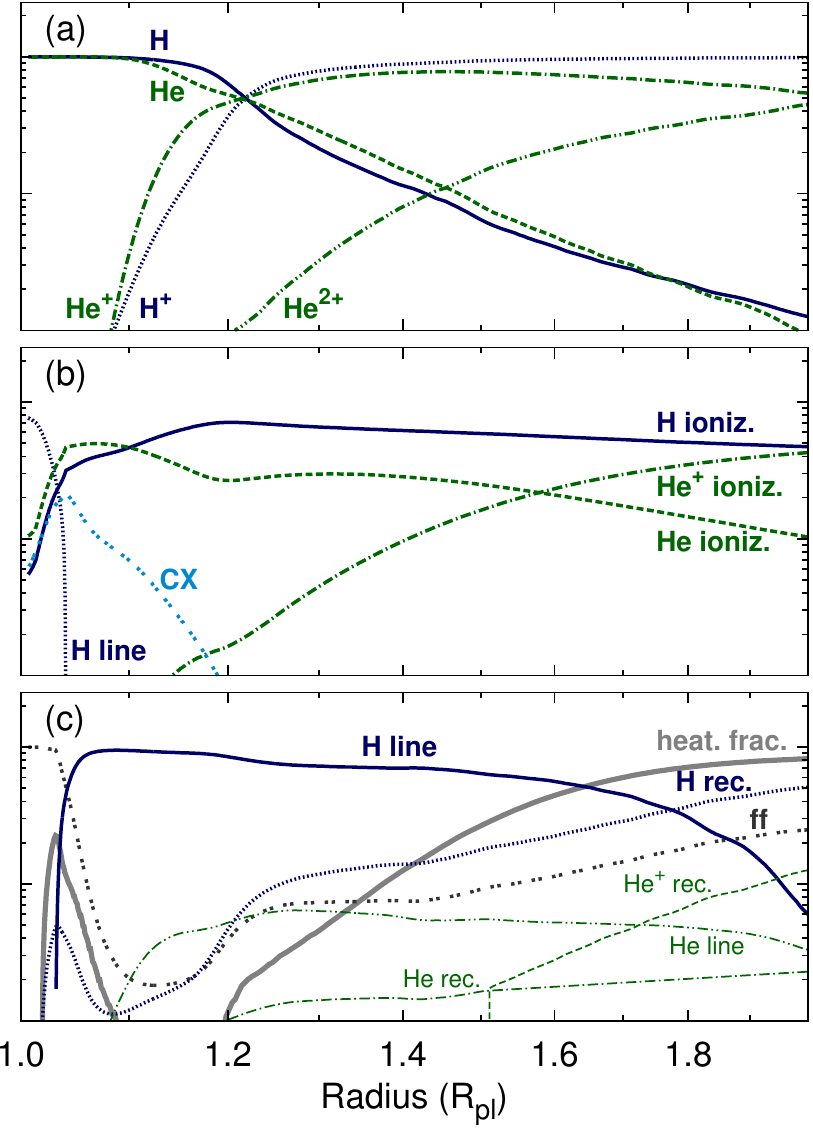}
  \caption{Abundances and heating/cooling agents in the atmospheres of HD\,209458\,b (H-only), 
           HD\,209458\,b (H+He), and HD\,189733\,b.
           Panels (a) depict the fractional abundances in the three atmospheres, (b)
           show the contribution of individual radiative heating agents to the total
           heating rates, and (c) show the same for the cooling rates.
           In panel (c) we also show the heating fraction by the solid thick
           gray line. Regions with little radiative cooling are shaded (heating
           fraction $>0.9$).
           We show the atmospheres up to 2~$R_{\mathrm{pl}}$.
           The atmosphere of HD\,189733\,b shows weak oscillations.
           Labels are:
           hydrogen and helium line cooling/heating (H/He line);
           ionization heating of hydrogen, helium, or ionized helium  (H/He/He$^+$ ioniz.);
           charge transfer between neutral hydrogen and ionized helium (CX);
           recombination cooling of hydrogen, helium, or ionized helium (H/He/He$^+$ rec.);
           free-free emission (ff); and heating fraction (heat. frac.).
           }
  \label{fig:comp_atmos}
\end{figure*}

To understand the general behavior of escaping planetary atmospheres, we compare two examples: HD\,209458\,b and HD\,189733\,b. For HD\,209458\,b we also juxtapose a simulation with only hydrogen as constituent (H-only). In Fig.~\ref{fig:hd2vshd1} we show the density, temperature, velocity, and specific heating rate throughout the atmospheres in the three simulations. Panel (e) further shows individual heating and cooling terms in the atmosphere of HD\,209458\,b with hydrogen and helium (H+He). Finally, Fig.~\ref{fig:comp_atmos} splits the radiative heating and cooling terms into the most important agents. In all our figures the righthand side of the plot is the top of the atmosphere, while the lefthand side is the lower boundary, which is located in the denser atmosphere close to the planetary photosphere.

The general structure of all planetary atmospheres is similar. The density in the atmospheres decreases with height; the density gradient in the thermosphere is shallow due to the high temperature, which results from the absorption of stellar XUV emission. The high temperature further leads to a persistent expansion of the atmospheric gas, which accelerates the atmosphere against the gravitational force of the planet. The initial velocity is small ($\sim 10^{-5}$~km\,s$^{-1}$) at the lower boundary, but reaches the sonic point within the simulation box, and the atmospheric material exits the simulated region as a supersonic wind. Thus, we simulate the transonic hydrodynamic planetary winds. At the Roche lobe height, all simulated atmospheres reach velocities between 10 and 20~km\,s$^{-1}$.

\subsubsection{Atmosphere of HD\,209458\,b}

First, we focus on the atmosphere of HD\,209458\,b, which is depicted by the green solid lines in Fig.~\ref{fig:hd2vshd1}. Panel (a) shows how the density decreases from the boundary density of $10^{14}$ to $10^{5}$~cm$^{-3}$. The extent of the Roche lobe for HD\,209458\,b is 4.2~$R_{\mathrm{pl}}$ above the substellar point and the sonic point is reached at 3.5~$R_{\mathrm{pl}}$. The optical depth for EUV photons (60~eV) reaches one at a height of 1.03~$R_{\mathrm{pl}}$. This height corresponds well with the atmospheric layer that experiences the maximum volume heating rate.

Panel (b) in Fig.~\ref{fig:hd2vshd1} shows that the absorption of XUV emission around the height of 1.03~$R_{\mathrm{pl}}$ increases the temperature. This raises the atmospheric scale height and the density gradient becomes more shallow above this height. The temperature reaches a maximum of 9100~K. The bulk velocity strongly increases along with the temperature in the lower atmosphere (see panel (c)). 

Panel (d) shows the net specific radiative heating rate throughout the atmosphere. We plot the specific heating rate, which is the heating rate per gram of material, because it clearly shows where the atmospheric material gains most of the energy for the escape. The net heating rate has a maximum close to the temperature maximum, but declines above 1.6~$R_{\mathrm{pl}}$ where the atmosphere is more strongly ionized. There is a distinct dip in the net heating rate at 1.15~$R_{\mathrm{pl}}$, which is caused by hydrogen line cooling. 
This can be seen in panel (e), where we show the radiative heating and cooling rates separately as thin, solid lines. Subtracting the radiative cooling rate from the heating rate gives the net radiative-heating rate (solid thick line), which is displayed in panel (d) and (e). Between 1.05 and 1.8~$R_{\mathrm{pl}}$ about 45\% of the radiative heating produced by the absorption of XUV radiation is counterbalanced by hydrogen line emission, which is mainly \lya{} cooling.

In panel (e) we further compare the radiative heating rate to the hydrodynamic sources and sinks of internal energy. In the atmosphere of HD\,209458\,b the advection of internal energy is a heat sink below 1.7~$R_{\mathrm{pl}}$, but acts as a heat source above this level, providing 50\% of the energy in the upper thermosphere. The adiabatic expansion of the gas is a heat sink throughout the atmosphere.
This term stands for the energy, which is converted into gravitational potential and kinetic energy. It now becomes clear why the temperature decreases in the upper thermosphere: Radiative heating becomes inefficient due to a high degree of ionization, but the adiabatic cooling term remains large. Therefore, the advected thermal energy is used to drive the expansion of the gas. Panel (e) also shows cooling due to thermal conduction, which is largest where the temperature increases most strongly, but it is not relevant in any of the simulations.

In Fig.~\ref{fig:comp_atmos} the H+He atmosphere of HD\,209458\,b is depicted in the middle column; panel (a) shows the fractional abundances. Ionized hydrogen is the major constituent above 1.6~$R_{\mathrm{pl}}$ (H/H$^+$ transition layer). This correlates with the height above which the radiative heating declines. The He/He$^+$ transition occurs slightly below, but charge exchange between H and He$^+$ couples the ionization heights of the two elements closely.
Panel (b) shows the individual radiative heating agents. Ionization of hydrogen is the most important heating agent, but also ionization of helium contributes throughout the atmosphere. Ionization of He$^+$ to He$^{2+}$ contributes less than 10\% of the heating rate in this atmosphere and is most important above the Roche lobe (not depicted in Fig.~\ref{fig:comp_atmos}). Charge exchange between H and He$^+$ is a heat source when hydrogen is weakly ionized.

Panel (c) of Fig.~\ref{fig:comp_atmos} shows the cooling agents and the heating fraction. Where the total radiative cooling rate is small ($<10$\% of the heating rate) the plot is shaded. The most important cooling agent is hydrogen line cooling between 1.06 and 1.7~$R_{\mathrm{pl}}$ (see the discussion of Fig.~\ref{fig:hd2vshd1}~(e) above). Beyond 1.8~$R_{\mathrm{pl}}$ radiative cooling is small; the main cooling agents in this region are hydrogen recombination and free-free emission of electrons (thermal bremsstrahlung).
Only at the bottom of the atmosphere ($<1.02~R_{\mathrm{pl}}$) free-free emission is stronger than the heating rate. At this level, the absorption of \lya{} radiation with following collisional de-excitation is the main local energy source, because ionizing radiation has been absorbed in higher atmospheric layers. The specific cooling at this height is negligible. Helium contributes less than 10\% to the radiative cooling rate in the atmosphere of HD\,209458\,b.

While our simulation shows that HD\,209458\,b produces a strong, transonic wind as a result of the absorption of XUV radiation, it also demonstrates that the hot atmosphere re-emits a significant fraction of the radiative energy input mostly by \lya{} line emission in intermediate thermospheric layers.

\subsubsection{The H-only atmosphere and the H/H$^+$ transition layer} \label{SectHonly}

If the atmosphere of HD\,209458\,b only consisted of hydrogen, the general structure would be similar to an atmosphere including hydrogen and helium (see Fig.~\ref{fig:hd2vshd1}). 
However, an H-only atmosphere of HD\,209458\,b produces a slightly stronger planetary wind. 
On average, including helium in our simulations decreases the mass-loss rate by a factor of two, which contradicts results from \citet{Garcia2007}. Since the involved \lya{} cooling is important for all atmospheres, we will explain the difference here in more detail.

The change in density throughout the atmosphere is larger than the change in velocity when including helium, so let's assume the velocity would remain constant.
Including helium increases the mean molecular weight of the atmosphere, therefore, a higher temperature is needed to sustain the atmospheric density structure and the planetary mass-loss rate. A higher temperature, however, increases the cooling rate by increasing the collisional excitation of hydrogen. \lya{} cooling is proportional to the occupation number of the second level; the relative occupation number can be approximated using the Boltzmann factor, $\exp(-h\nu/kT)$. 
\lya{} radiation mainly escapes from a depth around 1.15~$R_{\mathrm{pl}}$, where the temperatures are about 7900 and 8400~K in the H-only and H+He simulations respectively. The higher temperature in the H+He simulation raises the occupation number of the second level by a factor of two, which results in the increased cooling rate (see Fig.~\ref{fig:comp_atmos}~(c)). 
The simulations of \citet{Garcia2007} did not include \lya{} cooling, hence, he could not find this effect.

The H/H$^+$ transition in the H-only simulation occurs at a height of 2.3~$R_{\mathrm{pl}}$ in contrast to 1.6~$R_{\mathrm{pl}}$ in the H+He simulation. This higher transition layer results from the stronger mass-loss rate, because more neutral hydrogen is advected into the thermosphere. Additionally, the shift of the transition layer further reduces the \lya{} cooling rate by increasing the neutral hydrogen column density above the \lya{} emission layer. 

\citet{Koskinen2013a} found the H/H$^+$ transition to occur at 3.4~$R_{\mathrm{pl}}$ in the atmosphere of HD\,209458\,b. The height of this layer depends on the cooling in the lower thermosphere as seen by the comparison of the H-only and H+He simulations. This was also noted by \citeauthor{Koskinen2013a}; they compared their results with those of \citet{Yelle2004}, who found strong H$_3^+$ cooling and a transition layer height of 1.7~$R_{\mathrm{pl}}$. \citeauthor{Koskinen2013a} argued that H$_3^+$ is probably not formed if metals are present in the atmosphere, but this argument does not hold for \lya{} cooling. Unfortunately, the considerable uncertainties in this type of simulations (see Table~\ref{tabUncertainties}) do not allow a final statement regarding the transition layer height in the atmosphere of HD\,209458\,b.

\subsubsection{Impact of the gravitational potential on the mass loss: HD\,189733\,b versus HD\,209458\,b}\label{Sect:hd2vshd1}

Comparing the atmospheres of HD\,189733\,b and HD\,209458\,b, one finds three major differences:
The atmosphere of HD\,189733\,b is hotter, the expansion velocity is smaller below the Roche lobe, and the net radiative-heating rate is also smaller (see Fig.~\ref{fig:hd2vshd1}), although the XUV irradiation level is actually 16 times higher. The maximum temperature in the atmosphere of HD\,189733\,b is 2700~K higher than in HD\,209458\,b, which is also a result of less adiabatic cooling of the weaker planetary wind. The atmosphere reaches the sonic point just below the Roche lobe height, above which the material is accelerated freely toward the stellar surface in the 1D simulation. The temperature decreases more strongly in the upper thermosphere, because the degree of ionization is high and only little radiative heating remains.

In the case of HD\,189733\,b it is remarkable that the simulated wind is weaker in terms of the total mass-loss rate although the irradiation level is higher than in HD\,209458\,b. The weaker wind is a result of increased radiative cooling, which actually depends on the depth of the gravitational potential well of the planet. This can be demonstrated with a thought experiment: Let us start with a planet that hosts a strong wind and only weak radiative cooling. What happens if we increase the mass of the planet? 
\begin{itemize}

\item
The first reaction will be a contraction of the atmosphere, because the gravitational attraction was increased. Let us only slightly increase the planetary mass, so that the upward flux of the planetary wind does not collapse but only becomes weaker.

\item
The smaller mass flux has two effects: First, the adiabatic cooling decreases, because a smaller velocity means that the atmospheric expansion proceeds slower. This causes a higher atmospheric temperature. Second, the advection of neutral hydrogen into the upper atmosphere is weaker, which shifts the ionization front closer to the planetary surface.

\item
The increased temperature partially counterbalances the higher gravity and prevents the atmosphere to contract further, but it also increases radiative cooling (\lya{} and free-free emission). \lya{} cooling is further increased by the now smaller neutral hydrogen column density above the emission layer, which increases the escape probability for the radiation.

\item
The radiative cooling reduces the available energy for driving the planetary wind.

\item
The outflow settles at a lower mass-loss rate and the resulting atmosphere is hotter and more strongly ionized.
\end{itemize}
If the radiative cooling did not increase with the temperature, the down settling of the atmosphere in the first step would be a transient effect and the mass-loss rate would be restored to the original value.  Of course, we can drive this thought experiment further by again  increasing the mass of the hypothetical planet. At a certain mass the radiative energy input will be completely re-emitted by a hot and more compact thermosphere. This explains a transition in the atmospheres of hot exoplanets. Smaller planets host strong and cool planetary winds, and massive and compact planets host stable, hot, hydrostatic thermospheres.

The intermediate case is precisely what occurs in the atmosphere of HD\,189733\,b. Most of the EUV photons are absorbed at a height of 1.08~$R_{\mathrm{pl}}$. While the heating fraction is close to unity at this height in the atmosphere of HD\,209458\,b, in HD\,189733\,b it ranges from 5\% to 20\%, which results in only weak acceleration of the planetary wind. Hydrogen and helium are mainly ionized above 1.2~$R_{\mathrm{pl}}$ and helium is doubly ionized above 2.1~$R_{\mathrm{pl}}$ (see Fig.\ref{fig:comp_atmos}). 
The overall structure of the heating and cooling agents is similar to that in the atmosphere of HD\,209458\,b, but shifted toward smaller heights. Photoionization of He$^+$ to He$^{2+}$ is the main radiative heating agent above 2.0~$R_{\mathrm{pl}}$. However, the total radiative heating in the upper thermosphere ($>2.4~R_{\mathrm{pl}}$) is small compared to the advected energy. The lack of radiative heating in the upper thermosphere of HD\,189733\,b makes it cooler than that of HD\,209458\,b.

\subsubsection{Atmospheric structure at the thermobase}\label{Sect:onset_thermo}

\begin{figure}[tb]
  \centering
  \includegraphics[width=\hsize]{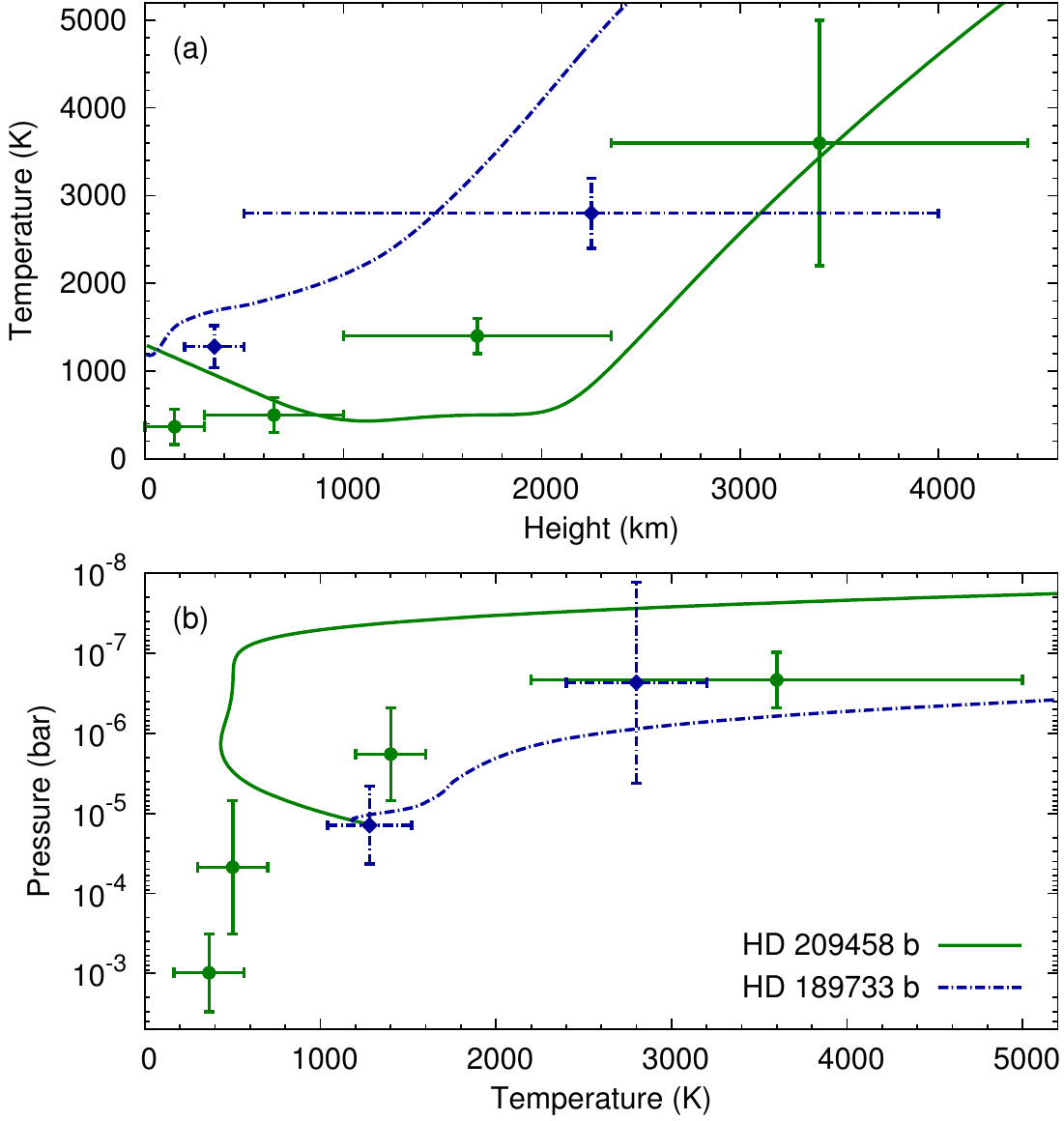}
  \caption{Atmospheric structure at the thermobase.
           Observationally deduced values are marked by
           dots, and the simulated structures by lines.
           Solid lines green lines show values of HD\,209458\,b
           and dash-dotted blue lines those of HD\,189733\,b.
           Panel (a) shows the temperature structure and panel
           (b) depicts the temperature-pressure profiles in the
           atmospheres.
           }
  \label{fig:onset_thermo}
\end{figure}

For HD\,209458\,b and HD\,189733\,b we can compare the temperature structure of the lower atmosphere with those derived from atmospheric absorption measurements in the sodium D lines \citep{Vidal2011, Vidal2011b, Huitson2012}.
In contrast, the inferred inverted or non-inverted temperature profiles obtained from Spitzer data or recently based on high-dispersion spectroscopy in the infrared (e.g., CHRIRES on the Very Large Telescope) are sensitive to atmospheric layers below the 10$^{-5}$~bar level and, thus, below our simulated region \citep{Line2014, Schwarz2015}.

Our atmospheric temperature structures show a reasonable agreement with the Na~D observations (see Fig.~\ref{fig:onset_thermo}~(a)). For the lower atmosphere of HD\,209458\,b, \citet{Vidal2011,Vidal2011b} found a temperature of $365\pm200$~K, which is below our adopted temperature of 1320~K derived from Spitzer observations \citep{Crossfield2012}. Decreasing the boundary temperature in our simulations slightly shifts the atmospheric profile toward smaller heights (see Fig.~\ref{fig:BCtemp}), so that we could certainly produce a better fit by adopting a lower boundary temperature. However, we simulated the substellar point at which the atmosphere is expected to be hotter than at the terminator, which is probed by the observations. The temperature minimum of 500~K in our simulated atmosphere of HD\,209458\,b agrees remarkably well with the low temperatures deduced from the observations.

While the observed temperature-pressure profile of HD\,189733\,b is reproduced by our simulations, the atmosphere of HD\,209458\,b shows more pronounced differences (see Fig.~\ref{fig:onset_thermo}~(b)). In our simulations the temperature rise at the thermobase occurs at lower pressure levels, thus higher in the planetary atmosphere than in the observations. This disagreement has been noted before for various other 1D escape models of this planetary atmosphere \citep{Vidal2011, Vidal2011b, Huitson2012}.
Our test simulations including molecules show a potential source for this discrepancy, namely H$^-$ dominates the atmospheric heating and cooling at the relevant heights (see Sect.~\ref{SectMetalsMolec}). While these test simulations are not fully converged and require a more detailed analysis, preliminary results reproduce the observed temperature-pressure profiles more closely in both HD\,209458\,b and HD\,189733\,b.

\subsection{The atmospheres of hot gas planets in comparison} \label{SectCompare}
                                                                                                                                                                       
\begin{figure*}[t]
  {\bf\footnotesize \hspace{24mm} low gravitational potential \hspace{57mm} high gravitational potential \\}
  \vspace{-6pt} \\    
  \centering                                                                                                                                                                
  \includegraphics[width=0.485\hsize]{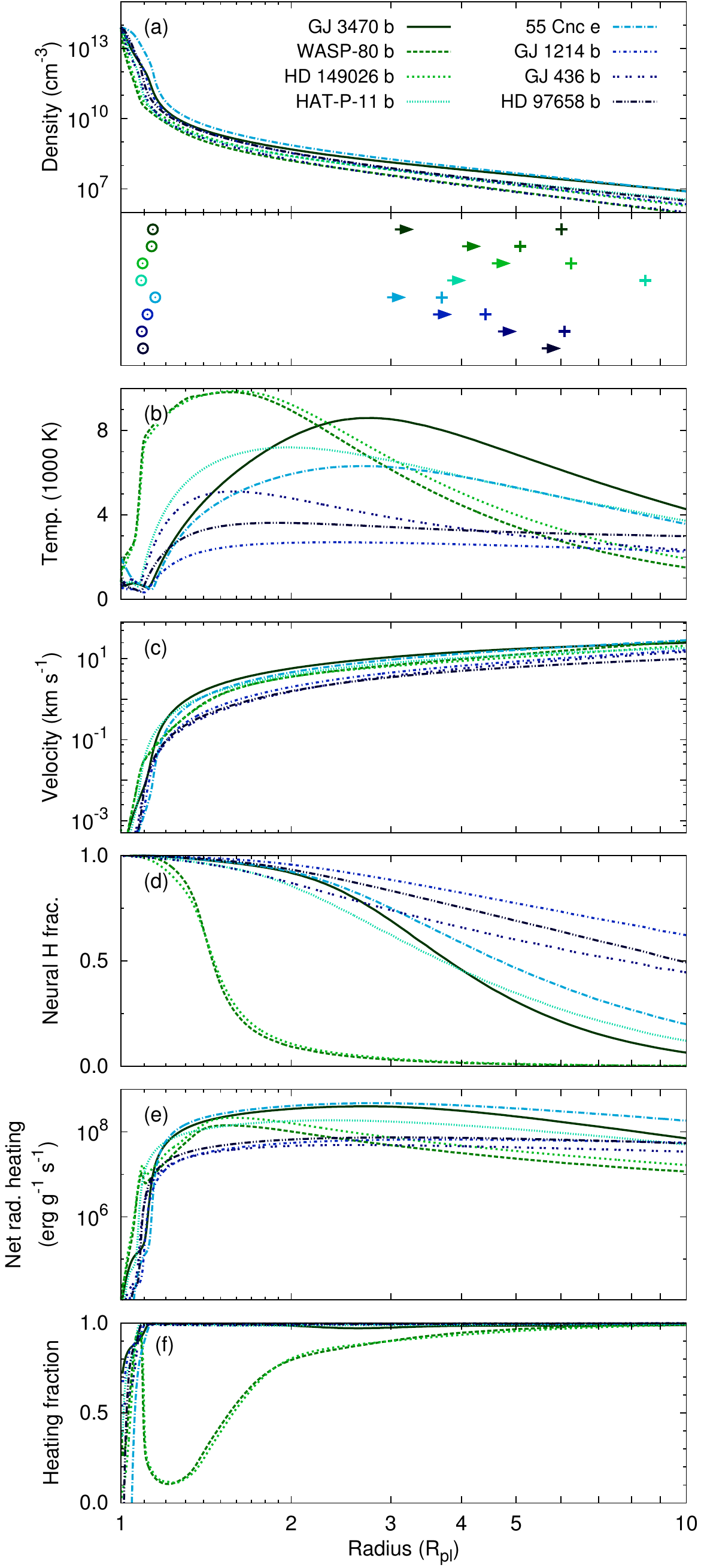}
  \hfill                                                                                                                                                                         
  \includegraphics[width=0.485\hsize]{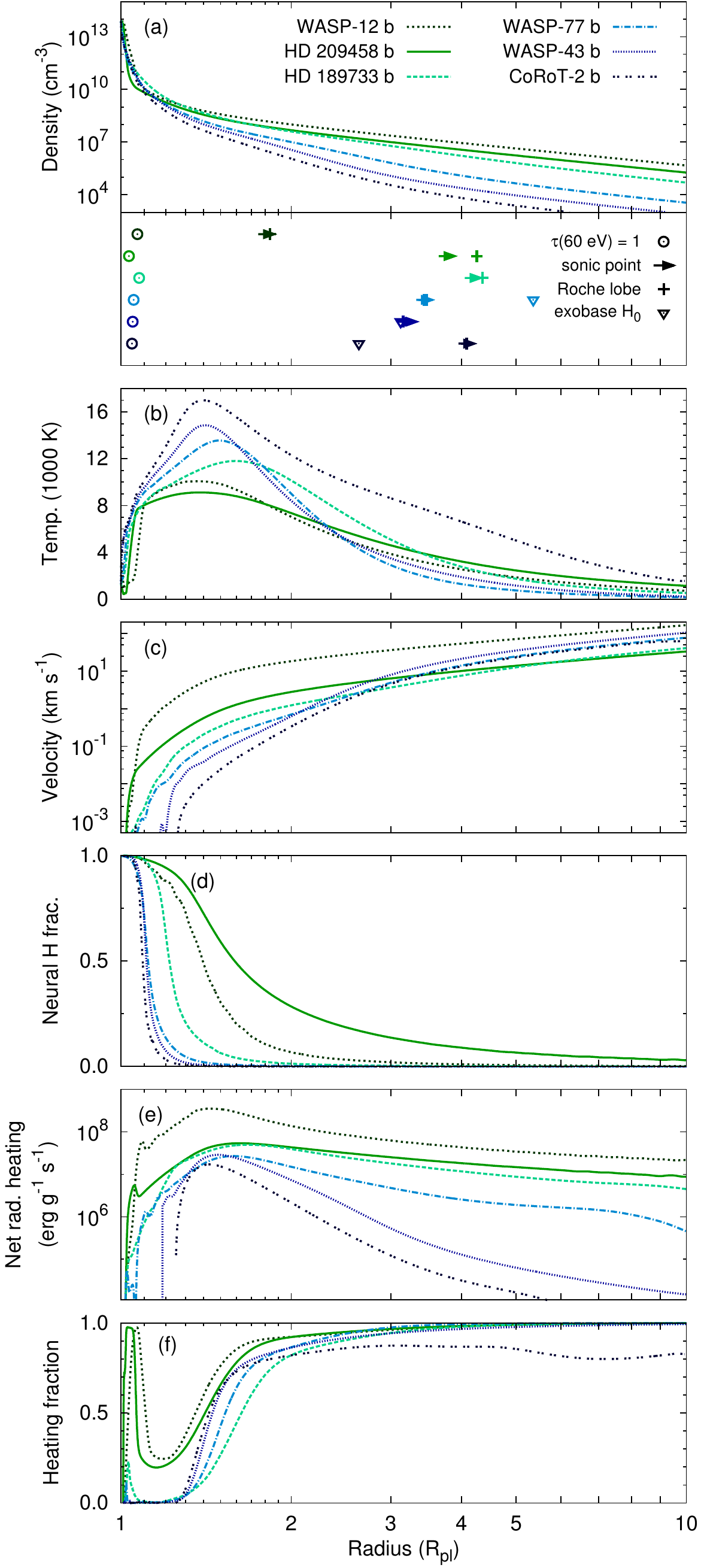}
  \caption{Atmospheres of the planets in the sample.                                                                                                                                               
           Low-potential planets
           are shown on the lefthand side and high-potential planets
           on the righthand side.
           Time averaged profiles are shown for high-potential planets.
           Symbols in the lower panel (a) are the same as in Fig.\ref{fig:hd2vshd1};
           additionally the height where the atmospheres become collisionless for
           neutral hydrogen are marked (triangles).
           On each side the planets are sorted from high (GJ\,3470\,b/WASP-12\,b) to low (HD\,97658\,b/CoRoT-2\,b) mass-loss rates,
           which is also referenced by a color scheme from green to blue.
           }
  \label{fig:sim_result}
\end{figure*}

We now compare the atmospheres of our sample of hot gas planets. HAT-P-2\,b, HAT-P-20\,b, WASP-8\,b, and WASP-10\,b are excluded, because these compact planets host stable thermospheres (see Sect.~\ref{SectJeans}). In the following, we discuss how the atmospheres depend on system parameters such as irradiation level, planetary density, or the planetary gravitational potential. We either compare two planets, which differ mostly in a single system parameter, or we compare two subsamples, for which we divide our systems loosely at a gravitational potential of $\log -\Phi_{\mathrm{G}} = \log GM_{\mathrm{pl}}/R_{\mathrm{pl}} = 13.0$ in units of erg\,g$^{-1}$. The cut allows us to explain differences in the atmospheres of high-potential and low-potential planets, which we use as a short term for the strength of the gravitational potential per unit mass.

The atmospheres of the complete sample of planets are shown in Fig.~\ref{fig:sim_result}, which is split into two columns: low-potential planets are depicted in the lefthand column and high-potential planets in the righthand column. In the figure each subsample is ordered with respect to the mass-loss rate. GJ\,3470\,b has the strongest mass-loss rate among the low-potential planets and WASP-12\,b is its counterpart among the high-potential planets. 
HD\,97658\,b and CoRoT-2\,b are the planets with the smallest mass-loss rates among the subsamples. The general behavior of all atmospheres follows our explanations in Sect.~\ref{SectGenStruc}. The density decreases with height, the temperature reaches several thousand degrees maximally and decreases in the upper thermosphere, and all atmospheres reach a sonic point between 3 and 5~$R_{\mathrm{pl}}$, except for the atmosphere of WASP-12\,b (see Sect.~\ref{Sect:wasp12}).

\subsubsection{Atmospheric temperature and the H/H$^+$ transition layer}

A comparison of both columns in Fig.~\ref{fig:sim_result} shows that high-potential planets have a lower average density in the thermosphere. The temperature maximum always occurs between 1.3 and 3~$R_{\mathrm{pl}}$ in the atmospheres, and rises from low-potential (2700~K in GJ\,1214\,b) to high-potential planets (16\,800~K in CoRoT-2\,b). The level of irradiation only weakly affects the maximal temperature in the atmospheres, e.g., the irradiation level of WASP-12\,b is 16 times higher than in HD\,209458\,b, but the maximal temperature is only about 10\,\% higher. Strong adiabatic cooling lowers the thermospheric temperature of WASP-12\,b.

Super-Earth-sized planets have atmospheres that are almost completely neutral throughout our simulation box. As the planets become more massive and compact, the H/H$^+$ transition layer moves closer to the planetary surface (see Fig.~\ref{fig:sim_result}~(d)), which is mainly a result of the higher temperature in the thermospheres of high-potential planets. To some degree the irradiation strength also affects the height of the transition layer, which can be demonstrated by comparing GJ\,436\,b with HAT-P-11\,b. 
Both planets have similar densities and gravitational potentials. Despite the stronger planetary wind of HAT-P-11\,b, the H/H$^+$ transition occurs at a height of 3.7~$R_{\mathrm{pl}}$ in its thermosphere compared to 7.8~$R_{\mathrm{pl}}$ in GJ\,436\,b. Thus, the higher XUV irradiation level of HAT-P-11\,b shifts the transition layer to smaller heights. The almost neutral atmospheres of GJ\,1214\,b and HD\,97658\,b result from
the small gravitational potentials of the exoplanets in combination with moderate XUV irradiation levels. These two thermospheres would also contain a significant fraction of H$_2$, which is neglected here (see the discussion about molecules in Sect.~\ref{SectMetalsMolec}).

The specific heating rate is more homogeneously distributed throughout the thermospheres of low-potential planets (see Fig.~\ref{fig:sim_result}~(e)), which is a result of a higher neutral hydrogen fraction in their atmospheres.
In contrast, the upper thermospheres of WASP-77\,b, WASP-43\,b, and CoRoT-2\,b experience negligible amounts of radiative heating, because both hydrogen and helium are almost completely ionized. For example, the fractional abundance of He$^{2+}$ is $>0.99$ above 2~$R_{\mathrm{pl}}$ in CoRoT-2\,b.

\subsubsection{Approaching radiative equilibrium: the effect of the gravitational potential}\label{Sect:Equi}

The $ \tau(60~\mathrm{eV}) = 1$ layer always occurs between 1.03 and  1.15~$R_{\mathrm{pl}}$ for all planets\footnote{
We need to keep in mind, that we plot the atmospheres versus the planetary radius, thus, the true size of the simulation box for WASP-12 is almost ten times larger than for 55\,Cnc\,e.}.
The planetary wind is strongly accelerated at this height in low-potential planets, but planets with a higher potential ($\log -\Phi_{\mathrm{G}} > 13.2$) show almost no acceleration in this atmospheric layer (see Fig.~\ref{fig:sim_result}~(c)).
Their atmospheres are more gradually accelerated between 1.1 and 2~$R_{\mathrm{pl}}$.
The small amount of acceleration in the lower thermospheres of high-potential planets is a result of strong radiative cooling. Consider, for example, CoRoT-2\,b with the strongest irradiation among our sample. This planet experiences the smallest net radiative-heating rate. Below 1.3~$R_{\mathrm{pl}}$ the atmosphere is close to radiative equilibrium (see Fig.~\ref{fig:sim_result}~(f)), so that \lya{} and free-free emission re-emit almost the complete radiative energy input of $1.55\times 10^5$~erg\,cm$^{-2}$\,s$^{-1}$. Radiative cooling of these two cooling agents becomes more and more important going from planets with a low gravitational potential to high-potential planets.

\lya{} cooling starts to re-emit larger fractions of the energy input at a height around 1.2~$R_{\mathrm{pl}}$ in the atmospheres of WASP-80\,b and HD\,149026\,b (see panels (f) in Fig.~\ref{fig:sim_result}). This cooling is small in the smallest planets, and only becomes important once the atmospheres are heated to more than 8000~K. For example, at a temperature of 5000~K the collisional excitation of the second level in hydrogen is reduced by a factor of 7000, thus, \lya{} cooling is small. In high-potential planets the $\tau(60~\mathrm{eV}) = 1$ layer occurs close to the bottom of the \lya{} emission layer. As a result, in HD\,209458\,b 80\% of the absorbed energy delivered by these photons is re-emitted, in HD\,189733\,b 99\% is re-emitted and in WASP-43\,b this fraction increases to about 99.8\%.

Radiation with higher energies penetrates deeper into the atmospheres and heats the atmosphere of HD\,209458\,b below the \lya{} emission layer. Free-free emission, the second major cooling agent, is weak in the atmosphere of HD\,209458\,b, which is reflected by a heating fraction of almost one around 1.04~$R_{\mathrm{pl}}$. This leads to heating and acceleration of the planetary wind by stellar irradiation with photon energies higher than 60~eV. The heating below the \lya{} emission layer becomes less important in the atmosphere of HD\,189733\,b and the heating fraction is reduced to 0.001 for WASP-43\,b, which is a result of increased free-free emission in hot, lower atmospheric regions. This temperature increase of the dense atmospheric layers close to the lower boundary in massive and compact planets can be seen in panel (b) of Fig.~\ref{fig:sim_result}.

\subsection{Mass loss rates}\label{Sect:mloss}

\begin{table*}[t]
\setlength{\tabcolsep}{5pt}
\small
\caption{Irradiation of the planets and results from the simulations. 
         Ranking according to the simulated mass-loss rates.}
\label{tabSim}      
\centering
\begin{tabular}{l @{\hspace{6pt}}l cccccccccccc}
\hline\hline\vspace{-5pt}\\
  System &  
   & 
  $\log L_{\mathrm{X}}$ & 
  Ref. & 
  $\log L_{\mathrm{Ly}\alpha}$  &
  $\log F_{\mathrm{Ly}\alpha}$\!\tablefootmark{c} & 
  Ref. & 
  $\log L_{\mathrm{EUV}}$  &
  $\log F_{\mathrm{XUV}}$\!\tablefootmark{d} & 
  $\log -\Phi_{\mathrm{G}}$ & 
  $\log \dot{M}^{\mathrm{sim}}$ & 
  $\dot{M}$ &
  $\dot{M}^{\mathrm{tot}}_{\mathrm{Y}}$ &
  \lya{} abs.  \\ 
  \vspace{-9pt}\\
   & 
   &  
  (erg\,s$^{-1}$) &
   & 
  (erg\,s$^{-1}$) & 
    &
    &
  (erg\,s$^{-1}$) &
    &
  (erg\,g$^{-1}$)  &
  (g\,s$^{-1}$) &
  (\textperthousand\,Ga$^{-1}$) &
  (\textperthousand) &
  (m\AA{}) \\
  \vspace{-7pt}\\ \hline\vspace{-5pt}\\ 
  \object{WASP-12 b}         &&             $<$\,27.58  & 7 &            $<$\,28.58 &            $<$\,$-14.66$ & 12 &            $<$\,28.35  &            $<$\,4.26 &
                     13.14 & 
                     11.60                  & 
                     4.8 &
                     \hspace{-0.5em}14.6 &
                     \hphantom{1}90 (690) \\                      
  \vspace{-10pt}\\
  \object{GJ 3470 b}        &&  \hphantom{$<$}\,27.63  & 7 & \hphantom{$<$}\,28.60 & \hphantom{$<$}\,$-12.41$ & 12 & \hphantom{$<$}\,28.37  & \hphantom{$<$}\,3.89 & 
                     12.33 &
                     10.66                  & 
                     \hspace{-0.5em}17.5 &
                     \hspace{-0.5em}13.2 &
                     180 (290) \\
  \vspace{-10pt}\\
  \object{WASP-80 b}         &&  \hphantom{$<$}\,27.85  & 1 & \hphantom{$<$}\,28.67 & \hphantom{$<$}\,$-12.97$ & 12 & \hphantom{$<$}\,28.46  & \hphantom{$<$}\,4.03 & 
                     13.02 &
                     10.55                  & 
                     1.1 &
                     1.5 &
                     320 \hphantom{1}(61) \\
  \vspace{-10pt}\\
  \object{HD 149026 b}      &&  \hphantom{$<$}\,28.60  & 7 & \hphantom{$<$}\,28.91 & \hphantom{$<$}\,$-12.97$ & 12 & \hphantom{$<$}\,28.80  & \hphantom{$<$}\,4.29 & 
                     13.00 &
                     10.43                  & 
                     1.2 &
                     0.9 &
                     \hphantom{1}87 \hphantom{1}(63) \\
  \vspace{-10pt}\\
  \object{HAT-P-11 b}        &&  \hphantom{$<$}\,27.55  & 7 & \hphantom{$<$}\,28.57 & \hphantom{$<$}\,$-12.65$ & 12 & \hphantom{$<$}\,28.33  & \hphantom{$<$}\,3.51 & 
                     12.55 &
                     10.29                  & 
                     4.0 &
                     8.1 &
                     150 (130) \\
  \vspace{-10pt}\\
  \object{HD 209458 b}      &&             $<$\,26.40  & 4 & \hphantom{$<$}\,28.77 & \hphantom{$<$}\,$-12.70$ &  8 &            $<$\,27.84  &            $<$\,3.06 & 
                     12.96 &
                     10.27                  & 
                     0.4 &
                     4.8 &
                     160 (200)   \\
  \vspace{-10pt}\\
  \object{55 Cnc e}         &&  \hphantom{$<$}\,26.65  & 4 & \hphantom{$<$}\,28.06 & \hphantom{$<$}\,$-11.98$ & 11 & \hphantom{$<$}\,27.66  & \hphantom{$<$}\,3.87 & 
                     12.41 &
                     10.14                  & 
                     9.0 &
                     \hspace{-0.5em}49.6 &
                     \hphantom{1}62 \hphantom{1}(30) \\
  \vspace{-10pt}\\
  \object{GJ 1214 b}        &&  \hphantom{$<$}\,25.91  & 5 &            $<$\,25.69 &            $<$\,$-14.62$ & 10 & \hphantom{$<$}\,26.61  & \hphantom{$<$}\,2.93 & 
                     12.19 &
                     \hphantom{1}9.68       & 
                     3.9 &
                     \hspace{-0.5em}18.8 &
                     200 (280) \\
  \vspace{-10pt}\\
  \object{GJ 436 b}         &&  \hphantom{$<$}\,25.96  & 4 & \hphantom{$<$}\,27.65 & \hphantom{$<$}\,$-12.46$ & 10 & \hphantom{$<$}\,27.14  & \hphantom{$<$}\,2.80 & 
                     12.54 &
                     \hphantom{1}9.65       & 
                     1.0 &
                     4.7 &
                     160 (210)  \\
  \vspace{-10pt}\\
  \object{HD 189733 b}      &&  \hphantom{$<$}\,28.18  & 4 & \hphantom{$<$}\,28.43 & \hphantom{$<$}\,$-12.22$ &  9 & \hphantom{$<$}\,28.61  & \hphantom{$<$}\,4.32 & 
                     13.26 &
                     \hphantom{1}9.61       & 
                     0.1 &
                     0.1 &
                     160 \hphantom{1}(22) \\
  \vspace{-10pt}\\
  \object{HD 97658 b}       &&  \hphantom{$<$}\,27.22  & 7 & \hphantom{$<$}\,28.46 & \hphantom{$<$}\,$-12.26$ & 12 & \hphantom{$<$}\,28.19  & \hphantom{$<$}\,2.98 & 
                     12.33 &
                     \hphantom{1}9.47       & 
                     2.0 &
                     5.9 &
                     110 \hphantom{11}(0) \\
  \vspace{-10pt}\\
  \object{WASP-77 b}         &&  \hphantom{$<$}\,28.13  & 1 & \hphantom{$<$}\,28.76 & \hphantom{$<$}\,$-13.26$ & 12 & \hphantom{$<$}\,28.59  & \hphantom{$<$}\,4.51 & 
                     13.42 &
                     \hphantom{1}8.79       & 
                     \hspace{-0.8em}$<$\,0.1 &
                     --- &
                     \hphantom{1}92 \hphantom{11}(0) \\
  \vspace{-10pt}\\
  \object{WASP-43 b}         &&  \hphantom{$<$}\,27.88  & 2 & \hphantom{$<$}\,28.68 & \hphantom{$<$}\,$-13.21$ & 12 & \hphantom{$<$}\,28.48  & \hphantom{$<$}\,4.82 & 
                     13.54 &
                     \hphantom{1}8.04       & 
                     \hspace{-0.8em}$<$\,0.1 &
                     --- &
                     \hphantom{1}92 \hphantom{11}(0) \\
  \vspace{-10pt}\\
  \object{CoRoT-2 b}         &&  \hphantom{$<$}\,29.32  & 6 & \hphantom{$<$}\,29.14 & \hphantom{$<$}\,$-13.80$ & 12 & \hphantom{$<$}\,29.13  & \hphantom{$<$}\,5.19 & 
                     13.61 &
                     \hphantom{1}7.63       & 
                     \hspace{-0.8em}$<$\,0.1 &
                     --- &
                     \hphantom{1}87 \hphantom{11}(0) \\
  \vspace{-10pt}\\
  \object{WASP-8 b}          &&  \hphantom{$<$}\,28.45  & 1 & \hphantom{$<$}\,28.86 & \hphantom{$<$}\,$-13.10$ & 12 & \hphantom{$<$}\,28.73  & \hphantom{$<$}\,3.66 & 
                     13.57 &
                     \hphantom{1}{\it 5.0}\hphantom{1}  & 
                     \hspace{-0.8em}$<$\,0.1 &
                     --- &
                     \hphantom{1}63 \hphantom{11}(0) \\
  \vspace{-10pt}\\
  \object{WASP-10 b}         &&  \hphantom{$<$}\,28.09  & 1 & \hphantom{$<$}\,28.74 & \hphantom{$<$}\,$-13.24$ & 12 & \hphantom{$<$}\,28.57  & \hphantom{$<$}\,4.08 & 
                     13.73 &
                     \hphantom{1}{\it 2.7}\hphantom{1}  & 
                     \hspace{-0.8em}$<$\,0.1 &
                     --- &
                     \hphantom{1}80 \hphantom{11}(0) \\
  \vspace{-10pt}\\                                                                                                                                                       
  \object{HAT-P-2 b}         &&  \hphantom{$<$}\,28.91  & 1 & \hphantom{$<$}\,29.01 & \hphantom{$<$}\,$-13.19$ & 12 & \hphantom{$<$}\,28.94  & \hphantom{$<$}\,4.12 & 
                     14.14 &
                     {\it \!\!$<$\,5.9}\hphantom{1}         & 
                     \hspace{-0.8em}$<$\,0.1 &
                     --- &
                     \!\!$<$\,70 \hphantom{11}(0) \\ 
  \vspace{-10pt}\\                                                                                                                                                    
  \object{HAT-P-20 b}        &&  \hphantom{$<$}\,28.00  & 1 & \hphantom{$<$}\,28.72 & \hphantom{$<$}\,$-13.05$ & 12 & \hphantom{$<$}\,28.53  & \hphantom{$<$}\,4.08 & 
                     14.18 &
                     {\it \!\!$<$\,4.5}\hphantom{1}         & 
                     \hspace{-0.8em}$<$\,0.1 &
                     --- &
                     \!\!$<$\,60 \hphantom{11}(0) \\
  \vspace{-10pt}\\
  \object{WASP-38 b}         &&             $<$\,28.04  & 1 &            $<$\,28.73 &            $<$\,$-13.43$ & 12 &            $<$\,28.55  &            $<$\,3.46 & 
                     13.65 &
                     ---                    & 
                     --- &
                     --- &
                     --- \\
  \vspace{-10pt}\\
  \object{WASP-18 b}         &&             $<$\,26.82  & 3 &            $<$\,28.34 &            $<$\,$-13.74$ & 12 &            $<$\,28.02  &            $<$\,3.99 & 
                     14.16 &
                     ---                    & 
                     --- &
                     --- &
                     --- \\
  \vspace{-10pt}\\
  \object{55 Cnc b}         &&  \hphantom{$<$}\,26.65  & 4 & \hphantom{$<$}\,28.06 & \hphantom{$<$}\,$-11.98$ & 11 & \hphantom{$<$}\,27.66  & \hphantom{$<$}\,2.14 & 
                     ---      &
                     ---                    & 
                     --- &
                     --- &
                     ---   \\
  \vspace{-6pt}\\ \hline
\end{tabular}                                                                                                                                                                    
\tablefoot{Explanation of the columns: 
           planetary name,
           X-ray luminosity (0.124-2.48~keV),
           Ly$\alpha$ luminosity \citep{Linsky2013},
           Ly$\alpha$ flux at Earth neglecting interstellar
           absorption \tablefoottext{c}{(erg\,cm$^{-2}$\,s$^{-1}$)},
           EUV lum. \citep[100-912~\AA{},][]{Linsky2014},
           XUV flux at planetary distance \tablefoottext{d}{($<$\,912~\AA{}, erg\,cm$^{-2}$\,s$^{-1}$)},
           mass-loss rate from the simulations 
           ($1/4 \hspace{0.05em}\times\hspace{0.05em} 4\pi R^2\,\rho\mathrm{v}$),
           total fractional mass loss in parts per thousand per Ga,
           fractional cumulative mass loss during first 100~Ma in parts per thousand,
           equivalent width of the \lya{} absorption signal and in brackets the additional equivalent width caused by unbound hydrogen (optical transit depth is subtracted).
           }     
\tablebib{(1) \citet[][]{Salz2015a}; 
          (2) \citet{Czesla2013};
          (3) \citet{Pillitteri2014};
          (4) \citet{Sanz2011};
          (5) \citet{Lalitha2014};
          (6) \citet{Schroeter2011};
          (7) prediction based on rotation period and mass \citep{Pizzolato2003};
          (8) \citet{Wood2005};
          (9) \citet{Bourrier2013};
          (10) \citet{France2013};
          (11) \citet{Ehrenreich2012};
          (12) rotation based prediction from \citep{Linsky2013}.
}                                                                                           
\end{table*}

In the simulations we evaluate the mass-loss rates using the formula $\dot{M}^{\mathrm{sim}}= 1/4 \hspace{0.05em}\times\hspace{0.05em} 4\pi R^2\,\rho\mathrm{v}$. The additional factor of $1/4$ results from averaging the irradiation over the planetary surface. In our spherical simulation the planet is irradiated from all sides, but in reality only the dayside is illuminated. The difference of the spherical surface and the surface of the planetary dayside disk is $1/4$. The factor is also supported by two-dimensional simulations \citep{Stone2009}. A sorted list of the calculated mass-loss rates is provided in Table~\ref{tabSim}. 
We also provide two more mass loss columns: $\dot{M}$ is the fractional mass loss per Ga, and $\dot{M}^{\mathrm{tot}}_{\mathrm{Y}}$ is an estimate for the total, fractional mass loss during the first 100~Ma after formation of the planetary system. This value is derived by linearly scaling the mass loss according to the expected irradiation from the young host stars. The XUV emission is derived in the same fashion as before, but assuming average X-ray luminosities for young stars between the T Tauri phase and the age of the Pleiades (100~Ma) from \citet{Stelzer2001}, i.e., G-type $\log L_{\mathrm{X}} \approx 29.7$, K-type $\log L_{\mathrm{X}} \approx 29.5$, and M-type $\log L_{\mathrm{X}} \approx 29.0$. 

In general, the current mass-loss rates of the planets in our sample are most strongly affected by the gravitational potential, the irradiation level and by the planetary density. The effect of the gravitational potential is directly visible in Fig.~\ref{fig:sim_result}, looking at the low thermospheric densities of high-potential planets.
The impact of the irradiation level can be demonstrated by coming back to our comparison of GJ\,436\,b with HAT-P-11\,b. HAT-P-11\,b experiences a five times higher irradiation level and its wind is 4.4 times stronger. The effect of the planetary density on the mass-loss rates can be seen by comparing HD\,149026\,b and WASP-80\,b, which both have almost the same gravitational potential. The irradiation of HD\,149026\,b is almost twice as strong as in WASP-80\,b, but the mass-loss rate is lower. This is a result of the higher planetary density.

Our simulations show that none of the planets is currently losing mass at a rate that would evaporate the complete planet over its lifetime. For WASP-12\,b this contradicts earlier mass loss estimates (see Sect.~\ref{Sect:wasp12}). 
Among our sample 55\,Cnc\,e is highly interesting, because the planetary density of 4~g\,cm$^{-3}$ suggests that the planet must consist of a significant amount of volatiles \citep[$\sim$\,20\%,][]{Gillon2012}. Our calculated mass-loss rate of $1.4\times10^{10}$~g\,s$^{-1}$ is higher than the upper limit set by \citet{Ehrenreich2012} and would erode all volatiles in 23~Ga. Nevertheless, the simulations do not contradict the observations of \citet{Ehrenreich2012} as we show in Sect.~\ref{Sect:Lya_compare}. Furthermore, the planet may have lost 5\,\% of its initial mass at young ages. We caution that these results are based on hydrogen-dominated atmospheres. The planetary composition has most likely been altered over the lifetime of 55\,Cnc\,e and although the planet can maintain a hydrogen-dominated atmosphere on long timescales, large amounts of volatiles are lost in the planetary wind.

\subsection{The extreme case of WASP-12\,b}\label{Sect:wasp12}

WASP-12\,b is an exception among all simulated planets. Its extreme proximity of only 2 stellar radii above the stellar photosphere reduces the size of the Roche lobe to only 1.8~$R_{\mathrm{pl}}$ above the substellar point. According to the derivation of \citet{Erkaev2007}, the gravitational potential of the planet is reduced by a factor of $K = 0.28$. This causes the strongest mass-loss rate among the planets in our sample, which is, however, 10\,000 times weaker than the proposed mass loss of \citet{Li2010} and still 400 times weaker than value of \citet{Lai2010}. In this case a multidimensional simulation can result in a higher mass-loss rate, because of the elongated shape of the planetary Roche lobe. The Roche lobe height perpendicular to the star-planet axis is $2/3 \hspace{0.05em}\times\hspace{0.05em} 1.8~R_{\mathrm{pl}} = 1.2~R_{\mathrm{pl}}$ \citep{Li2010}, which results in a gravitational reduction factor of $K = 0.04$. However, this is not sufficient to explain the differences of our result and the published values.

For their mass loss estimates, both \citeauthor{Li2010} and \citeauthor{Lai2010}  derived a density of $3\times10^{-12}$~g\,cm$^{-3}$ at $1.2~R_{\mathrm{pl}}$ and assume a bulk flow with the speed of sound.
The density in our simulation decreases more strongly with height, due to a temperature minimum of 1000~K at 1.014$~R_{\mathrm{pl}}$, resulting in a density that is a factor of 600 lower. Thus, the authors assumption of an isothermal atmosphere with a temperature of 3000~K is incorrect. Furthermore the sound speed is reached at the L1 point and at 1.2$~R_{\mathrm{pl}}$ it is a factor 10 lower than the sound speed. These two effects cause the large differences between our simulated mass-loss rate and the estimates of \citet{Li2010} and \citet{Lai2010}. While according to \citeauthor{Li2010} the planet would have a lifetime of only 10~Ma, our simulation results in about $0.28/0.04 \times 0.48\%=3.4\,\%$ fractional mass loss per Ga. In respect to the planetary mass loss, this resolves the problem of the small chance to discover the planet in a brief period of its lifetime.

\subsection{Exobase and stable thermospheres}\label{SectJeans}

Hydrodynamic simulations are only valid as long as collisions are frequent enough and a Maxwellian velocity distribution is maintained. This is not always true in planetary atmospheres because of the exponential density decrease with height. At a certain height collisions become too infrequent and atmospheric particles move without collisions on parabolic or hyperbolic paths. This height is called the exobase. Energetic atoms from the tail of the Maxwellian velocity distribution can escape from the exobase, a process called Jeans escape.
Jeans escape appears on all planetary bodies, but the escape rates are quite small. For expanded thermospheres of hot gas planets Jeans escape is typically one order of magnitude smaller than hydrodynamic escape \citep{Tian2005, Garcia2007}. Thus, hydrodynamic escape is the dominant escape process, if the hydrodynamic approximation is valid. We verify the validity of the hydrodynamic simulations in retrospect by determining the height of the exobase, which is defined as the height above which the collision probability for atmospheric constituents is unity in the outward direction \citep[e.g.,][]{Tian2005}.

The exobase height crucially depends on the collision cross-section. For planetary atmospheres a typical collision cross-section often used is $\sigma = 3.3\times10^{-15}$~cm$^2$ \citep{Chamberlain1963, Hunten1973, Tian2005, Garcia2007}. The calculations of \citet{Dalgarno1960} confirm that this value also holds for collisions of neutral hydrogen, $\sigma = 4.3\times10^{-15}$~cm$^2$. However, some authors argue for a collisionless regime closer to the planetary photosphere. They use an H-H collision cross-section of $\sigma = 3\times10^{-17}$~cm$^2$ \citep{Ekenbaeck2010, Bourrier2013, Kislyakova2014}, which places the exobase deeper in the atmosphere. 
However, the atmospheres of these close-in planets are always highly ionized at the exobase level. Thus, charge exchange between H-H$^+$ with a cross-section of $\sigma\sim2\times10^{-15}$~cm$^2$ \citep{Ekenbaeck2010, Kislyakova2014} becomes the dominant process for determining the height at which neutral hydrogen enters a collisionless regime \citep{Guo2011}. Note that the cross-section for proton-proton scattering is yet a factor of 100 larger \citep[e.g.][]{Spitzer1978, Murray2009}, so that the ionized wind stays collisional to even larger heights above the planetary photosphere. This result is in agreement with considerations based on collisional relaxation times of protons \citep[using the equations in][]{Goedbloed2004}.

The exobase level as defined by \citet{Tian2005} is marked in all plots of the simulated atmospheres for neutral hydrogen and protons, if it is inside the simulated region. For most of our planets the hydrodynamic simulations are valid within the simulation box. In the atmospheres of CoRoT-2\,b, WASP-43\,b, and WASP-77\,b neutral hydrogen decouples from the ionized species between 2.6 and 5.5~$R_{\mathrm{pl}}$ (see Fig.~\ref{fig:sim_result}). 
The atmosphere of HD\,189733\,b becomes collisionless for neutral hydrogen at a height of 12~$R_{\mathrm{pl}}$. Neutral hydrogen above this height can be accelerated out of the ionized planetary wind by radiation pressure in the \lya{} line (see Sect.~\ref{SectRadPres}).
Nevertheless, the thermospheres are collisional for ionized particles. Above 2~$R_{\mathrm{pl}}$ ionization of He$^+$ is the main heat source for the wind of CoRoT-2\,b and this ion is coupled to H$^+$ and electrons, thus, an ionized planetary wind can form.

\begin{figure}[t]
  \centering
  \includegraphics[width=\hsize]{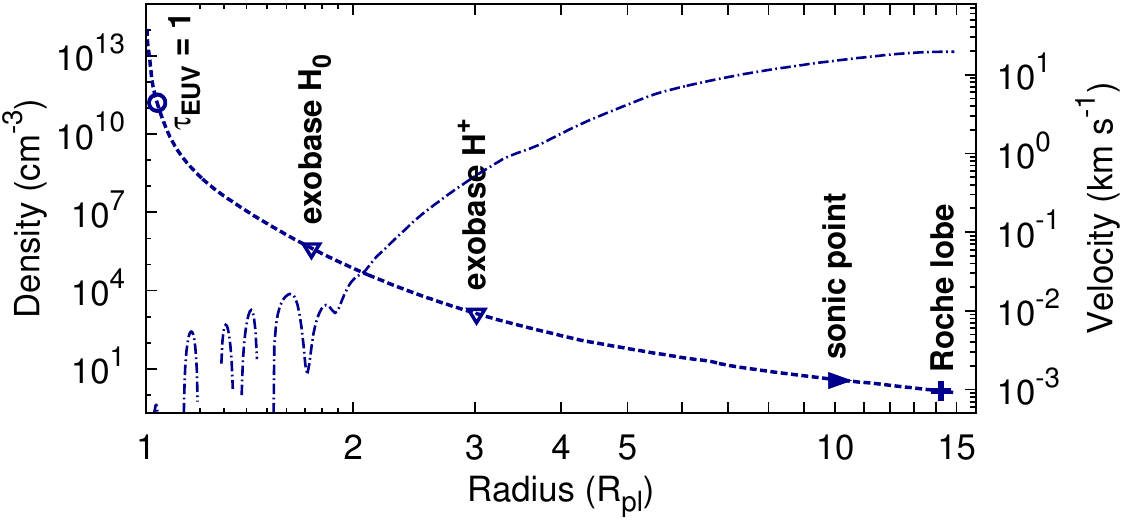}
  \caption{Mean density and velocity structure in the hydrodynamic simulation of
           the atmosphere
           of WASP-8\,b. The density is depicted by the dashed line
           and the velocity by the dash-dotted one. The atmospheric layer where the bulk EUV emission is absorbed is marked by a circle. The triangles mark the radii, where the atmosphere becomes collisionless for neutral hydrogen and protons. The arrow marks the sonic point and the plus sign the Roche lobe size.
           Oscillations are visible in the velocity structure of the lower atmosphere.
           The hydrodynamic simulation is only valid below the exobase at 3~$R_{\mathrm{pl}}$. This planet hosts a stable and compact thermosphere.}
  \label{fig:collisionless}
\end{figure}

More importantly, we find that the atmospheres of the massive and compact planets HAT-P-2\,b, HAT-P-20\,b, WASP-10\,b, and WASP-8\,b become collisionless before the atmospheric gas is significantly accelerated. Figure~\ref{fig:collisionless} shows this for WASP-8\,b. The EUV radiation is absorbed deep in the thermosphere, but does not lead to strong acceleration because of efficient radiative cooling. The density gradient is steep and the atmosphere becomes collisionless for neutral hydrogen at 1.7~$R_{\mathrm{pl}}$ and for protons at 3.0~$R_{\mathrm{pl}}$. 
The strongest acceleration takes place above these layers, where the hydrodynamic simulation is invalid and internal energy cannot be converted into a bulk flow. The simulations of these four planets result in very low mass-loss rates ($<3\times10^5$~g\,s$^{-1}$), but these are unphysical and cannot be interpreted as valid predictions. Even if we consider the uncertainties in our simulations, we conclude that these planets do not support a hydrodynamic wind, but rather compact and stable thermospheres.
WASP-38\,b and WASP-18\,b are similarly compact and massive as WASP-10\,b or WASP-8\,b, thus, these two hot Jupiters also host stable thermospheres. Naturally, these thermospheres are again subject to Jeans escape and non-thermal escape processes like ion pick-up.

\subsection{Ram pressure of the winds}\label{Sect:Ram}

In general, the atmospheres of the compact planets WASP-77\,b, WASP-43\,b, and CoRoT-2\,b are collisional so that planetary winds can form. However, the ram pressures of the planetary winds are only between 0.2 and 0.01~pbar at the sonic point. If we compare these values to a stellar wind pressure of about 10~pbar for a hot Jupiter at 0.05~au \citep{Murray2009}, it is most likely that these planetary thermospheres are confined over a large portion of the planetary surface. 
\citet{Murray2009} argued that a suppressed dayside wind could be compensated for by a stronger wind from the planetary nightside, because the radiative energy input must be spent in some way. Our simulations allow a different scenario. Confined planetary atmospheres exhibit increased thermospheric temperatures, which can lead at least to a partial re-emission of the radiative energy input via \lya{} and free-free emission (see Sect.~\ref{Sect:Equi}). This would reduce the overall planetary mass-loss rates and not only the dayside wind.

Because of stellar wind confinement, the mass-loss rates of WASP-77\,b, WASP-43\,b, and CoRoT-2\,b are likely even smaller than the results from our simulations. The same argument holds for HD\,189733\,b with a ram pressure of only 1.0~pbar at the sonic point. In contrast, the wind of HD\,209458\,b has a ram pressure of 4.3~pbar and that of WASP-80\,b reaches 16~pbar. The wind of WASP-80\,b is probably strong enough to blow against the stellar wind pressure over the full dayside of the planet. The ram pressures of the other planets range from 6.5~pbar (GJ\,436\,b) to 480~pbar (WASP-12\,b) at the sonic point.

\subsection{Radiation pressure}\label{SectRadPres}

The absorption of stellar \lya{} emission by neutral hydrogen in a planetary atmosphere leads to a net acceleration that can exceed the stellar gravitational attraction. If neutral hydrogen is transported beyond the planetary Roche lobe and if it is in a collisionless regime, it can be accelerated to velocities of about 100~km\,s$^{-1}$ \citep{Bourrier2013}. However, our hydrodynamic simulations show that escaping atmospheres stay collisional to large distances above the planetary photosphere (see Fig.~\ref{fig:sim_result}). 
Hence, when considering radiation pressure, we must take into account that neutral hydrogen is coupled to the ionized planetary wind through charge exchange \citep{Guo2011}. This reduces the radiation pressure according to the ionization fraction of the atmosphere. The coupling of the ionized and neutral planetary wind constituents should be accounted for in simulation that model the escape with direct simulation Monte Carlo codes \citep[e.g.,][]{Bourrier2013, Kislyakova2014}.

Radiation pressure is included in our simulations as described in \citet{Salz2015b}, but in the standard simulations the resolution of the SED does not resolve the stellar \lya{} line. While the line flux is correct, the peak is about a factor 10 too small, which also decreases the maximal acceleration by a factor of 10. 
To evaluate the relevance of radiation pressure, we performed a simplified CLOUDY simulation of a hydrogen cloud with similar conditions to those at the top of the atmosphere of HD\,209458\,b. We ran the test with a fixed ionization fraction of zero and 90\%, using the same SED as in the standard simulation, but with an increased resolution at the \lya{} line to fully resolve the central emission line strength.

\begin{figure}[tb]
  \centering
  \includegraphics[width=\hsize]{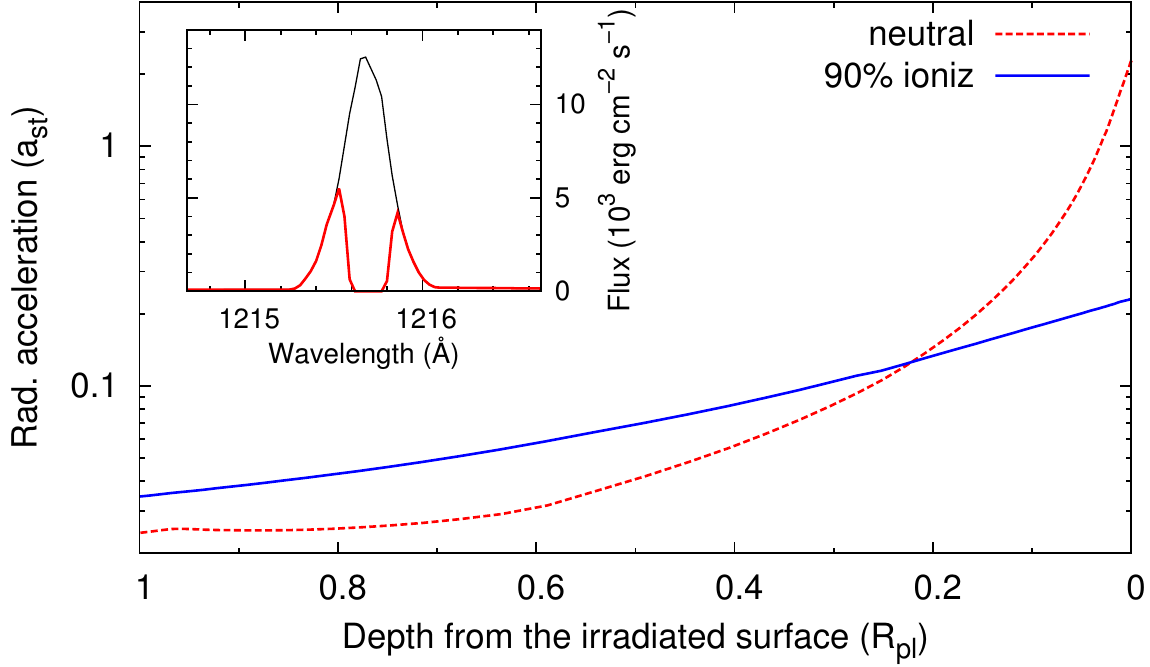}
  \caption{Radiative acceleration caused by \lya{} absorption in an
           upper atmospheric layer of the atmospheres of HD\,209458\,b.
           The acceleration is given in units
           of the stellar gravitational acceleration. A neutral (dashed, red)
           and a 90\% ionized atmosphere (solid, blue) are compared.
           The insert shows the \lya{} flux at the distance of the planet 
           (thin) and the transmitted flux behind this atmospheric
           layer (thick line).
           Only in a neutral atmosphere the radiative acceleration
           can exceed the stellar gravity.}
  \label{fig:radaccel}
\end{figure}

Figure~\ref{fig:radaccel} shows the resulting radiative acceleration. The neutral atmosphere experiences 2.5 times the gravitational acceleration of the host star ($a_{\mathrm{st}} = GM_{\mathrm{st}}/a$, $a$: semimajor axis) at the upper most layers of the atmosphere. Shadowing effects lead to a rapid decrease of the acceleration below the irradiated surface. In principle neutral hydrogen could be accelerated away from the planetary thermosphere, which is consistent with the results from \citet{Bourrier2013}.
However, the upper atmosphere of HD\,209458\,b is about 90\% ionized and collisional. This reduces the radiative acceleration by a factor of ten as depicted in Fig.~\ref{fig:radaccel}. In the escaping atmosphere of HD\,209458\,b the radiative acceleration reduces the gravitational attraction of the host star only slightly. The radiation pressure in the \lya{} line is more than a factor of 30 smaller than the ram pressure of the planetary wind at the Roche lobe of HD\,209458\,b. The wind from the dayside of the planet cannot be stopped by radiation pressure, it will rather collide with the stellar wind as described in the previous section.

Among all simulated planets, HAT-P-11\,b and HD\,97658\,b experience the strongest radiative acceleration due to a favorable combination of the distance from the host star, the host star's \lya{} luminosity, and the neutral hydrogen fraction in the upper thermosphere. In a thin layer (width $<0.5~R_{\mathrm{pl}}$) at the top of the simulated atmosphere these two planets experience a radiative acceleration of half of the gravitational acceleration of the host star. The reduced stellar attraction close to our upper boundary does not affect the structure of the thermospheres below 14~$R_{\mathrm{pl}}$. In all other systems the radiative acceleration is less than a tenth of the gravitational acceleration of the host star.

In summary, acceleration due to radiation pressure in the \lya{} line can be neglected in 1D simulations of the planetary wind formation. It only reduces the effective stellar gravitational attraction in the uppermost thermospheric layers. 3D Models simulating the interaction of the evaporating atmosphere with the stellar wind and the stellar radiation pressure should consider the coupling of neutral hydrogen to the ionized planetary wind instead of removing ionized planetary wind particles from their simulations \citep[e.g.,][]{Bourrier2013, Kislyakova2014}.

\section{Planetary Ly$\mathbf{\alpha}$ absorption and emission signals}\label{Sect:AbsSignals}

The systems simulated in this paper were selected based on a prediction of the host stars' \lya{} fluxes at Earth, so that all host stars are potentially bright enough to detect the escaping planetary atmospheres in absorption. We first calculate \lya{} absorption signals from the simulated atmospheres, which are used to identify the most promising targets for future transit spectroscopy. For systems with observed signals, we compare our results with the measurements. Second, we estimate the planetary \lya{} {\it emission} line strength and consider the detectability of these emission signals.

\subsection{\lya{} absorption signals}\label{Sect:ABS}

Our simulations provide information on the possible strength of \lya{} absorption during transit observations, because the more neutral hydrogen is transported into the planetary thermosphere and beyond, the more absorption occurs.
However, the observed planetary \lya{} absorption signals show relatively large radial velocity offsets and an asymmetric blue shift \citep[e.g.,][]{Lecavelier2012, Ehrenreich2015}. Among others, direct simulation Monte Carlo codes have been successfully used to model the observed signals by including interactions with the stellar wind and the \lya{} radiation pressure \citep{Bourrier2013, Kislyakova2014}. It is clear that the complex interaction region between the stellar and planetary winds cannot be approximated with our simplified 1D models. However, also in more complex models the neutral hydrogen mass-loss rate is adjusted to fit the measured planetary absorption signal and larger mass-loss rates result in stronger \lya{} absorption \citep{Bourrier2013, Kislyakova2014}.

In addition to the planetary neutral hydrogen mass-loss rate, the measured absorption signals are affected by several other system parameters. The stellar high energy irradiation level affects the lifetime of neutral hydrogen and the planet-to-star size ratio influences the contrast of the signals. Stellar wind parameters and \lya{} radiation pressure affect the production of energetic neutrals and the shape of the escaping atmosphere. Both processes cause enhanced absorption in the \lya{} line wings, where the stellar flux is less affected by interstellar absorption. Furthermore, strong planetary magnetic fields can shield the planetary atmosphere and affect the \lya{} absorption signals \citep{Trammell2014, Kislyakova2014}.

To estimate the strength of the planetary \lya{} absorption,
we compute the absorption signals from our spherically symmetric planetary atmospheres (see App.~\ref{Sect:Method}). The resulting values reflect the combined impact of the strength of the planetary mass-loss rate, the lifetime of neutral hydrogen in the upper atmosphere, and the planet-to-star size ratio.
The absorption is first calculated below the Roche lobe height and second for the complete simulated atmospheres (up to a height of 12/15~$R_{\mathrm{pl}}$). This twofold approach was chosen, because the signal from below the Roche lobe is caused by the rather unperturbed planetary wind, which can be approximated by our simulations. The signal of the full atmospheres is a measure of the neutral hydrogen transported beyond the Roche lobe. In Table~\ref{tabSim} the equivalent width of both signals is given; the values in brackets are the additional equivalent width of unbound hydrogen. In App.~\ref{Sect:Method} we further show absorption profiles and simulated transmission images.
In the following, we do not use the term exosphere, because according to our estimates the escaping atmospheres are collisional up to large distances above the planetary limb (see Sect.~\ref{SectJeans}), rather we distinguish between bound and unbound hydrogen.

\subsubsection{Absorption signals from nearby systems}\label{SectAbsorption}

Our results show that in general low-potential planets produce deeper absorption signals, because they host strong and neutral winds (see Figs.~\ref{figLyaTransLmass}\,\&\,\ref{figLyaTransHmass}). All of the low-potential planets qualify for \lya{} transit measurements in terms of the expected absorption signal.  WASP-80\,b is one of the most promising systems in the search for yet undetected expanded atmospheres. The planet's size is large compared to its host star and the Roche lobe has an extent of 0.85~$R_{\mathrm{st}}$. Thus, the planetary atmosphere can cover almost the complete host star. This results in the strongest predicted absorption signal with an equivalent width of 320~m\AA{} not including excess absorption of unbound hydrogen.

The simulated absorption signal of GJ\,3470\,b is even stronger than that of WASP-80\,b if unbound hydrogen is included, thus, strong absorption in the post-transit phase could be present. Although GJ\,1214\,b also produces a large absorption signal, the host star was not detected in a \lya{} observation with the Hubble Space Telescope \citep{France2013}, so that the absorption probably cannot be measured today. 

HD\,97658\,b produces the smallest total absorption signal of the low-potential planets due to the unfavorable planet to star size ratio. We do not predict absorption of unbound hydrogen for this planet, because the Roche lobe is larger than our simulation box. Our computed signal is slightly stronger than that of 55\,Cnc\,e, where evaporation was not detected during a dedicated observation campaign (see below). HD\,97658\,b is highly interesting, because among those systems with detectable evaporation, it is probably the closest proxy to the young Venus, which is believed to have been enshrouded in a dense protoatmospheric hydrogen envelope \citep{Lammer2008}. The mass of this planet is only 8 Earth-masses and the irradiation level is similar to that of Venus at a solar age of 100~Ma \citep[$\sim$1000~erg\,cm$^{-2}$\,s$^{-1}$,][]{Ribas2005}. 

The large mass-loss rate of WASP-12\,b results in strong absorption of unbound hydrogen. In principle, the strong absorption agrees with the detection of excess absorption in metal lines \citep{Fossati2010}, but \lya{} emission cannot be measured in this system due to its large distance. In the atmospheres of WASP-43\,b, WASP-77\,b, or CoRoT-2\,b neutral hydrogen remains confined within the Roche lobe. The equivalent width of the absorption signals for these three planets remains lower than that of 55\,Cnc\,e.
The hot and stable thermospheres of HAT-P-2\,b, HAT-P-20\,b, WASP-10\,b, and WASP-8\,b produces about 1\% of excess absorption in the \lya{} line, which is challenging for a detection. The same applies to the planets WASP-18\,b and WASP-38\,b, which were not simulated.

\subsubsection{Comparison with measured absorption signals}\label{Sect:Lya_compare}

\begin{figure*}[t]
  \centering
  {\bf\footnotesize \hspace{5mm} HD\,209458\,b \hspace{43mm} HD\,189733\,b \hspace{46mm} GJ\,436\,b \\}
  \vspace{6pt}
  \includegraphics[width=0.33\hsize]{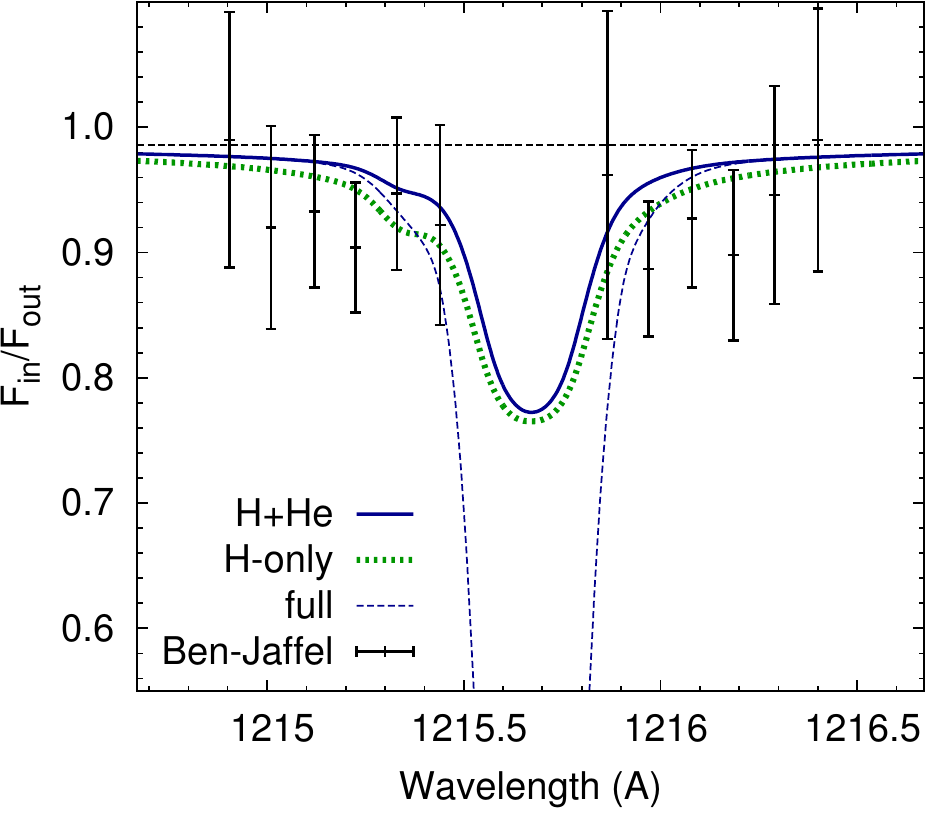}
  \hfill
  \includegraphics[width=0.309\hsize, trim=0.6cm 0cm 0cm 0cm, clip=true]{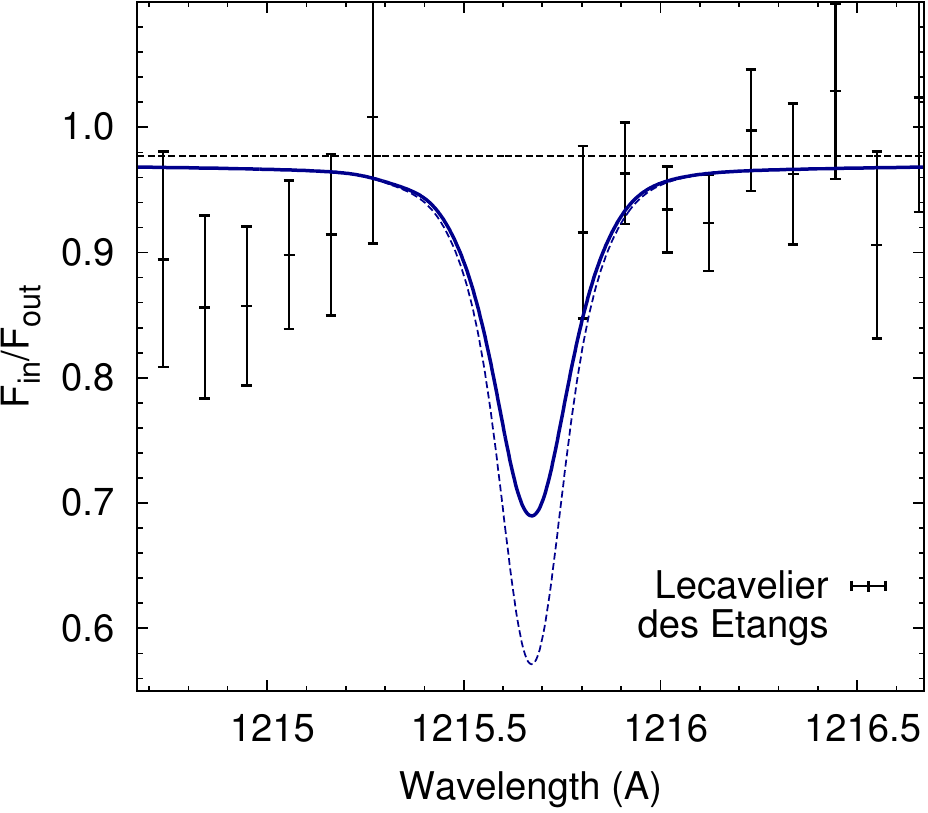}
  \hfill
  \includegraphics[width=0.309\hsize, trim=0.6cm 0cm 0cm 0cm, clip=true]{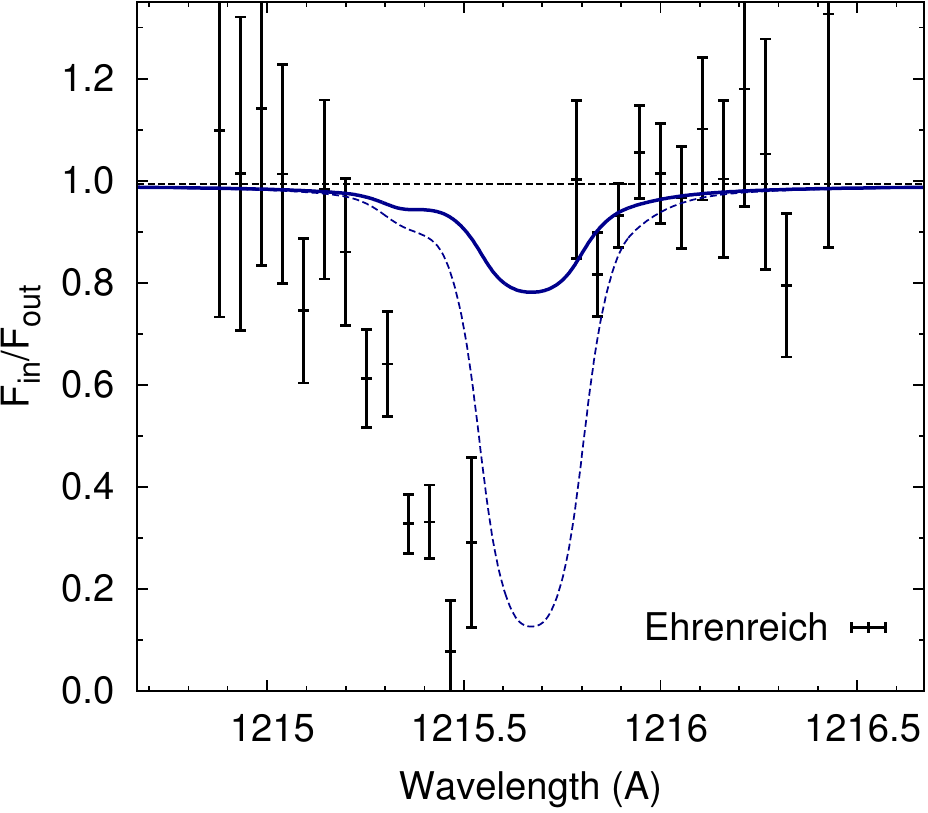}
  \\ \vspace{9pt}
  \hfill
  \includegraphics[width=0.25\hsize]{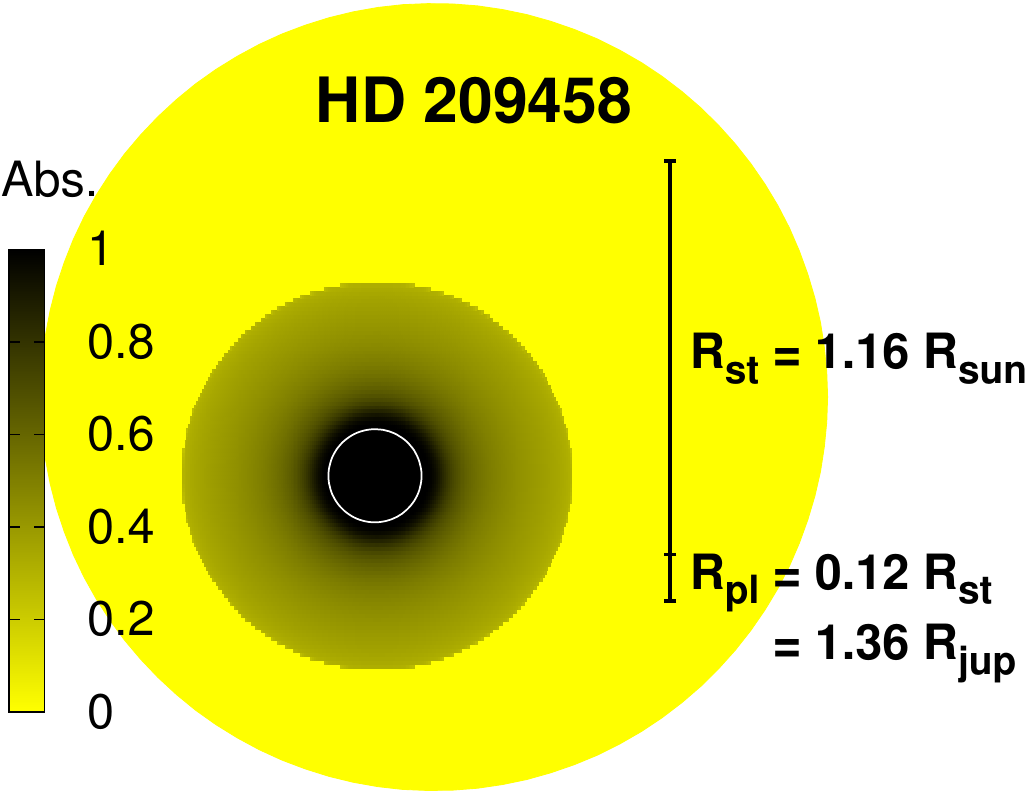}
  \hfill
  \includegraphics[width=0.25\hsize]{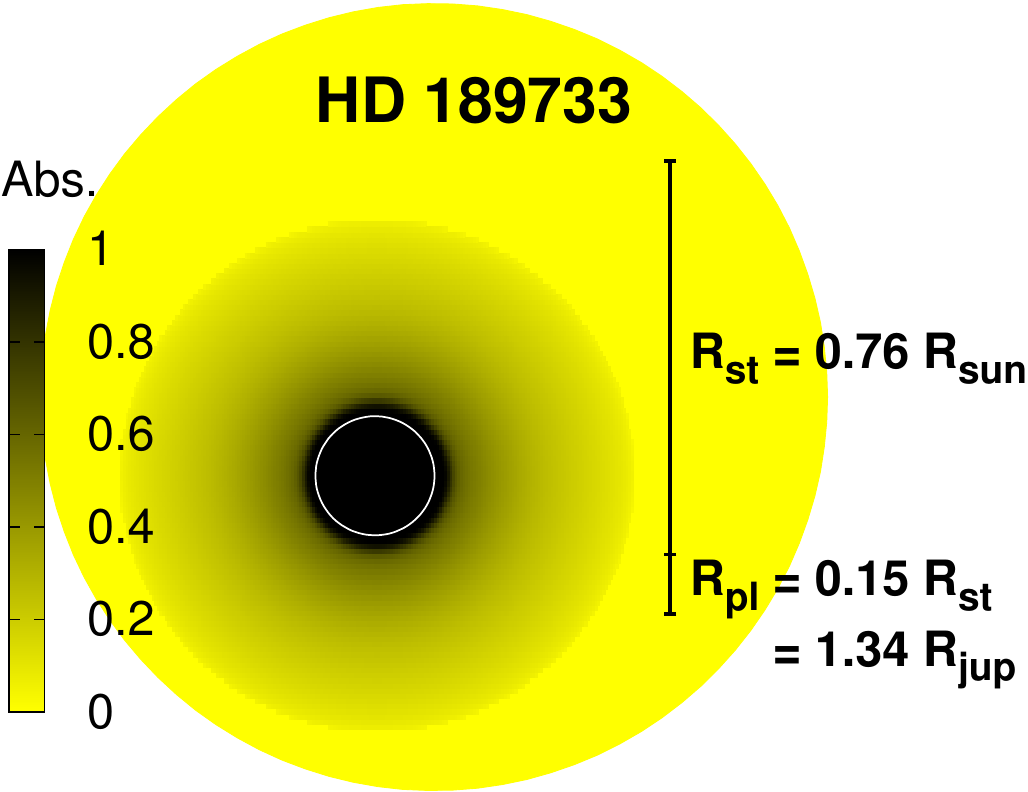}
  \hfill
  \includegraphics[width=0.25\hsize]{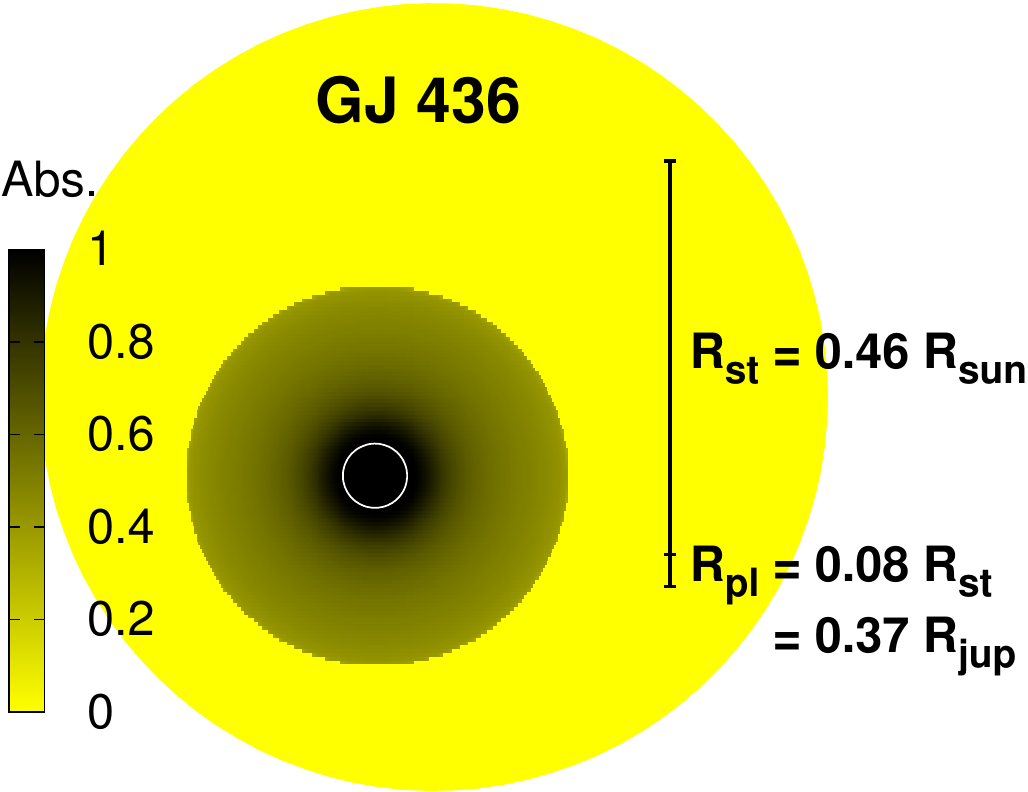}
  \hfill
  \caption{Comparison of the measured \lya{} absorption signals
           with the simulation results.
           The upper panels show the in- and out-of-transit \lya{} flux ratios.
           Computed transmission images are depicted at the bottom; the
           size of the atmospheres is given by the height of the Roche lobe.
           The central part of the \lya{} line affected by geocoronal
           emission is omitted.
           For HD\,209458\,b we show the flux ratio from the analysis of \citet{Ben2008}, for GJ\,436 the data from \citet{Ehrenreich2015}, and for HD\,189733\,b of \citep{Lecavelier2012}.
           The horizontal dashed lines represent the optical transit depth;
           solid thick lines show the absorption from the atmosphere
           below the Roche lobe; the dashed thin lines depict the absorption
           from the complete simulated atmosphere including unbound hydrogen.
           For the system HD\,209458\,b, we also depict the absorption from the 
           Roche lobe of the H-only simulation by the green dotted line.
           }
  \label{fig:lya_abs}
\end{figure*}

\begin{figure}[t]
  \centering
  \includegraphics[width=0.47\hsize]{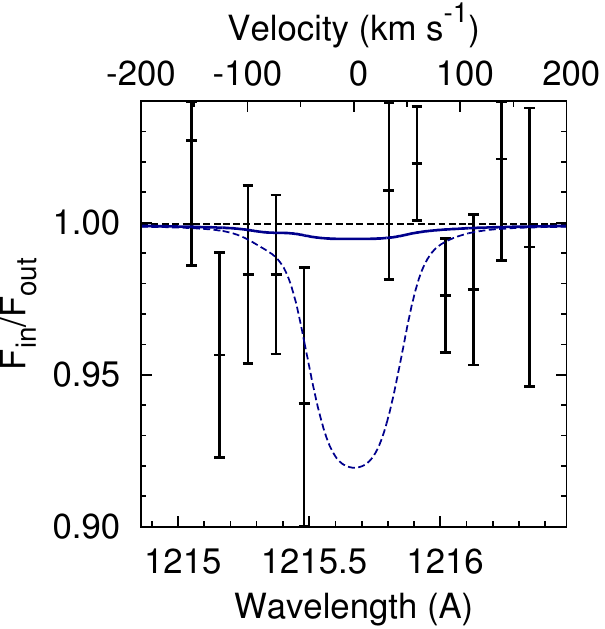}
  \hfill
  \raisebox{18pt}{\includegraphics[width=0.49\hsize]{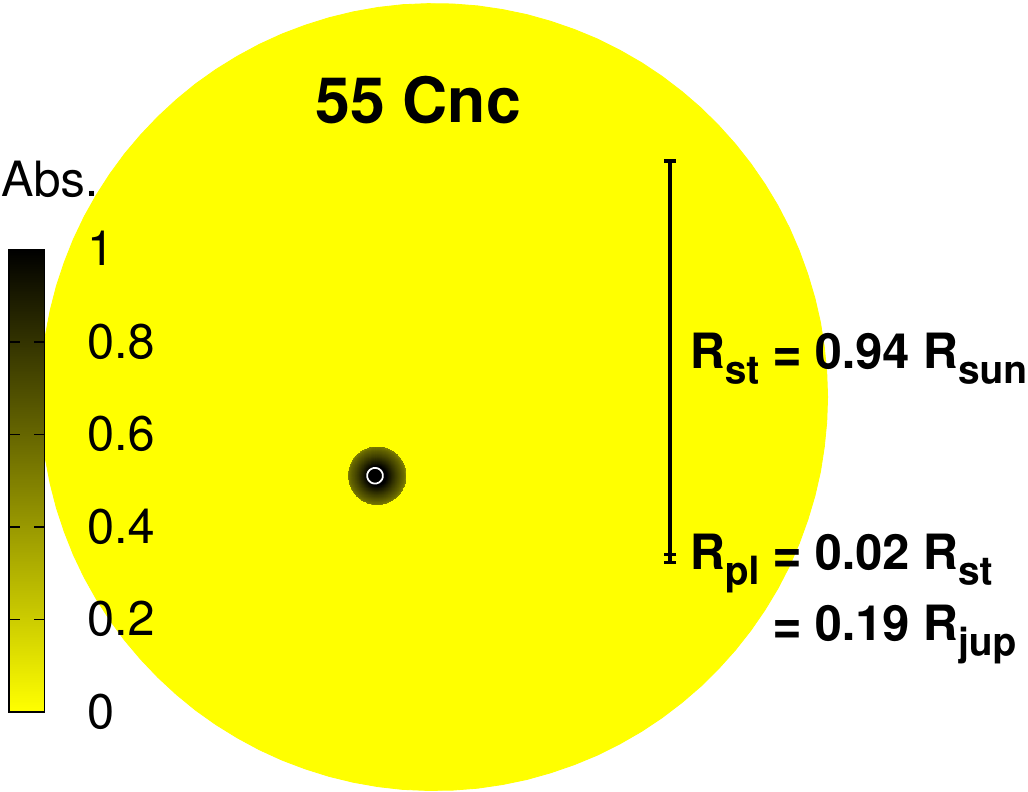}}
  \caption{Simulated \lya{} transmission for 55\,Cnc\,e.
           Symbols are the same as in
           Fig.~\ref{fig:lya_abs}.
           The transit data is from \citet{Ehrenreich2012}.
           The non-detection of a planetary absorption signal
           agrees with the result from our simulation.}
  \label{fig:55cnc}
\end{figure}

In HD\,209458\,b, HD\,189733\,b, GJ\,436\,b, and for the non-detection in 55\,Cnc\,e, we can compare the computed absorption signals with observational results. The equivalent widths of the measured and simulated absorption signals of the four systems are provided in Table~\ref{tabMeasAbs}. Planets with strong neutral winds produce large absorption signals in our simulations, which is qualitatively consistent with the observations. Fig.~\ref{fig:lya_abs} shows the spectral shape of the simulated absorption signals in comparison with the measurements.
Note that the strong absorption in the line center cannot be detected because it would be extinguished by interstellar absorption. The figure also contains calculated transmission images, which correlate with the absorption signals as follows: The strong absorption signal in the line center is produced by the thin upper atmospheric layers that cover a large percentage of the stellar disk, and the absorption in the line wings is caused by the dense, lower thermosphere. The results for 55\,Cnc\,e are given in Fig.~\ref{fig:55cnc}.

\begin{table}[t]
  \caption{Absorption signals from observations and simulations.}
  \label{tabMeasAbs}
  \centering
  \begin{tabular}{l ccc }
    \hline\hline\vspace{-5pt}\\
                    &  Measured       & Simulated   & \\
      System        &  \lya{} abs.    & \lya{} abs. & Ref.\\
                    & (m\AA{})         & (m\AA{})        & \\
    \vspace{-7pt}\\ \hline\vspace{-5pt}\\ 
      GJ\,436\,b      & $206\pm36$            & 370 & 3 \\
      HD\,209458\,b   & $105\pm38$            & 360 & 1 \\
      HD\,189733\,b   & \hphantom{1}$97\pm28$ & 188 & 2 \\
      55\,Cnc\,e      & \hphantom{11}$9\pm13$ & \hphantom{1}92 & 4 \\
   \vspace{-7pt}\\ \hline 
   \end{tabular}
\tablefoot{The columns are:
           planet name,
           absorption equivalent width (excluding geocoronal contamination),
           predicted absorption of the full atmosphere.
          }
\tablebib{(1) \citet{Ben2008}; 
          (2) \citet{Lecavelier2012};
          (3) \citet{Ehrenreich2015};
          (4) \citet{Ehrenreich2012}.
}
\end{table}

The top panel in the left column of Fig.~\ref{fig:lya_abs} shows the absorption signal of HD\,209458\,b. The system exhibits the most symmetric absorption signal measured so far. The absorption was discussed in detail by \citet{Murray2009} and by \citet{Koskinen2013a}. HD\,209458\,b is the only system, where the absorption signal can be solely explained by a spherical atmosphere, depending on the height of the H/H$^+$ transition layer. 
In the model of \citeauthor{Koskinen2013a} the transition occurs at a height of 3.4~$R_{\mathrm{pl}}$, so that their predicted absorption fits the transit observations. Our simulation results in an H/H$^+$ transition layer at a height of 1.6~$R_{\mathrm{pl}}$, which does not produce sufficient absorption in the line wings. A twice as strong planetary wind already shifts the ionization front to 2.3~$R_{\mathrm{pl}}$ as seen in our H-only simulation (see Sect.~\ref{SectGenStruc}), but the absorption remains insufficient (see Fig.~\ref{fig:lya_abs}).
The fact that changes within the uncertainty of the current simulations decisively affect the absorption signal renders it difficult to determine its origin. For example, \citet{Trammell2014} reproduce the observed signal with an atmosphere that is largely stabilized through a strong planetary magnetic field; in contrast, \citet{Kislyakova2014} reproduce the signal with an exospheric model including interactions with the stellar wind and radiation pressure. The current data is likely insufficient to distinguish between the different models.

In the case of HD\,189733\,b it is impossible to explain the spectral shape of the measured absorption signal with a spherically expanding atmosphere. The absorption of HD\,189733\,b is found at radial velocity shifts of up to $-250$~km\,s$^{-1}$ and around $+100$~km\,s$^{-1}$ (see Fig.~\ref{fig:lya_abs}). Even considering all uncertainties, a spherical wind will not produce such a signal.

GJ\,436\,b shows similar absorption signals during and after the planetary transit, which indicates a cometary tail like structure of the escaping atmosphere \citep{Kulow2014, Ehrenreich2015}. Again, the signal cannot be explained by a spherical atmosphere, but interaction with the stellar wind or radiation pressure is required to reproduce the detected radial velocity shift and the temporal evolution of the absorption \citep{Ehrenreich2015}. In our simulations, GJ\,436\,b has one of the largest amounts of unbound neutral hydrogen, which agrees with the detection of a cometary tail.
Although both HD\,209458\,b and GJ\,436\,b produce strong neutral winds, the detected \lya{} absorption of HD\,209458\,b is much weaker and this planet shows no post-transit absorption.
There are several possibilities to explain these differences. If our estimate of the XUV irradiation level is too low, HD\,209458\,b would host a stronger but also more ionized planetary wind, which reduces the amount of unbound neutral hydrogen. Also a partially confined atmosphere or the interaction with the stellar wind could reduce the amount of unbound neutral hydrogen in HD\,209458\,b. Finally, \citet{Bourrier2015} explain such differences based on their radiation pressure driven exosphere model.

Excess absorption  was not detected during the transit of 55\,Cnc\,e. There is a weak signal in the blue line wing, but the false alarm probability is 89\% \citep{Ehrenreich2012}. 
The absorption produced by the simulated escaping atmosphere including unbound hydrogen is no larger than 8\% in the line center (see Fig.~\ref{fig:55cnc}). Although the planet can host a strong hydrogen-dominated wind, the simulated absorption signal remains small, mainly because the planet is small compared to its host star, but also because of the short lifetime of neutral hydrogen in the highly irradiated atmosphere.
The simulated signal is close to the detection limit and, considering additional interstellar absorption, the simulation results agree with the non-detection. Therefore, the observations by \citet{Ehrenreich2012} do not exclude the existence of a hydrogen-dominated thermosphere that currently undergoes strong evaporation.

\subsection{\lya{} emission signals}\label{Sect:lya_emit}

Our simulations show that planetary atmospheres are strongly cooled by \lya{} emission. Especially high-potential planets re-emit large amounts of the XUV energy input through \lya{} emission. This poses the question whether the planetary \lya{} emission is detectable from Earth. Based on the simulations we can predict the planetary \lya{} emission line strength, which can be compared with the stellar line fluxes (see Sect.~\ref{SectSED}). 

We compute the ratio between the planetary and stellar \lya{} line fluxes $\phi_{\mathrm{pl}}/\phi_{\mathrm{st}}$, similar to the approach of \citet{Menager2013}. 
Our photoionization solver does not spectrally resolve the stellar and planetary \lya{} lines, hence, in a first step we scale our values to derive the flux in the line center. For the stellar emission lines we used a ten times broader line width than usually measured for stellar line profiles (see Sect.~\ref{SectSED}), thus, the central stellar \lya{} line flux will be a factor 10 higher than in our reconstructions. The planetary emission is given by the photoionization solver with a bin width of $\Delta\lambda = 6.1$~\AA{}. We derive the central emission line strength by assuming a Gaussian shape. In the models of \citet{Menager2013} the planetary emission has a FWHM of about 0.13~\AA{}. Assuming an atmospheric temperature of 10\,000~K we derive a similar value of
\begin{align}
  \mathrm{FWHM} = 2\,\sqrt{\ln 2} \, b
              &= 2\,\sqrt{\ln 2} \sqrt{v^2_{\mathrm{th}}+v^2_{\mathrm{micro}}} \notag\\
              &= 2\,\sqrt{\ln 2} \sqrt{2kT/m_u+5kT/3m_u} \notag \\
              &= 0.17~\AA{} \label{eq:fwhm} .
\end{align}
Here, $b$ is the Doppler parameter, $v_{\mathrm{th}}$ the thermal Doppler velocity, and for the microturbulence $v_{\mathrm{micro}}$ we use the sound speed. Comparing the two results, we regard a FWHM of 0.15~\AA{} a reasonable value for the planetary line profile. 

The central emission line flux from the planet is given by conserving the integrated line flux:
\begin{align}
  &\phi_{\mathrm{pl, CLOUDY}}\,\Delta\lambda = \phi_{\mathrm{pl}} \,\mathrm{FWHM}/(2\sqrt{2\ln 2})\, \sqrt{2\pi} \notag \\
  &\Rightarrow\;\; \phi_{\mathrm{pl}} \approx 38\,\phi_{\mathrm{pl, CLOUDY}} .
\end{align}
Here $\phi_{\mathrm{pl, CLOUDY}}\,\Delta\lambda$ is the integrated planetary flux given by the photoionization solver and the righthand side of the equation is the integrated flux from the assumed Gaussian line profile. Finally, we scale the flux ratios $\phi_{\mathrm{pl}}/\phi_{\mathrm{st}}$ with the relative sizes of planetary and stellar disks, where we assume that only the hot dayside of the planet emits \lya{} radiation. We adopt a height of 1.2~$R_{\mathrm{pl}}$ for the \lya{} emission layer in the planetary atmospheres, which is indicated by the dip in the heating fractions (see panels (f) in Fig.~\ref{fig:sim_result}).

The derived values for HD\,209458\,b and HD\,189733\,b can be compared to those of \citet{Menager2013}, who find flux ratios of $2.1\times10^{-3}$ and $1.5\times10^{-3}$ respectively. Our value of $1.8\times10^{-4}$ for HD\,209458\,b is one order of magnitude smaller. However, the two results are based on different atmospheric models. \citet{Menager2013} used the atmospheric models from \citet{Koskinen2013a} with a 1200~K hotter thermosphere and a larger H/H$^+$ transition layer height, which is a possible reason for the differences.
Our flux ratio of $1.3\times10^{-3}$ for HD\,189733\,b agrees with the results from \citet{Menager2013}, however, these authors used a solar irradiation level, which is about 20 times weaker than our value. Since a higher XUV irradiation increases the flux ratios, it seems that our simulations generally produce lower planetary \lya{} luminosities than those obtained by \citet{Menager2013}. Yet, the reasons for the differences can only be assessed with a detailed analysis based on an identical atmospheric model.

We now focus on the maximum flux ratios
$(\phi_{\mathrm{pl}}/\phi_{\mathrm{st}})_{\mathrm{max}}$,
which approximately occur when the planetary emission experiences the maximal radial velocity shift through the orbital motion of the planet at phases 0.25 and 0.75.
This causes radial velocity shifts between 0.3 and 0.9~\AA{} in our sample of planets. The reconstructed stellar \lya{} line profiles from \citet{Wood2005} show that the line flux drops by a factor of 10 compared to the central emission line strength if the line shift exceeds 0.5~\AA{}. Only for stars with unusually wide line profiles this assumption breaks down. Therefore, we apply a factor of 10 to the flux ratios, assuming that the observations are placed at the corresponding phases. 

\begin{figure}[tb]
  \centering
  \includegraphics[width=\hsize]{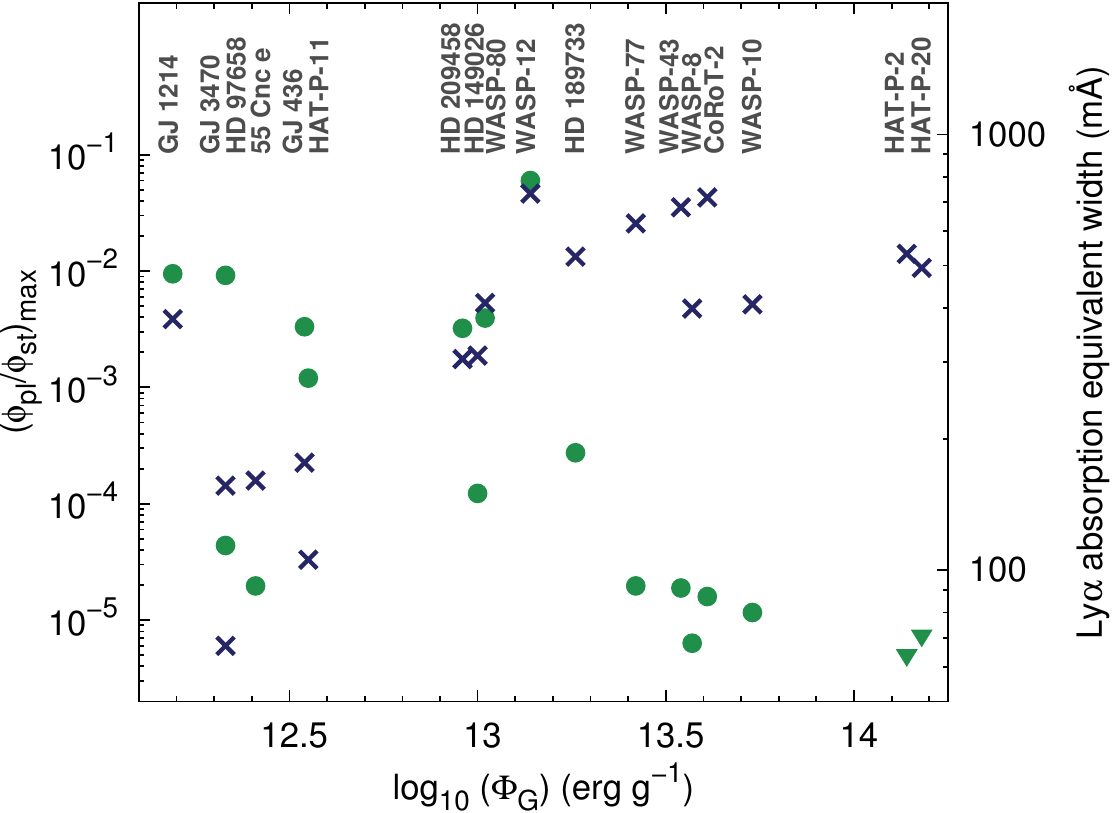}
  \caption{Planetary to stellar \lya{} flux ratios plotted
           versus the planetary gravitational potential (blue crosses).
           The values correspond to the phases 0.25 and 0.75 during
           the planetary orbit.
           The green circles depict the equivalent width of the planetary \lya{}
           absorption signals during transit observations.
           }
  \label{fig:lya_flux_ratio}
\end{figure}

The calculated flux ratios depend on several parameters. Naturally, a larger size of the planet increases the ratio and a smaller semimajor axis also increases the ratio through higher irradiation levels. Figure~\ref{fig:lya_flux_ratio} shows the flux ratios for the simulated planets versus the planetary gravitational potential. 
Massive and compact planets show larger flux ratios, because their thermospheres are close to radiative equilibrium and convert the XUV energy input efficiently into \lya{} radiation. The total planetary \lya{} line fluxes at the top of the planetary atmospheres reach maximally 30 to 40\% of the combined irradiating stellar \lya{} and XUV flux. The maximal flux ratios at orbital phases 0.25 and 0.75 are smaller than 10\%. The strong emission of WASP-12\,b is caused by the extreme proximity of this planet to its host star. GJ\,1214\,b shows a relatively large flux ratio, because the host star emits little \lya{} radiation, but the planet converts part of the XUV energy input into \lya{} emission. The \lya{} emission of WASP-43\,b and WASP-77\,b reach 3.5\% and 2.6\% of the stellar flux in the line wing. These signals may be detectable, depending on the stellar line flux at Earth and the specific stellar and planetary line profiles.

The presented results are somewhat rough estimates because of the described limitations. They remain valid if metals are included according to our test simulation of HD\,209458\,b with solar metal abundances (see Sect.~\ref{SectMetalsMolec}). In principle, one could assume that the \lya{} emission of planetary thermospheres containing metals is significantly reduced, because metals are efficient cooling agents, but in HD\,209458\,b the emission was only reduced by a factor of two. This general factor should not impact the comparison between the individual systems.

A detection of such \lya{} emission from planets would substantiate our finding that massive and compact hot gas planets host stable thermospheres. Only the conversion of stellar XUV flux into \lya{} emission in the hot and stable thermospheres boosts the flux ratios to detectable values. Furthermore, the mechanism allows a distinction between two classes of thermospheres in hot gas planets. Figure~\ref{fig:lya_flux_ratio} shows that small planets with strong and cool winds create large absorption signals and only weak \lya{} emission; massive and compact planets with close to stable thermosphere create little absorption but re-emit the XUV energy input partially via \lya{} emission. 
Therefore, planetary \lya{} absorption and emission are to some extent mutually exclusive and the onset of strong planetary winds around $\log -\Phi_{\mathrm{G}} = 13.2$~erg\,g$^{-1}$ provides an observational test of the hydrodynamic escape model. According to our simulations, WASP-43 and HD\,189733 are equally suited for an observational campaign aimed at the detection of planetary \lya{} emission. While the expected signal is a factor 2.7 larger in the system WASP-43, the predicted \lya{} flux is a factor 10 smaller than that in HD\,189733, hence, the signal-to-noise ratio should be similar.

\section{Summary and conclusion}\label{Sect:Conclude}

We present coupled photoionization-hydrodynamics simulations of the planetary winds from hot gas planets in the solar neighborhood. The most promising candidates in terms of \lya{} transit spectroscopy were selected for the simulations, which were performed with our TPCI code. We have described the behavior of the escaping atmospheres in detail with a focus on the differences between winds from small planets and massive, compact planets. The use of a photoionization and plasma simulation code, which gives a precise solution for the radiative cooling, has proven to be crucial for the results presented here.
In particular, we find that compact and massive hot Jupiters host stable thermospheres, because of strong radiative cooling in their hot thermospheres. Radiative cooling is weak in smaller planets, because adiabatic cooling in the rapidly expanding atmospheres efficiently reduces the thermospheric temperatures. In our simulations of hydrogen and helium atmospheres, \lya{} and free-free emission are the most important cooling agents. Specifically, we found stable thermospheres in the planets HAT-P-2\,b, HAT-P-20\,b, WASP-8\,b, WASP-10\,b, WASP-38\,b, and WASP-18\,b.

The simulated thermosphere of HD\,189733\,b reproduces the temperature-pressure profile reconstructed from measured sodium absorption signals, but in HD\,209458\,b our simulations and the observations show significant differences that currently remain unexplained.

\lya{} absorption signals of the spherically symmetric atmospheres are presented to assess the systems' potential for transit spectroscopy. The calculations reproduce large absorption signals, where strong \lya{} absorption has been detected.
Our simulations show that smaller planets tend to cause larger \lya{} absorption signals, because they produce strong and cool winds that transport large amounts of neutral hydrogen into the upper thermospheres. 
In contrast, the stable thermospheres of massive and compact planets are hot and highly ionized and, thus, only create small \lya{} absorption signals. 
Instead, such hot thermospheres are strong \lya{} emitters, and the planetary emission could be detectable with current instrumentation.
The distinction between planets with strong \lya{} absorption but no emission and planets with little \lya{} absorption but strong emission can be used to distinguish between these two classes of thermospheres.
Based on the predicted strong absorption signals, WASP-80\,b and GJ\,3470\,b are the most promising targets among our sample in the search for yet undetected escaping atmospheres.

\begin{acknowledgements}
MS acknowledges support through Verbundforschung (50OR 1105) and
the German National Science Foundation (DFG) within the Research Training College 1351.
PCS acknowledges support by the
DLR under 50 OR 1307 and by an ESA Research Fellowship.
This research has made use of the Exoplanet Orbit Database and the Exoplanet Data Explorer at exoplanets.org.
\end{acknowledgements}

\bibliographystyle{aa}
\setlength{\bibsep}{0.0pt}
\bibliography{salz_hdyn_esc}

\begin{appendix}
\section{Calculation of the \lya{} absorption signals}\label{Sect:Method}

Here, we describe the method used to calculate the \lya{} absorption in the simulated atmospheres. Although we use the term \lya{} absorption, we note that the actual physical process is scattering of photons out of our line of sight. Absorption of \lya{} with collisional de-excitation only occurs deep in the simulated atmospheres (see Fig.~\ref{fig:comp_atmos}). We do not account for any emission or line transfer in the computation of the absorption signals, or for scattering of photons into our line of sight. We further assume a homogeneous brightness of the stellar disk in \lya{} emission.

\begin{figure}[h]
  \sidecaption
  \includegraphics[width=0.35\hsize]{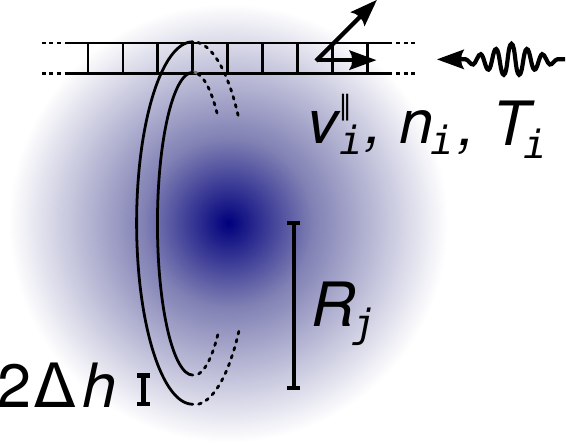}
  \caption{Computation of the transmission signals and images.
           The absorption is calculated by shooting rays through the planetary
           atmosphere. Density, temperature, and tangential velocity in
           each cell are taken into account.}
  \label{fig:2D-transmission}
\end{figure}

To compute the absorption signals we shoot rays through the spherical atmospheres (see Fig.~\ref{fig:2D-transmission}). The wavelength dependent transmission $T^{\lambda}$ in a single ray with index $j$ is given by the product of the absorption in each cell:
\begin{equation}\label{eq:transmission}
  T^{\lambda}_j = \prod_i \exp (-\Delta \tau^i_{\mathrm{abs}} )
  = \prod_i e^{-\sigma^i_{\lambda}n_i\Delta s_i} \,.
\end{equation}
Here, $n_i$ is the neutral hydrogen density in cell $i$ with the cell width $\Delta s_i$.
$\sigma^i_{\lambda}$ is the standard \lya{} absorption cross-section with a Voigt-profile, shifted by the local tangential velocity of the gas.
The Doppler width is given by the sum of the local thermal and turbulent widths (see Eq.~\ref{eq:fwhm}); the latter is given by the speed of sound. Deuterium absorption is included with a deuterium fraction of $(\mathrm{D/H}) = 2.2\times10^{-5}$ according to measurements from Jupiter \citep{Lellouch2001}. 

The total transmission is given by:
\begin{align}
  &F_{\mathrm{in}}/F_{\mathrm{out}}({\lambda}) = T^{\lambda}_{tot} \notag\\
  &= \left(1-\left(\frac{R_n+\Delta h}{R_{st}}\right)^2\right)
      + \sum^n_{j=1} \frac{(R_j+\Delta h)^2-(R_j-\Delta h)^2}{R^2_{st}} T^{\lambda}_j\,.\label{eq:tot_trans}
\end{align}
Here, $n\approx 100$ is the number of rays and $\Delta h$ is half of the vertical resolution. In this equation, the first term accounts for the fraction of radiation that bypasses the simulated atmospheres (transmission equals one). In most of the systems this term is zero, because the simulated atmospheres cover the full stellar disk. The sum adds the transmission of each ray weighted with the fractional area that a ring with the given height $R_j$ above the planetary photosphere covers of the stellar disk. The transmission of the planetary disk is zero and does not add to the total transmission ($R_1 = R_{\mathrm{pl}}$).
For each planet we compute the equivalent width of the absorption signal over the full line ($\pm 500$~km\,s$^{-1}$)
\begin{equation}\label{eq:equiwidth}
  W_{\lambda} = \int (1-F_{\rm in}/F_{\rm out})\, d\lambda .
\end{equation}
These values are given in Table~\ref{tabSim} for the atmosphere below the Roche lobe and the values for unbound atmospheric material above the Roche lobe is given by the value in brackets.

\begin{figure*}[t]
  \begin{minipage}[b]{0.48\textwidth}
    \centering
    \includegraphics[width=0.425\hsize, trim=0cm 0.9cm 0cm 0cm, clip=true]{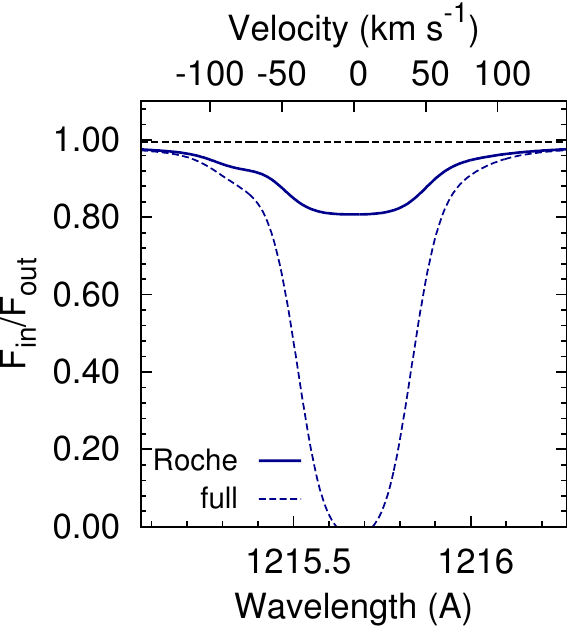}
    \hfill
    \raisebox{2pt}{\includegraphics[width=0.42\hsize]{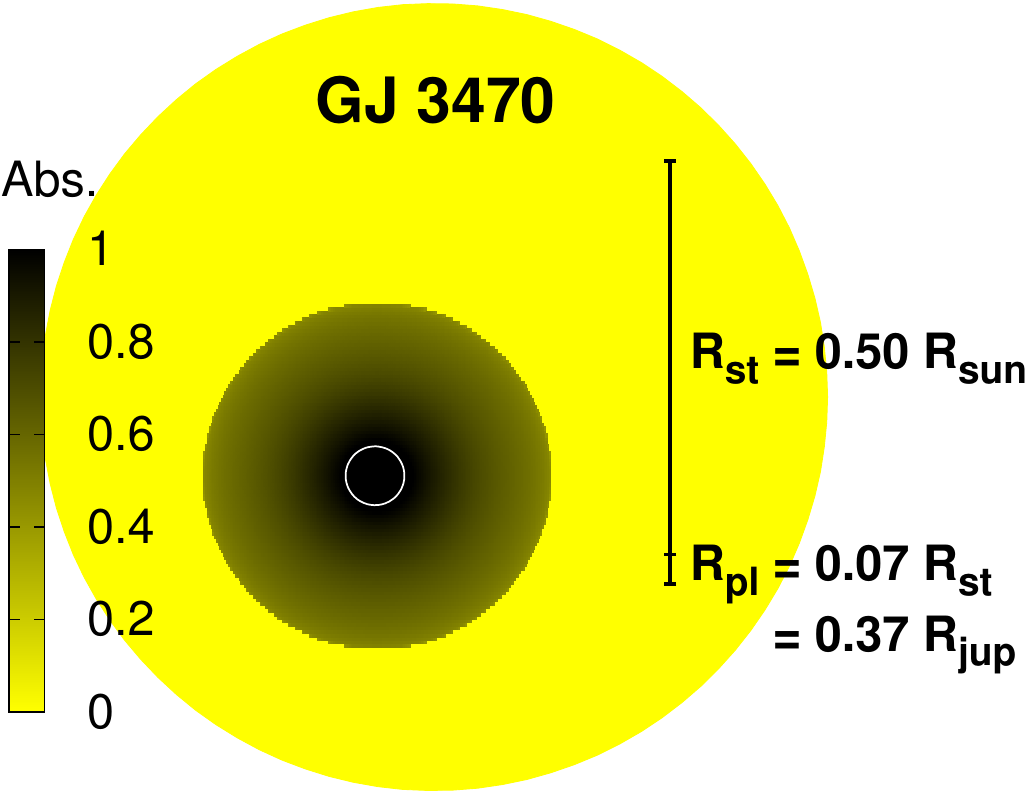}}
    \\ \vspace{5pt}
    \includegraphics[width=0.425\hsize, trim=0cm 0.9cm 0cm 0.9cm, clip=true]{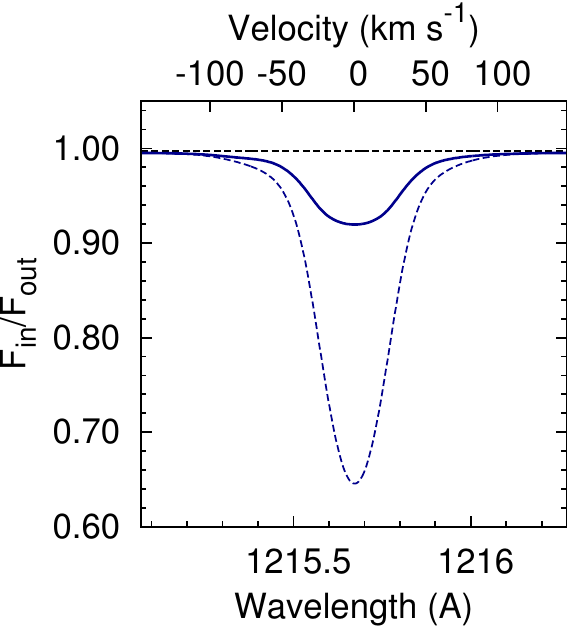}
    \hfill
    \raisebox{2pt}{\includegraphics[width=0.42\hsize]{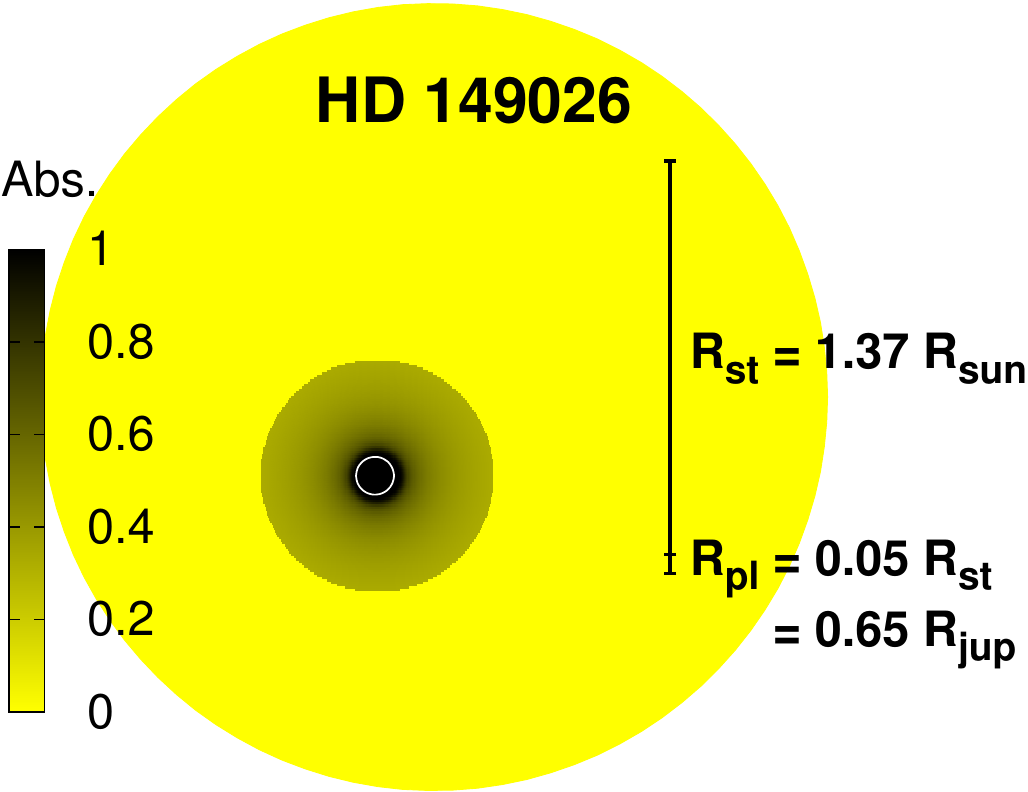}}
    \\ \vspace{5pt}
    \includegraphics[width=0.425\hsize, trim=0cm 0.9cm 0cm 0.9cm, clip=true]{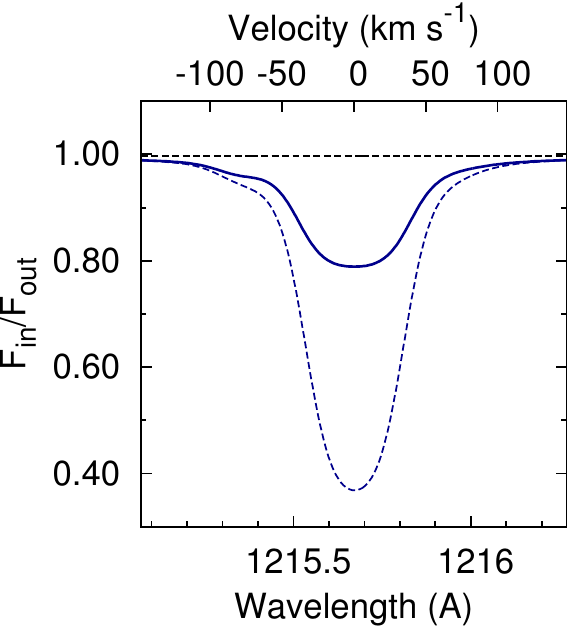}
    \hfill
    \raisebox{2pt}{\includegraphics[width=0.42\hsize]{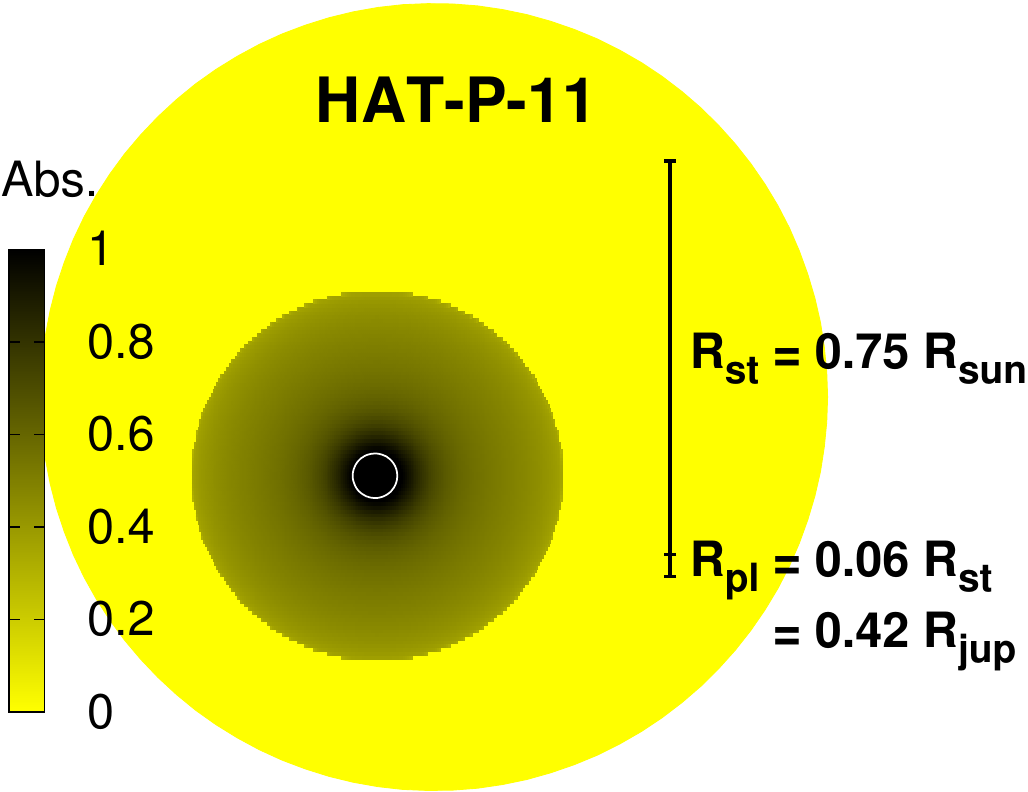}}
    \\ \vspace{5pt}
    \includegraphics[width=0.425\hsize, trim=0cm 0.9cm 0cm 0.9cm, clip=true]{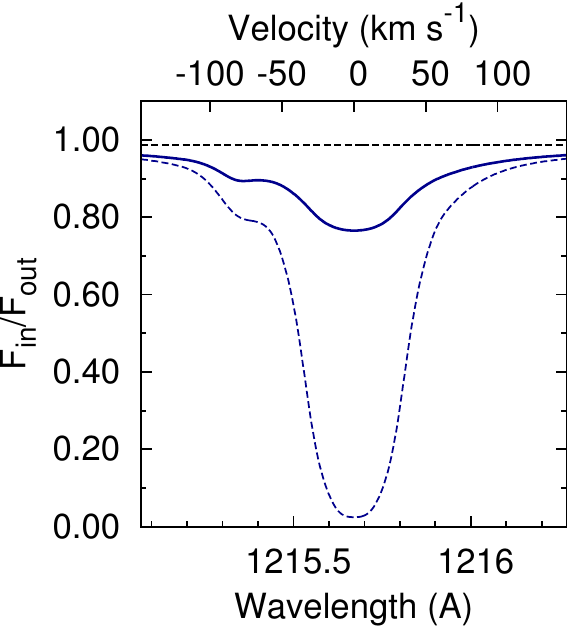}
    \hfill
    \raisebox{2pt}{\includegraphics[width=0.42\hsize]{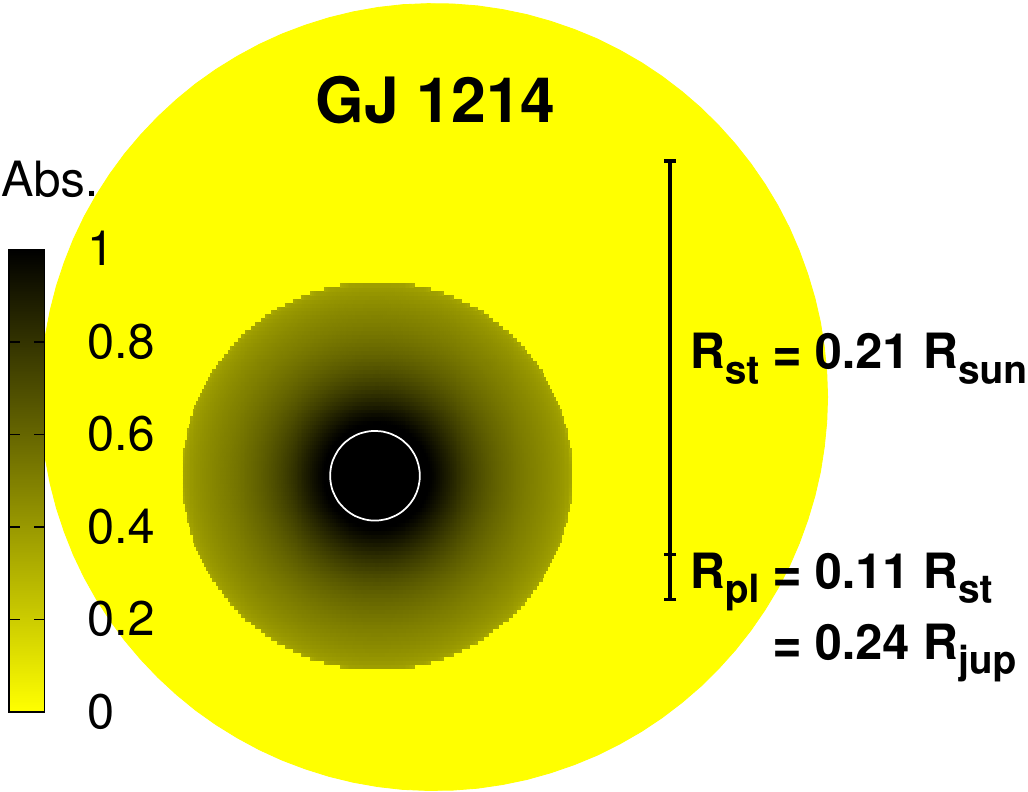}}
    \\ \vspace{5pt}
    \includegraphics[width=0.425\hsize, trim=0cm 0cm 0cm 0.9cm, clip=true]{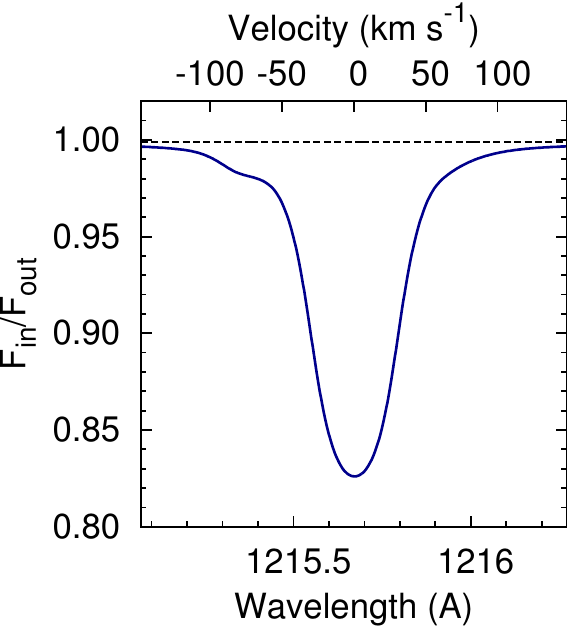}
    \hfill
    \raisebox{18pt}{\includegraphics[width=0.42\hsize]{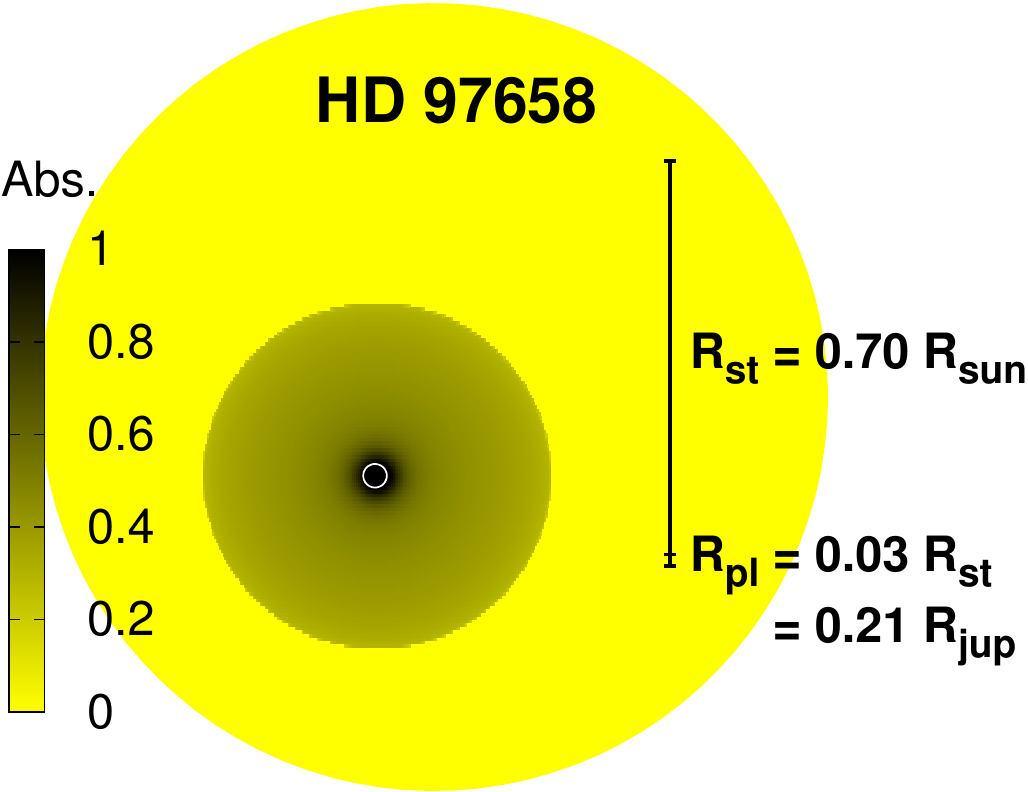}}
    \caption{Simulated \lya{} transmission for low-potential planets.
             For each system we show the absorption in the \lya{}
             line from below the Roche lobe (solid line) and from unbound
             hydrogen (dashed, blue). 
             Black dashed lines indicate the optical transit depth.
             Calculated transmission images are depicted next
             to each spectrum.
             The mass-loss rate decreases from top to bottom.
             }
    \label{figLyaTransLmass}
  \end{minipage}
  \hfill
  \begin{minipage}[b]{0.48\textwidth}
    \centering
    \includegraphics[width=0.425\hsize, trim=0cm 0.9cm 0cm 0cm, clip=true]{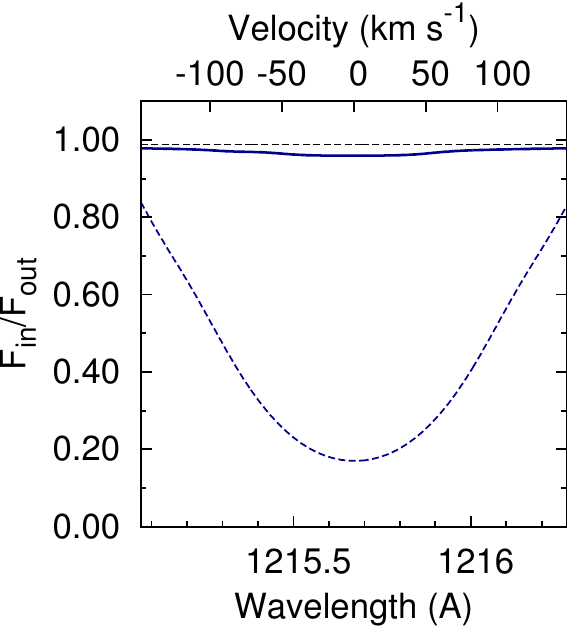}
    \hfill
    \raisebox{2pt}{\includegraphics[width=0.42\hsize]{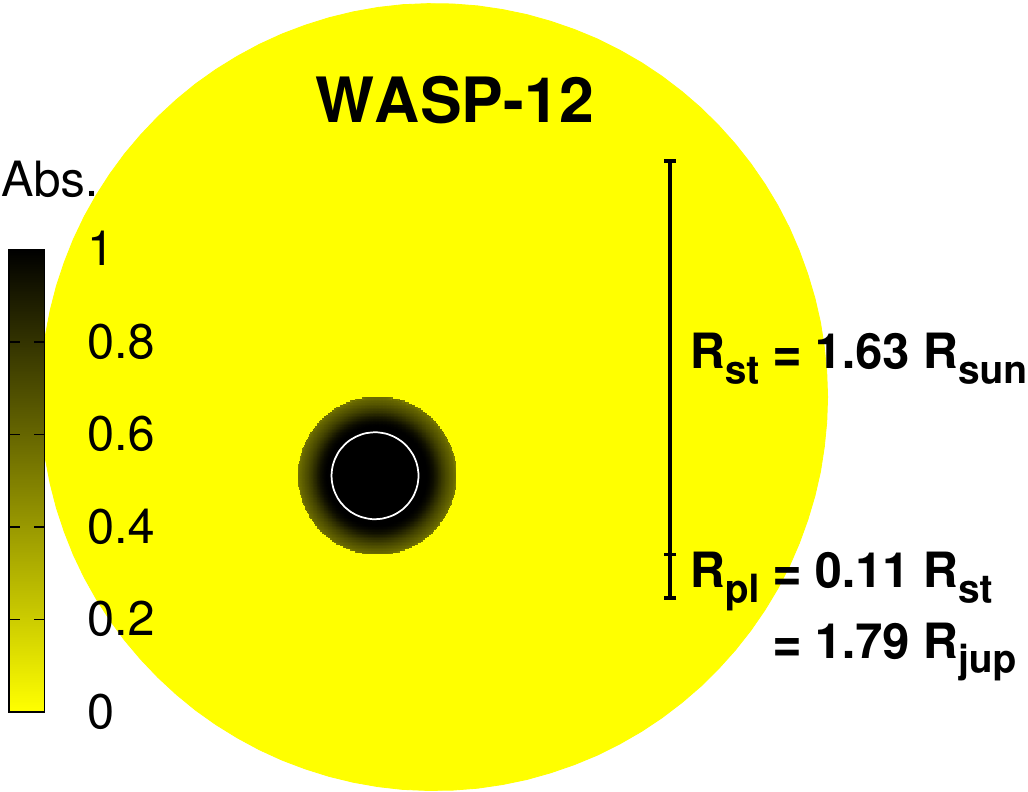}}
    \\ \vspace{5pt}
    \includegraphics[width=0.425\hsize, trim=0cm 0.9cm 0cm 0.9cm, clip=true]{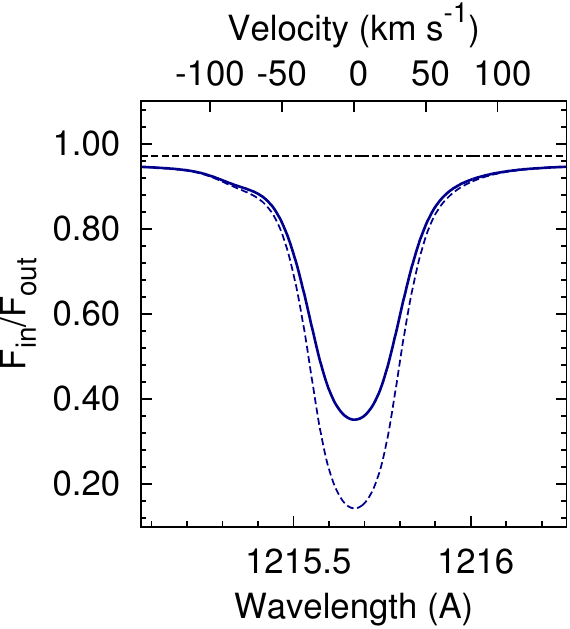}
    \hfill
    \raisebox{2pt}{\includegraphics[width=0.42\hsize]{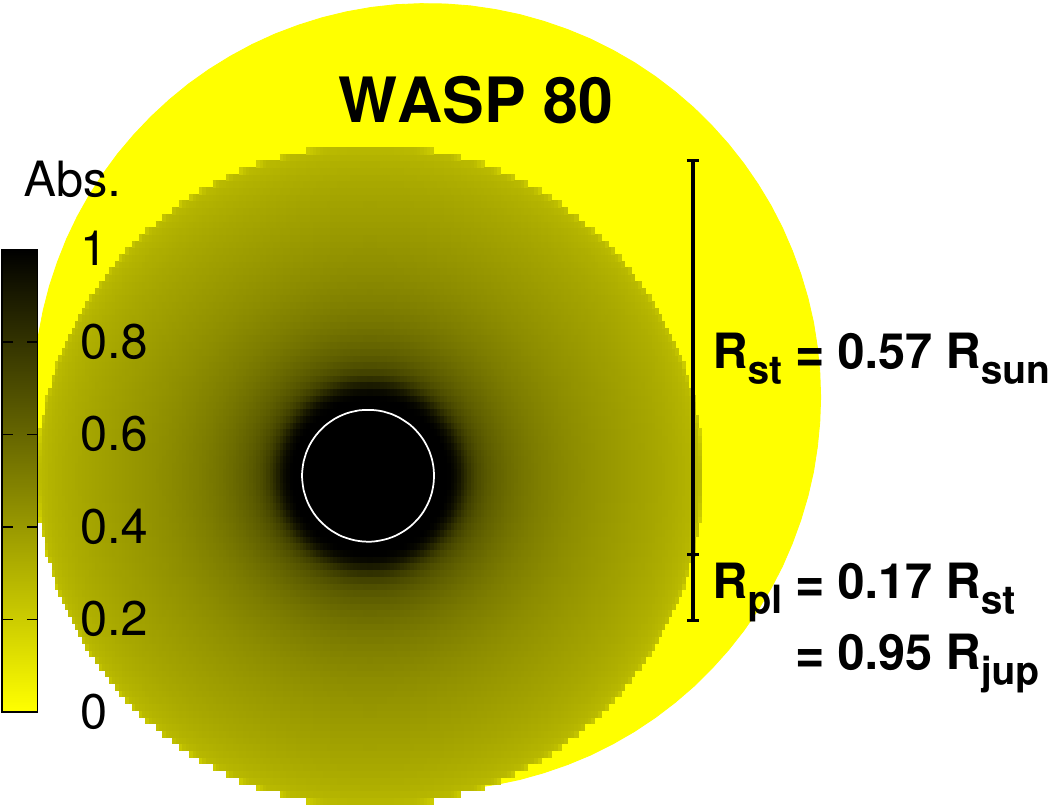}}
    \\ \vspace{5pt}
    \includegraphics[width=0.425\hsize, trim=0cm 0.9cm 0cm 0.9cm, clip=true]{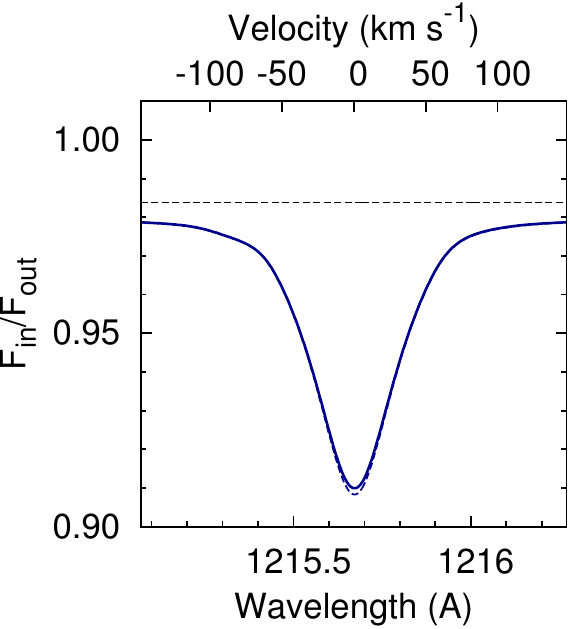}
    \hfill
    \raisebox{2pt}{\includegraphics[width=0.42\hsize]{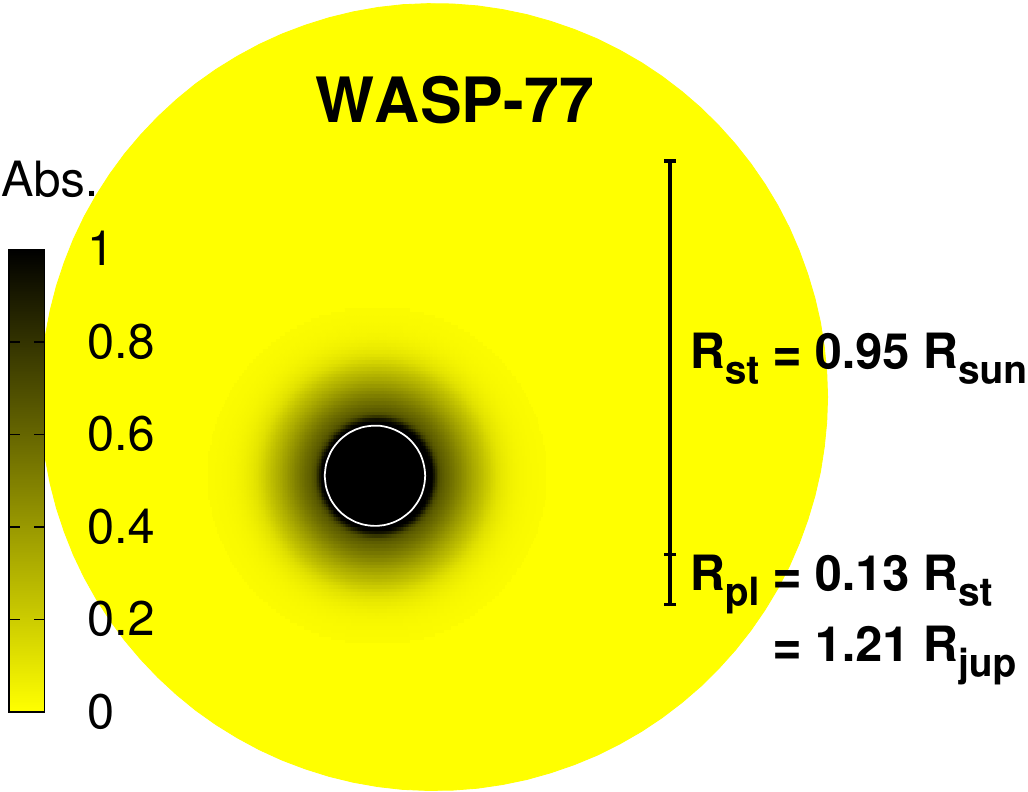}}
    \\ \vspace{5pt}
    \includegraphics[width=0.425\hsize, trim=0cm 0.9cm 0cm 0.9cm, clip=true]{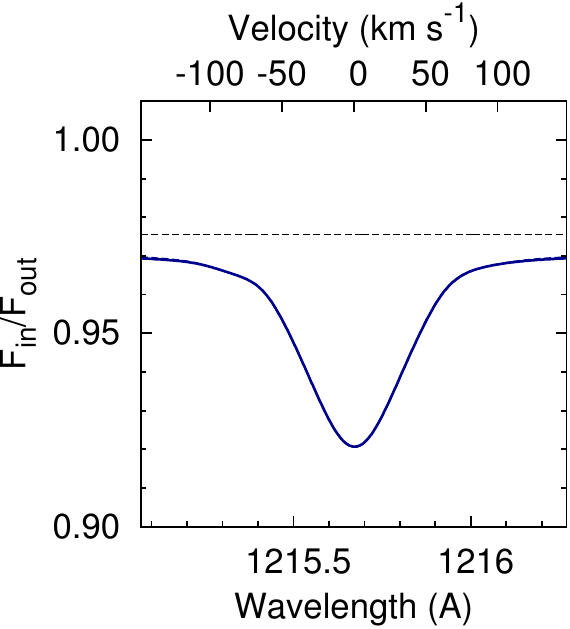}
    \hfill
    \raisebox{2pt}{\includegraphics[width=0.42\hsize]{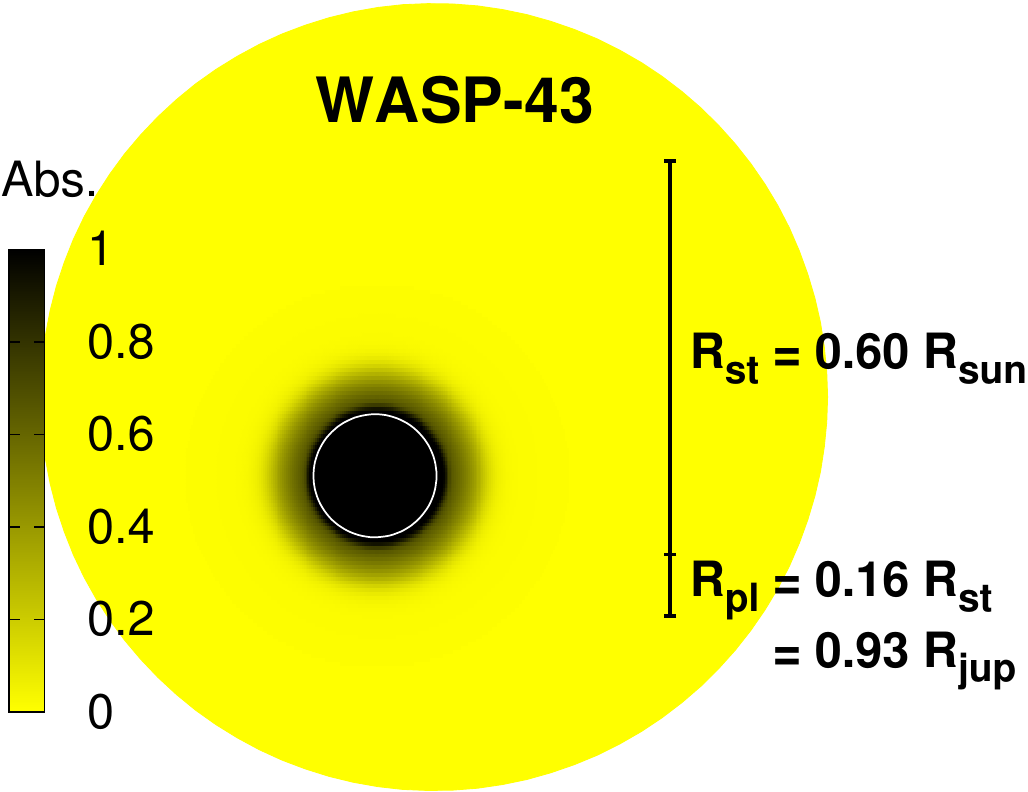}}
    \\ \vspace{5pt}
    \includegraphics[width=0.425\hsize, trim=0cm 0cm 0cm 0.9cm, clip=true]{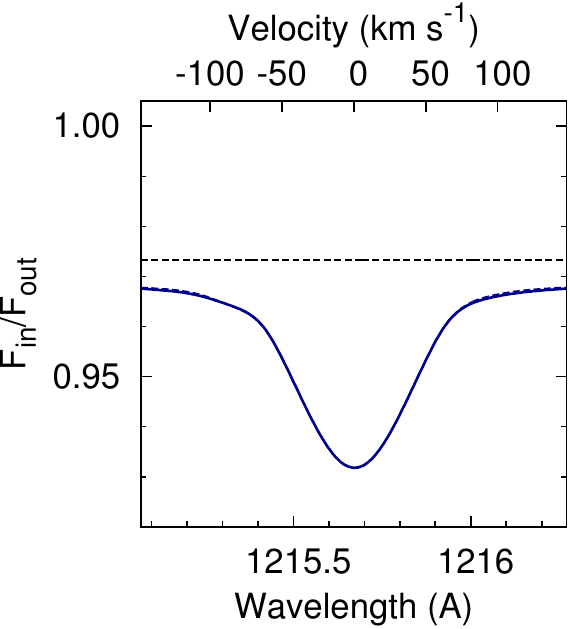}
    \hfill
    \raisebox{18pt}{\includegraphics[width=0.42\hsize]{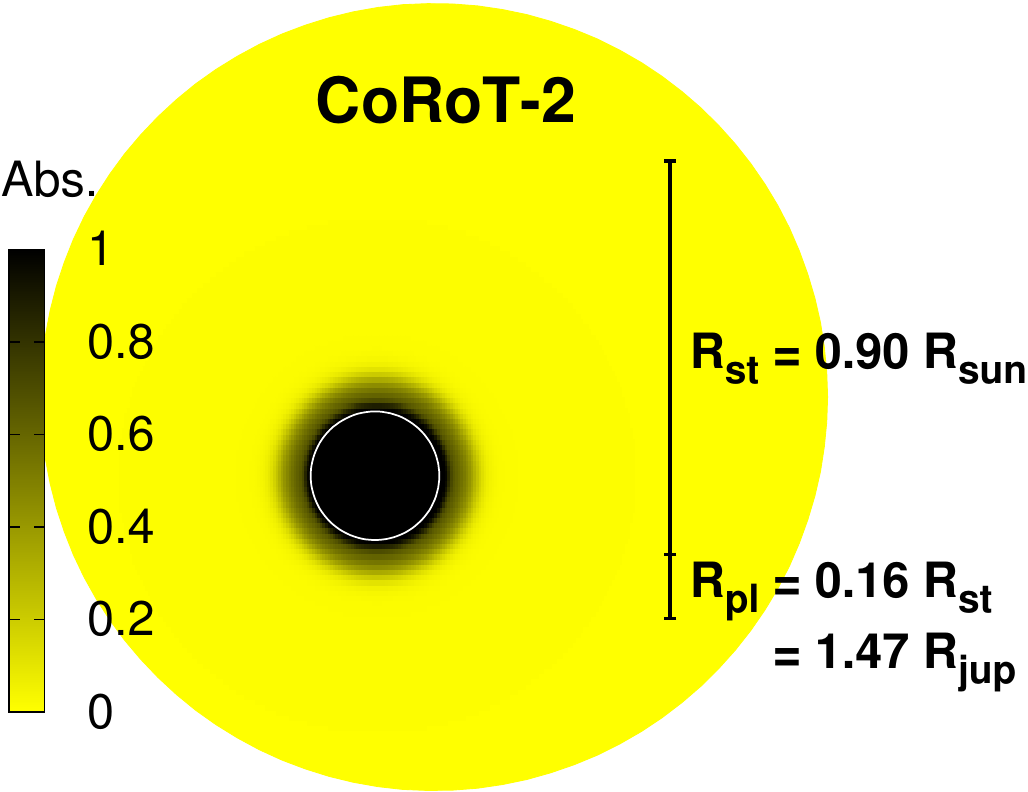}}
    \caption{Simulated \lya{} transmission spectra and images
             for high-potential planets plus
             WASP-80\,b (same as Fig.~\ref{figLyaTransLmass}).
             With the exception of WASP-12\,b (see text) and WASP-80\,b,
             these planets produce small absorption signals.
             WASP-80\,b produces the strongest absorption signal of all planets.
             \newline
             \mbox{}
             }
    \label{figLyaTransHmass}
  \end{minipage}
\end{figure*}

We further fold transmission signal with the line spread function of the 52\arcsec{}\,$\times$\,0.1\arcsec{} aperture of the Space Telescope Imaging Spectrograph on board the Hubble Space Telescope. This reduces the sharpness of the predicted absorption signals and provides results comparable with measurements. The spectral shapes are shown in Figs.~\ref{figLyaTransLmass} \& \ref{figLyaTransHmass}.
Our procedure is similar to the description in \citet{Koskinen2010}, but we do not include interstellar absorption in the calcuations, because the main result is the potential \lya{} absorption of the atmsosphere. For example, interactions with the stellar wind can cause large radial velocity shifts of the absorbing neutral hydrogen, which will result in absorption in the \lya{} line wings rather than in the line center. Such signals are less affected by interstellar absorption.
For visualization we also compute transmission images, which show how the planets would be observed during transit locally in the systems (without interstellar absorption).
To create these images we apply the wavelength dependent transmission to an average stellar \lya{} profile with a double Gaussian shape \citep{Salz2015a} and then integrate the total absorption of stellar \lya{} flux for each ray.
Plotting the individual atmospheric rings over the stellar disk creates the transit images.

\end{appendix}

\end{document}